\documentclass[]{aa}  

\usepackage[export]{adjustbox}[2011/08/13]
\usepackage{acronym}
\usepackage{amsmath}
\usepackage{amssymb}
\usepackage{array}
\usepackage{caption}
\usepackage{color, colortbl}
\usepackage{graphicx}
\usepackage[hidelinks]{hyperref} 
\usepackage{multicol}
\usepackage{multirow}
\usepackage{natbib}
\usepackage{orcidlink}
\usepackage{physics}
\usepackage{pifont}
\usepackage{subcaption}
\usepackage{txfonts}
\usepackage[normalem]{ulem}
\usepackage{placeins}
\newcommand{\cmark}{\ding{51}}
\newcommand{\xmark}{\ding{55}}
\renewcommand{\arraystretch}{1.5}
\usepackage{pifont}
	
\definecolor{Gray}{gray}{0.95}

\hypersetup{
    colorlinks,
    linkcolor={red},
    citecolor={blue},
}

\bibpunct{(}{)}{;}{a}{}{,}

\newcommand{\refapp}[1]{Appendix~\ref{#1}}
\newcommand{\refsec}[1]{Sect.~\ref{#1}}
\newcommand{\reftab}[1]{Table~\ref{#1}}
\newcommand{\refeq}[1]{Eq.~(\ref{#1})}
\newcommand{\reffig}[1]{Fig.~\ref{#1}}
\newacro{BH}{black hole}
\newacro{BNS}{binary neutron star}
\newacro{CE}{Cosmic Explorer}
\newacro{DEC}{declination}
\newacro{EM}{electromagnetic}
\newacro{ET}{Einstein Telescope}
\newacro{EOS}{equation of state}
\newacro{GRB}{gamma-ray burst}
\newacro{GW}{gravitational wave}
\newacro{KN}{kilonova}
\newacro{LVKI}{LIGO-Hanford, LIGO-Livingston, Virgo, KAGRA, LIGO-India}
\newacro{NS}{neutron star}
\newacro{RA}{right ascension}
\newacro{SNR}{signal-to-noise ratio}
\newacro{SVD}{singular value decomposition}
\begin{document} 

   \title{Prospects for optical detections from binary neutron star mergers with the next-generation multi-messenger observatories}

   \author{E. Loffredo\thanks{\email{eleonora.loffredo@inaf.it}}\orcidlink{0000-0001-7888-9733}
         \inst{1,3,2},  
            N. Hazra\thanks{\email{nandini.hazra@gssi.it}}\orcidlink{0000-0002-3870-1537}    
            \inst{1,3,2},
            U. Dupletsa\thanks{\email{ulyana.dupletsa@gssi.it}}\orcidlink{0000-0003-2766-247X}     
            \inst{1,2},
            M. Branchesi \orcidlink{0000-0003-1643-0526}     
            \inst{1,2},            
            S. Ronchini \orcidlink{0000-0003-0020-687X}
            \inst{4,2},
            F. Santoliquido \orcidlink{0000-0003-3752-1400}    
            \inst{1,2},
            A. Perego \orcidlink{0000-0002-0936-8237}
            \inst{5,6},
            B. Banerjee \orcidlink{0000-0002-8008-2485}
            \inst{1,2},
            S. Bisero \orcidlink{0009-0005-6643-1473}
            \inst{7},
            G. Ricigliano \orcidlink{0000-0003-1626-1355}
            \inst{8},
            S. Vergani
            \inst{7},
            I. Andreoni
            \inst{9,10,11,12},
            M. Cantiello \orcidlink{0000-0003-2072-384X}
            \inst{3},
            J. Harms
            \inst{1,2},
            M. Mapelli
            \inst{13},
            G. Oganesyan
            \inst{1,2}.
          }

   \institute{Gran Sasso Science Institute (GSSI), I-67100 L’Aquila, Italy
            \and INFN, Laboratori Nazionali Del Gran Sasso,  I-67100 Assergi, Italy
            \and  INAF – Osservatorio Astronomico d’Abruzzo, 64100 Teramo, Italy 
            \and Department of Astronomy and Astrophysics, The Pennsylvania State University, University Park, PA 16802, USA
            \and Dipartimento di Fisica, Università di Trento, Via Sommarive 14, 38123 Trento, Italy          
            \and INFN-TIFPA, Trento Institute for Fundamental Physics and Applications, Via Sommarive 14, I-38123 Trento, Italy
            \and GEPI, Observatoire de Paris, PSL University, CNRS, Meudon, France.
            \and Institut für Kernphysik, Technische Universität Darmstadt, Schlossgartenstr 2, Darmstadt 64289, Germany
            \and Department of Physics and Astronomy, University of North Carolina at Chapel Hill, Chapel Hill, NC 27599-3255, USA
            \and Joint Space-Science Institute, University of Maryland, College Park, MD 20742, USA
            \and Department of Astronomy, University of Maryland, College Park, MD 20742, USA
            \and Astrophysics Science Division, NASA Goddard Space Flight Center, Mail Code 661, Greenbelt, MD 20771, USA
            \and Institut für Theoretische Astrophysik, ZAH, Universität Heidelberg, Albert-Ueberle-Straße 2, 69120 Heidelberg, Germany
 }

\date{}
 
\abstract
{Next-generation gravitational wave (GW) observatories, such as the Einstein Telescope (ET) and Cosmic Explorer, will observe binary neutron star (BNS) mergers across cosmic history, providing precise parameter estimates for the closest ones. Innovative wide-field observatories, such as the Vera Rubin Observatory, will quickly cover large portions of the sky with unprecedented sensitivity to detect faint transients.} 
{This study aims to assess the prospects for detecting optical emissions from BNS mergers with next-generation detectors, considering how uncertainties in neutron star (NS) population properties and microphysics may affect detection rates, while developing realistic observational strategies by ET operating with the Rubin Observatory.}
{Starting from BNS merger populations exploiting different NS mass distributions and equations of state (EOSs), we modelled the GW and kilonova (KN) signals based on source properties. We modelled KNe ejecta through numerical-relativity informed fits, considering the effect of prompt collapse of the remnant to black hole and new fitting formulas appropriate for more massive BNS systems, such as GW190425. We included optical afterglow emission from relativistic jets consistent with observed short gamma-ray bursts. We evaluated the detected mergers and the source parameter estimations for different geometries of ET, operating alone or in network of current or next-generation GW detectors. Finally, we developed target-of-opportunity strategies to follow up on these events using Rubin and evaluated the joint detection capabilities.} 
{ET as a single observatory enables the detection of about ten to a hundred KNe per year by the Rubin Observatory. This improves by a factor of $\sim 10$ already when operating in network with current GW detectors. Detection rate uncertainties are dominated by the poorly constrained local BNS merger rate, and depend to a lesser extent on the NS mass distribution and EOS.}
{}

\titlerunning{---}
\authorrunning{---}
\maketitle

%%%%%%%%%%%%%%%%%%%%%%%%%%%%%%%%%%%%%%%%%%%%%%%%%%%%%%%%%%%%%%%%%%%%%%%%%%%%%

\section{Introduction}

Coalescencing \acp{BNS} are primary astrophysical sources of \acp{GW} detectable by ground-based detectors. The matter ejected during BNS mergers is neutron-rich and provides the optimal conditions to trigger rapid neutron-capture nucleosynthesis (r-process), giving origin to heavy unstable nuclei \citep[e.g.][]{Eichler1989Natur, Freiburghaus1999ApJ}.
The radioactive decay of these nuclei powers a quasi-thermal optical transient known as \ac{KN}.  
KN emission is intrinsically faint with peak luminosity around $\rm 10^{39}-10^{42}~ erg ~s^{-1}$, rapidly evolving within days, and characterised by a distinct colour evolution from ultraviolet/blue to red/near-infrared (NIR), \citep[e.g.][]{Li1998ApJ, Kulkarni2005astro, Metzger2010MNRAS, Kasen2013ApJ, Barnes2013ApJ, Grossman2014MNRAS, Metzger2019}.
The epochal discovery of the GW signal from the BNS merger GW170817 and the detection of its associated optical counterpart \citep{Abbott2017PhRvLgw170817, Abbott2017ApJgw170817}, AT2017gfo, provided the first direct evidence that BNS mergers produce KN emission and firmly demonstrated that these events are one of the main channels of heavy element formation in the Universe \citep[e.g.][]{Pian2017, Smartt2017Natur}. The detection of AT2017gfo made it possible to identify the BNS host galaxy. The galaxy recession velocity used along with the distance to the source estimated from the GW signal provided a new way to estimate the local expansion rate of the Universe \citep{LVK2017Nature}.
The \ac{EOS} of dense matter has a direct imprint on the \ac{GW} signal, as well as on the properties of the ejected material. The latter eventually shapes the KN emission making the joint observation of \acp{GW} and KN signals a powerful tool for unveiling the nature of matter in \acp{NS} \citep[see e.g.][]{Margalit2017,Dietrich2020Sci,Breschi2021MNRAS,Breschi:2024qlc}.
Several factors played a key role in the successful detection of AT2017gfo and its identification as the counterpart of the GW signal; the precise localisation in the sky ($\rm \sim 28~{\rm deg^2}$) provided by the LIGO and Virgo network, the relatively small distance (40 Mpc) that enabled to identify candidate host galaxies% \citep{Cantiello2018ApJ}
, the timely and effective follow-up by ultraviolet, optical and NIR telescopes, and finally the extensive photometric and spectroscopic observational campaign to characterize the candidate KN. The characterization campaign identified distinct colour evolution and broad absorption features consistent with the presence of neutron-capture elements.  

In 2019, a second GW signal from the inspiral phase of a BNS merger was detected, named GW190425 \citep{LVK2020GW190425}. Despite the extended follow-up efforts, no \ac{EM} counterparts were identified mainly due to the poorly constrained sky localisation of the event ($\simeq 10^4~{\rm deg^{2}}$).
This event was particularly interesting in itself because of its chirp mass ($1.44\pm0.02 M_\odot$) and total mass between $3.3 M_\odot $ and $3.7 M_\odot$, which are significantly larger than that of any other BNS system in the Milky Way. Several theoretical studies have since explored the expected EM emission from events similar to GW190425, either considering high-mass BNS or neutron star-black hole (NS-BH) binaries with a low-mass \ac{BH} \citep{Kyutoku2020ApJ, Raaijmakers2021ApJ, Barbieri2021A&A, Camilletti2022MNRAS, Dudi2022PhRvD, Radice2024MNRAS}.

Although GW170817 has shown the great potential and impact that such multi-messenger events have on several fields of physics, from relativistic astrophysics to cosmology and nuclear physics, it remains a one-occurrence event. No significant low-latency BNS candidates have been reported in the fourth run of LIGO, Virgo and KAGRA \citep{Akutsu2019CQGra} (LVK) observations. However, it is noteworthy to report that two long \acp{GRB}, GRB 211211A \citep{Rastinejad2022,Troja2022,Mei2022} and GRB 230307A \citep{Levan2024,Yang2024}, at a distance within the reach of the current GW detectors, were detected showing KN emission signatures that demonstrated their origin from the coalescence of NS-NS or NS-\ac{BH} binaries.
They were observed while GW detectors were undergoing upgrades.

Upcoming observational campaigns of the Advanced GW detectors LIGO, Virgo, KAGRA, and LIGO-India \citep{Bailes2021} are expected to detect between a few up to several hundred BNS mergers per year up to redshift $z \sim 0.2$ \citep{Abbott2020LRR_senscurves, Petrov2022}.
The next-generation GW observatories, such as \ac{ET} and \ac{CE}, are expected to significantly increase the detection rates, identifying up to $10^5$ BNS mergers per year, and expand the observable horizon to $z \sim 4$ with \ac{ET}  \citep{Branchesi2023} and $z \sim 10$ with \ac{CE} \citep{Evans2021,Evans2023,Gupta2023}, respectively. Importantly, \ac{ET} and \ac{CE} will offer precise estimates of source parameters, including sky localisation, luminosity distance, and inclination angle \citep{Grimm2020PRD, Evans2021, Branchesi2023}.
In the optical and NIR bands, an unprecedented observatory, the Vera Rubin Observatory \citep{ivezic19}, is expected to start collecting data in the Fall of 2025. Its large field of view, very fast slewing time and deep sensitivity provide unique capabilities to detect KNe coincident with \acp{GW} coming from increasingly distant GW sources, which will still be poorly localised compared to observatories with arc-minute field of view.
Additionally, innovative sensitive spectroscopic telescopes, such as the Extremely Large Telescope \citep[ELT,][]{Liske2014tmt, Marconi2022SPIE}, will enable us to characterize well-localised counterparts and the Wide-field Spectroscopic Telescope project \citep[WST,][]{Mainieri2024arXiv240305398M} to identify and characterize GW source counterparts.

Optimal strategies for KNe detections using the Vera Rubin Observatory have been explored by considering either serendipitous KN discovery \citep{Scolnic2018ApJ,Bianco2019PASP,Setzer2019MNRAS,Andreoni2019PASP,andreoni22_serend,Ragosta2024} or the more effective target-of-opportunity (ToO) searches triggered by GW events detected by the current generation of GW detectors \citep{Margutti2018arXiv,Cowperthwaite2019ApJ,andreoni22_too}. 
Rubin's ToO strategy to follow-up GW signals detected by LVK has been recently defined and approved\footnote{The official documentation from the Rubin Survey Cadence Optimizing Committee is available at \href{https://pstn-056.lsst.io/}{this link}.}. Predictions on multi-messenger detections of KNe and short GRBs associated with BNS mergers have been evaluated over several spectral bands for the LVK O4 and O5 runs by \cite{Colombo2022ApJ, Patricelli2022, Petrov2022, Bhattacharjee2024, Shah2024}. \cite{Sarin2022PASA} evaluated the prospects for KN counterpart detections by Rubin observing together with the newly proposed Neutron Star Extreme Matter Observatory (NEMO), which is a high-frequency interferometer proposed to be operative in the late 2020s and early 2030s in Australia \citep{Ackley2020PASA}. 
Moving on to the next generation of \ac{GW} detectors, studies on multi-messenger perspectives have been performed considering high-energy satellites \citep{Ron2022A&A...665A..97R, Hendriks2023}, very-high-energy observatories using pre-merger alerting \citep{Banerjee2023}, and optical telescopes 
\citep{chen21, Branchesi2023}. In particular, \cite{chen21} explored the prospects of ToO search by Rubin in synergy with Advanced LIGO plus, Voyager, and CE for multi-messenger cosmology, while \cite{Branchesi2023} evaluated the prospects as part of a study to compare different ET designs. All the above studies focusing on the optical band used a KN modelling based on the AT2017gfo observation and numerical relativity simulations calibrated on the BNS properties mainly extracted by the GW170817 signal. 

In this paper, we used a population synthesis code to build catalogues of BNS mergers covering ten years of observations, providing a sufficiently large statistical sample to cover the full parameter space and avoiding being biased by small statistics. The catalogues were calibrated to reproduce the current GW observations. We then evaluated the GW detection and parameter estimation capabilities considering several scenarios where ET operates as a standalone observatory or as part of a network of GW detectors, including current and next-generation detectors and taking into account two possible ET designs, triangular shape and 2L, as described in \refsec{sec:GWsimul}. 
We present an improved modelling of KN signals for population studies by extending state-of-the-art fitting formulas to cover newly explored regimes. Specifically, we incorporated the results of GW190425-targeted BNS merger numerical simulations and accounted for the impact of the prompt collapse of the merger remnant into a \ac{BH} on the ejecta.
We added the emission expected from the relativistic jet of short GRBs.
Finally, we simulated Rubin's observing scenarios that are as realistic as possible, taking into account the Rubin accessible sky, slewing time, filters, seeing, and the night-day duty cycle.
We assess the impact of different GW detector networks on the number of KN multi-messenger detections, as well as the influence on the results of our limited knowledge of the BNS merger rate, NS mass distribution, and NS \ac{EOS}.

The paper is structured as follows. \refsec{sec:ModelAnalysis} describes the details of our modelling and simulations, from the population of BNS mergers to the modelling of GW and \ac{EM} emissions. It also includes a description of the analysed GW networks and the methodology followed to estimate the number of detections and the parameters of the sources. \refsec{sec:Rubin} describes Rubin's observational strategies to follow up the GW events. \refsec{sec:ResDisc} presents the results for the GW detections (\refsec{sec:GWresults}) and the joint optical/GW detections discussing them in terms of joint detection efficiency (\refsec{sec:efficiency}), 
different GW networks and ET designs (\refsec{sec:network} and \refsec{sec:ETdesign}), uncertainty on BNS merger rate, NS mass distribution, EOSs, and different observational strategies (\refsec{sec:impacts}). \refsec{sec:ResDisc} also presents a comparison between the results obtained with the new generation GW detectors and possible future scenarios of the current detector network and their upgrade (\refsec{sec:current}). Finally, \refsec{sec:ResDisc} provides an assessment of the gain from using deeper Rubin observations (\refsec{sec:deeperexp}). \refsec{sec:summary} summarises the results and draws conclusions. Throughout the article, we use a $\Lambda$CDM cosmology with Hubble constant $\rm H_0 = 67.66\, {\rm km}\, {\rm s}^{-1}\, {\rm Mpc}^{-1}$ and present matter fraction $\rm \Omega_m = 0.31$, from {\em Planck-2018} \citep{Planck:2018vyg}.

\section{Modelling and simulations}
\label{sec:ModelAnalysis}
\subsection{Population of binary neutron star mergers}\label{sec:population}
The population of BNS mergers was generated using the population-synthesis code {\sc{sevn}}\footnote{{\sc{sevn}} is publicly available at \href{https://gitlab.com/sevncodes/sevn}{this link}.} \citep{Spera2019MNRAS, Mapelli2020ApJ, Iorio2023MNRAS}. 
 {\sc{sevn}} combines the evolution of single stars and binary systems by interpolating a set of pre-calculated tracks of stellar evolution
and uses analytic and semi-analytic formalism for the binary processes.
To obtain the BNS merger rate density as a function of redshift, 
we used the {\sc{cosmo$\mathcal{R}$ate}} code \citep{Santoliquido2020ApJ}, which adopts the star formation rate density and average metallicity evolution of the Universe from \cite{Madau2017ApJ}, along with a metallicity spread of $\sigma_{Z} = 0.2$.

In this study, we adopted the fiducial model where all {\sc{sevn}} parameters are set to their default values, as described in \cite{Iorio2023MNRAS}. To determine whether a core-collapse supernova produces a \ac{NS} or a \ac{BH}, we adopted the rapid supernova model as in \cite{Fryer2012ApJ}.  
We drew \ac{NS} natal kicks following the formalism by \cite{Giacobbo2020ApJ}:
\begin{equation}
V_{\rm kick}=f_{\rm H05}\,{}\frac{\langle{}M_{\rm NS}\rangle{}}{M_{\rm rem}}\,{}\frac{M_{\rm ej}}{\langle{}M_{\rm ej}\rangle},
\end{equation}   
where $\langle M_{\rm NS} \rangle$ and $\langle M_{\rm ej} \rangle$ are the average \ac{NS} mass and ejecta mass from single stellar evolution, respectively, while $M_{\rm rem}$ and $M_{\rm ej}$ are the remnant compact object mass and the ejecta mass. The term $f_{\rm H05}$ is a random number drawn from a Maxwellian distribution with one-dimensional root mean square $\sigma_{\rm{kick}} = 265$ km s$^{-1}$ \citep{Hobbs2005MNRAS}.

As shown in \cite{Santoliquido2021MNRAS}, the common envelope ejection efficiency parameter, $\alpha$, is one
of the main sources of uncertainty in the determination of the number of BNS mergers per year.
In order to evaluate the impact of the uncertainty on the BNS merger rate normalization 
on our results, we generated two catalogues of BNS mergers assuming $\alpha = 0.5$ and $1.0$. 
The mass of the progenitor primary star was drawn from a Kroupa mass function \citep{Kroupa2001MNRAS}, in the mass range [5 - 150] $M_\odot$.
The initial mass ratios, orbital periods, and eccentricities were determined based on the distributions inferred by \cite{Sana2012Science}.

Under these assumptions, the total population consists of approximately $[0.7,~ 3.6] \times 10^4$ BNS mergers at redshifts $z < 1$ per year (as observed from Earth by an ideal instrument), corresponding to $\alpha = 0.5$ and $1.0$, respectively. The local BNS merger rates are $\mathcal{R}_{\rm{BNS}} = [23,~ 107]$ Gpc$^{-3}$ yr$^{-1}$ for $\alpha = 0.5$ and $1.0$, respectively. These rates are in agreement with the most recent 90\% credible interval provided in GWTC-3 by the Advanced LIGO and Virgo detectors ($\mathcal{R}_{\rm{BNS}} \in [10 - 1700]$ Gpc$^{-3}$ yr$^{-1}$, \citealt{Abbott2023popGWTC3}) and the current absence of firm detections of BNS signals during the fourth run of observations which began in May 2023 and is ongoing. The population corresponding to $\alpha = 1.0$ is considered our fiducial population, while the one obtained with $\alpha = 0.5$ is consistent with the lower limit of the BNS merger rate and is named through the paper as pessimistic population.

We drew the component masses of the \ac{NS} binaries, $M_1$ and $M_2$, 
from two different mass distributions: Gaussian and uniform. The Gaussian distribution is centred at $1.33~M_\odot$ with a standard deviation of 0.09 $M_\odot$. This model is derived from a fit to the masses of NSs in binary systems in the Milky Way \citep{Ozel2012ApJ, Kiziltan2013ApJ, Ozel2016ARAA}. The uniform mass distribution ranges in $[1.1 M_\odot,~ M_{\rm max}]$, where $M_{\rm max}$ depends on the selected \ac{EOS} (see \refsec{sec:NS_EOS}). This model is more consistent with the \ac{GW} observations which favour a broad distribution for the NS mass in binaries \citep{Abbott2023popGWTC3}. The tidal deformabilities for each component of the BNS, $\Lambda_1$ and $\Lambda_2$, were assigned based on the chosen \ac{EOS} and binary masses (see ~\refsec{sec:NS_EOS}). We assumed non-spinning \acp{NS}, since \acp{NS} in binary systems are expected to carry negligible spins \citep{Burgay2003Nat}.

The BNS mergers are distributed isotropically in the sky, and the inclination of the orbital plane with respect to the line of sight is randomly distributed. This was obtained uniformly sampling \ac{RA} in the interval $[0, 2\pi]$ and \ac{DEC} in cosine in $[-\pi/2,+\pi/2]$. The inclination angle, $\iota$, follows a uniform in sine distribution between $[0,\pi]$. The polarization, $\Psi$, and phase angles are both uniformly distributed in the range $[0, 2\pi]$. To have enough statistics and appropriately sample all the parameter space of the BNS merger properties and their \ac{GW} and \ac{EM} signals, we built and analysed catalogues of BNS mergers observed in 10 years from Earth by an ideal instrument.

\subsection{Neutron star equations of state}\label{sec:NS_EOS}
Since the \ac{NS} \ac{EOS} affects both the \ac{GW} and \ac{EM} signals expected from BNS mergers, we considered two different \acp{EOS}, namely the APR4 \citep{Akmal1998PhysRevC, Baym1971ApJ, Douchin2001A&A} and BLh \citep{Bombaci2018AAP, Logoteta2021AAP} microscopic \acp{EOS}. The APR4 and BLh \ac{EOS}s are based on ab initio calculations; specifically, APR4 is obtained within the variational approach, while BLh is constructed in the framework of the Brueckner-Bethe-Goldstone many-body theory \citep[e.g.][]{DayRevModPhys.39.719}. Although both the \acp{EOS} are fully compatible with the present astrophysical constraints on the \ac{NS} maximum mass, the \ac{NS} radius, and the dimensionless tidal deformability as derived from GW170817 \citep{Antoniadis2013Science, Lattimer2014APJ, Oertel2017RvMP, LIGO2018PhRvL_eos, Riley2019ApJ, Miller2019ApJ, Raaijmakers2021APJL, Miller2021ApJL}, they bracket present uncertainties in the predicted NS compactness.

APR4 and BLh predict $M_{\rm max} = 2.2M_\odot$ and $M_{\rm max} = 2.1 M_\odot$ for the maximum mass $M_{\rm max}$ of a cold non-rotating \ac{NS}, respectively. These values were used to set the maximum mass for our populations (see \refsec{sec:population}).
While the maximum masses values are similar, APR4 produces significantly more compact \acp{NS}, with larger compactness 
\begin{equation}\label{eq:compactness}
    \mathcal{C} = \dfrac{G~M_{\rm NS}}{c^2~R_{\rm NS}}
\end{equation}
and smaller dimensionless tidal deformability
\begin{equation}\label{eq:NS_tidal_def}
    \Lambda = \dfrac{2}{3} k_2 \left(\dfrac{c^2 R_{\rm NS}}{G M_{\rm NS}}\right)^5
\end{equation}
compared to BLh, where $k_2$ is the second tidal Love number \citep{Hinderer2008ApJ, Damour2012PrD, Favata2014PhRvL} and $R_{\rm NS}$ is the NS radius. These differences influence the amount of ejected mass, which is typically larger for BLh than APR4, changing the \ac{KN} emission properties (see \refsec{sec:ejecta_properties} for details).
In particular, APR4 and BLh differ in the behaviour of the nuclear incompressibility, $K$, which describes the response of nuclear matter pressure to baryon density variations. \cite{Perego2022PhRvL} show that the nuclear incompressibility at the maximum Tolman–Oppenheimer–Volkoff (TOV) baryon density, $K_{\rm max}$, together with the maximum TOV mass, influences the threshold for prompt collapse for arbitrary mass ratios. In \refsec{sec:kn_modeling}, we show how our modelling of the \ac{KN} emission is sensitive to this difference. In \reftab{tab:eos_prop}, we summarise the properties of the two \acp{EOS} relevant to our analysis.\\
Given a BNS with masses $M_{1,2}$ and EOS-dependent tidal deformabilities $\Lambda_{1,2}$, we computed the reduced tidal deformability of the binary as
\begin{equation}
    \tilde{\Lambda} = \dfrac{16}{3} \dfrac{(M_1 + 12 M_2) M_1^4 \Lambda_1 + (M_2 + 12 M_1) M_2^4 \Lambda_2}{(M_1 + M_2)^5} .
\end{equation}
The distributions of the properties (component mass, mass ratio, chirp mass, and tidal deformability) of the BNS populations obtained for the two considered EOSs are shown \refapp{app:A}.  

\begin{table}
  \caption[]{Properties of the \acp{EOS} used in this work and the corresponding cold, spherically symmetric \acp{NS}.}
  \label{tab:eos_prop}
  \centering
  \begin{tabular}{ccccccc}
  \hline \\ [-1.5ex]
  EOS & $K_{\rm max}$ & $M_{\rm max}$ & $R_{\rm max}$ & $\mathcal{C}_{\rm max}$ & $R_{1.4}$ & $\Lambda_{1.4}$\\
    & [GeV] & [$M_\odot$] & [km] & [-] & [km] & [-]\\
  \hline \\ [-1.5ex]
  APR4 & 28.88& $2.20$ & $9.92$ & $0.328$ & $11.12$ & $256.81$\\
  BLh   & 17.20& $2.10$ & $10.46$ &  $0.297$ & $12.43$  & $431.22$\\
  \hline \\ [-1.5ex]
  \end{tabular}
  \tablefoot{$K_{\rm max}$ is the \ac{EOS} incompressibility corresponding to the maximum TOV baryon density, while $M_{\rm max}$, $R_{\rm max}$, and $C_{\rm max}$ are the gravitational mass, radius, and compactness of the maximum TOV \ac{NS}. $R_{1.4}$ and $\Lambda_{1.4}$ are the radius and the dimensionless tidal deformability for a \ac{NS} mass of $1.4~ M_{\odot}$.}
\end{table}

\subsection{Gravitational-wave simulation}
\label{sec:GWsimul}
In order to analyse the detection prospects for the \ac{GW} signals emitted by the populations of BNS mergers described in \refsec{sec:population}, we considered different detector networks including the current generation and the next generation of \ac{GW} detectors. In the following, we provide details about the different sets of GW injections, the different combinations of \ac{GW} detectors, and how the signal analysis was carried out.

Starting from the populations of ten years BNS mergers described in \refsec{sec:population}, we analysed all the mergers up to $z=1$ and we associated to each merger a coalescence time, $t_{\rm c}$, uniformly sampled in the years between the year 2035 and the year 2045, which is roughly the period when the next generation of ground-based observatories is expected to be operative. For each BNS merger, we injected a \ac{GW} signal described by IMRPhenomD\_NRTidalv2~\citep{Khan:2015jqa, Dietrich:2019kaq}, a state-of-the-art waveform including tidal effects. 

Given the two values of $\alpha$ ($0.5$ and $1.0$), the two mass distributions (uniform and Gaussian), and the two EOSs (BLh and APR4), we have a total of eight different population sets, which constitute our injections for the gravitational signal analysis. For each of these datasets, we considered the following \ac{GW} detector configurations:
\begin{enumerate}
    \item \ac{ET} in its triangular design of 10\,km arms, located in Sardinia, alone or operating together with:
    \begin{itemize}
        \item the current ground-based network \ac{LVKI};
        \item one L-shaped \ac{CE} with $40$\,km arms, located in the USA;
        \item 2 \acp{CE}, both with $40$\,km arms, one in the USA and one in Australia;
    \end{itemize}
        \item \ac{ET} in its 2L-shaped interferometer configuration of 15\,km arms misaligned at 45\,deg \citep{Branchesi2023} (one located in Sardinia and the other in the Netherlands); we considered the same networks as above, using the 2L-configuration instead of the triangular one.
\end{enumerate}

We thus have eight different detector networks (see \reftab{tab:networks_inj}), giving a total of 64 simulations to be performed. For \ac{ET}, in both triangular and 2L-configuration, we used the sensitivity curve corresponding to the xylophone configuration which consists of two interferometers, one tuned towards high-frequencies and the other tuned towards low-frequencies working at cryogenic temperatures \citep{Branchesi2023}\footnote{The sensitivity curves are available at \href{ https://apps.et-gw.eu/tds/ql/?c=16492}{this link}.}. For \ac{CE}, we used the sensitivity curve given by \cite{Evans2021} for the 40 km detector. For the current ground-based interferometers (\ac{LVKI}), we used the optimal sensitivity curves expected for the future fifth run of observations (O5)  \citep{Abbott2020LRR_senscurves}. The starting frequency for \ac{ET}, both triangle and 2L, is at $2$\,Hz, whereas all the other detectors start from $8$\,Hz\footnote{It is important to note that while the data analysis extended to the mentioned starting frequencies, it does not imply that the detectors are sensitive down to these thresholds. \reffig{fig:sens_curves} in \refapp{app:GW_detectors} shows the rapid decrease of sensitivity at lower frequencies, especially for current detectors.}. We report all the main characteristics of the detectors employed in this work and we show the corresponding sensitivity curves in \refapp{app:GW_detectors}.
\begin{table}[ht]
\caption{List of detector networks and injections data sets used in the present analysis.}
\centering
{\small
\begin{tabular}{l || r r c c}
\hline
\textbf{NETWORKS} &\multicolumn{4}{c}{\textbf{INJECTION SETS}}\\
&$\boldsymbol{\alpha}$ &\textbf{$\#$ events} &\textbf{EOS} &\textbf{Mass distr.}\\
\hline
\textbf{ET-$\Delta$}  &\multirow{4}{*}{$0.5$} &\multirow{4}{*}{$71338$} &\multirow{2}{*}{APR4} &Gaussian \\
ET-$\Delta$ \textbf{+} LVKI  &\multirow{4}{*}{} &\multirow{4}{*}{} &\multirow{2}{*}{} &Uniform \\
ET-$\Delta$ \textbf{+} 1CE  &\multirow{4}{*}{} &\multirow{4}{*}{} &\multirow{2}{*}{BLh} &Gaussian \\
ET-$\Delta$ \textbf{+} 2CE &\multirow{4}{*}{} &\multirow{4}{*}{} &\multirow{2}{*}{} &Uniform \\
\textbf{ET-2L}  &\multirow{4}{*}{$1.0$} &\multirow{4}{*}{$363560$} &\multirow{2}{*}{APR4} &Gaussian \\
ET-2L \textbf{+} LVKI &\multirow{4}{*}{} &\multirow{4}{*}{} &\multirow{2}{*}{} &Uniform \\
ET-2L \textbf{+} 1CE  &\multirow{4}{*}{} &\multirow{4}{*}{} &\multirow{2}{*}{BLh} &Gaussian \\
ET-2L \textbf{+} 2CE  &\multirow{4}{*}{} &\multirow{4}{*}{} &\multirow{2}{*}{} &Uniform \\
\hline
\end{tabular}
}
\tablefoot{ We considered eight different detector networks and eight different injection data sets for a total of 64 simulations. For each network, an injection set was obtained by a combination of choices of common envelope efficiency parameter $\alpha$, NS \ac{EOS}, and NS mass distribution, yielding eight variations for each choice of network. The number of ejected events refers to the BNS mergers up to redshift $z=1$ observed in ten years
from Earth by an ideal instrument.}
\label{tab:networks_inj}
\end{table}

The parameter estimation of the injected GW signals by the various detector networks was obtained through the Fisher matrix software \texttt{GWFish}\footnote{\texttt{GWFish} is publicly available on \href{https://github.com/janosch314/GWFish}{GitHub}.} \citep{Dupletsa2023AaC}. The Fisher analysis method approximates the likelihood with a multivariate Gaussian distribution, which is valid in the high \ac{SNR} limit. \texttt{GWFish} simulates various detector networks in both the time and frequency domains, relying on all the available waveforms from \texttt{LALSimulation} \citep{LIGO2018lalsuite}. It takes into account the motion of the Earth, which improves the sky-localisation capabilities for long-lasting signals, such as the ones from BNS mergers.

For our analysis, a detection SNR threshold of 8 was applied. All the parameters [${\mathcal{M}_c}$, ${q}$, $d_L$, $\iota$, RA, DEC, $\Psi$, phase, $t_{\rm c}$, $\Lambda_1$, $\Lambda_2$] were considered for the Fisher matrix derivation, where $\mathcal{M}_c$ and $q$ are combinations of $M_1$ and $M_2$, called respectively chirp mass and mass ratio. The uncertainties on parameters coming from the covariance matrix (the inverse of the Fisher matrix) are given at $1\sigma$. The sky-localisation uncertainty, $\Omega_{90}$, is given as $90\%$ credible region\footnote{The sky-localisation region at $1\sigma$ was obtained using the uncertainties on RA and DEC. To obtain the $X\%$ level of sky localisation one should multiply the value obtained by \texttt{GWFish} by $2 \ln \left(1-X/100 \right)$. This results in a multiplicative factor of $\sim 4.6$ when considering $90\%$ confidence level.}.
We implemented a duty cycle of $85\%$ for each of the L-shaped detectors, and in each of the three nested detectors composing the triangle, so that we take into consideration a more realistic scenario, where detectors are not operative all the time as occurs with current \ac{GW} detectors. The results of the \ac{GW} simulations obtained with \texttt{GWFish} are given in \refsec{sec:GWresults} and are used as input for the \ac{KN} analysis as described in the following sections.

\subsection{Kilonova emission modelling}
\label{sec:kn_modeling}
To model the \ac{KN} emission and its evolution with time (light curves), we assumed two-component ejecta. The first component (C1) comprises dynamically ejected matter (dynamical ejecta) and ejections via spiral-wave winds \citep{Nedora2019ApJ,Nedora:2020hxc}. These ejecta are expected to be fast and to show a significant degree of anisotropy due to neutrino irradiation, especially in the presence of a long-lived remnant. The second component (C2) involves matter unbound from the disc (secular ejecta), expelled on longer timescales and characterised by more uniform property distribution, especially if a BH has formed in the centre \citep{Wu2016MNRAS, Siegel2017PhRvL}. Using the masses $M_1$ and $M_2$ and the \ac{EOS} of the simulated binaries, we calculated the properties of the ejected matter using numerical relativity-informed fitting formulas, whose details are provided in \refsec{sec:ejecta_properties}. Since the characteristics of the ejected matter are significantly affected by the collapse time of the merger remnant, our model includes the possibility of prompt collapse to a \ac{BH} by defining a prompt-collapse mass threshold $M_{\rm PC}$.
In \refsec{sec:ejecta_properties}, we discuss how we compute $M_{\rm PC}$ and how we model the effect of prompt collapse on the ejecta properties. 
Based on this input, we defined the properties of the two ejecta components.
Finally, we computed \ac{KN} light curves using the  \texttt{xkn} code\footnote{The \texttt{xkn} framework is publicly available on  \href{https://github.com/GiacomoRicigliano/xkn}{GitHub}.} \citep{Ricigliano2024MNRAS}.
We describe the input parameters and the setup of \texttt{xkn} for our \ac{KN} light curve modelling in \refsec{sec:xkn}.

\subsubsection{Ejecta properties}\label{sec:ejecta_properties} 
Our BNS merger populations, covering the masses of GW170817 and GW190425-like events, explore a wide range of the $M_1 - M_2$ parameter space and include asymmetric and massive binaries. This large $M_1-M_2$ space yields two scenarios based on the total mass $M_{\rm tot} = M_1 + M_2$ of the binary system. While binaries with $M_{\rm tot}$ exceeding $M_{\rm PC}$ undergo prompt collapse, those with $M_{\rm tot} < M_{\rm PC}$ do not.
To compute the amount and velocity of the dynamical ejecta, $m_{\rm dyn}$ and $v_{\rm dyn}$, and the mass of the remnant disc, $m_{\rm disc}$, we modelled these two scenarios, named as PC and non-PC, and we smoothly interpolated between them through the function
\begin{equation}
    T = 0.5 + 0.5 \tanh\left(~2\pi~(M_{\rm tot}-M_{\rm PC})~\right).
\end{equation}
Specifically, we adopted the following prescriptions:
\begin{equation}\label{eq:mdyn}
    m_{\rm dyn} = \left(1 - T \right)~m_{\rm {dyn,~non-PC}} + T~m_{\rm dyn,~PC} \, ,
\end{equation}
\begin{equation}\label{eq:vdyn}
    v_{\rm dyn} = \left(1 - T\right)~v_{\rm {dyn,~non-PC}} + T~v_{\rm dyn,~PC} \, ,
\end{equation}
\begin{equation}\label{eq:mdisc}
    m_{\rm disc} = \left(1 - T \right)~m_{\rm {disc,~non-PC}} + T~m_{\rm disc,~PC} .
\end{equation}
We determined $M_{\rm PC}$ using the fitting formulas provided in \cite{Perego2022PhRvL} and \cite{Kashyap2022PhRvD}, which take into account the properties of the \ac{EOS}, specifically the maximum compactness, maximum TOV mass, and maximum incompressibility, as well as the binary mass ratio. \cite{Perego2022PhRvL} demonstrate that the prompt-collapse mass threshold is significantly affected by the binary mass ratio $q$, with $M_{\rm PC}(q) < M_{\rm PC}(q = 1)$ for $q \lesssim \tilde{q} = 0.7 - 0.75$, regardless of the \ac{EOS}. This implies that asymmetric binaries have a smaller mass threshold for prompt collapse. In our model, we set $\tilde{q} = 0.725$ and we note that APR4 consistently predicts smaller $M_{\rm PC}$ values compared to BLh.

To model the non-PC scenario, we used numerical relativity fitting formulas available in the literature. These formulas are calibrated using the results from BNS merger numerical simulations, typically targeted to GW170817, and provide fits of the merger ejecta characteristics starting from the source properties, such as the binary masses and NS EOS \citep[see e.g.][]{Dietrich2017CQGra, Dietrich2020Sci, Radice2018ApJ, Kruger2020PhRvD, Nedora2022CQGra}. In particular, we computed $m_{\rm dyn, ~non-PC}$ and $v_{\rm dyn, ~non-PC}$ using the fitting formulas from \cite{Radice2018ApJ}, and $m_{\rm disc, ~non-PC}$ using the fitting formula from \cite{Kruger2020PhRvD}.

\begin{figure}[h!]
    \includegraphics[scale = 0.5]{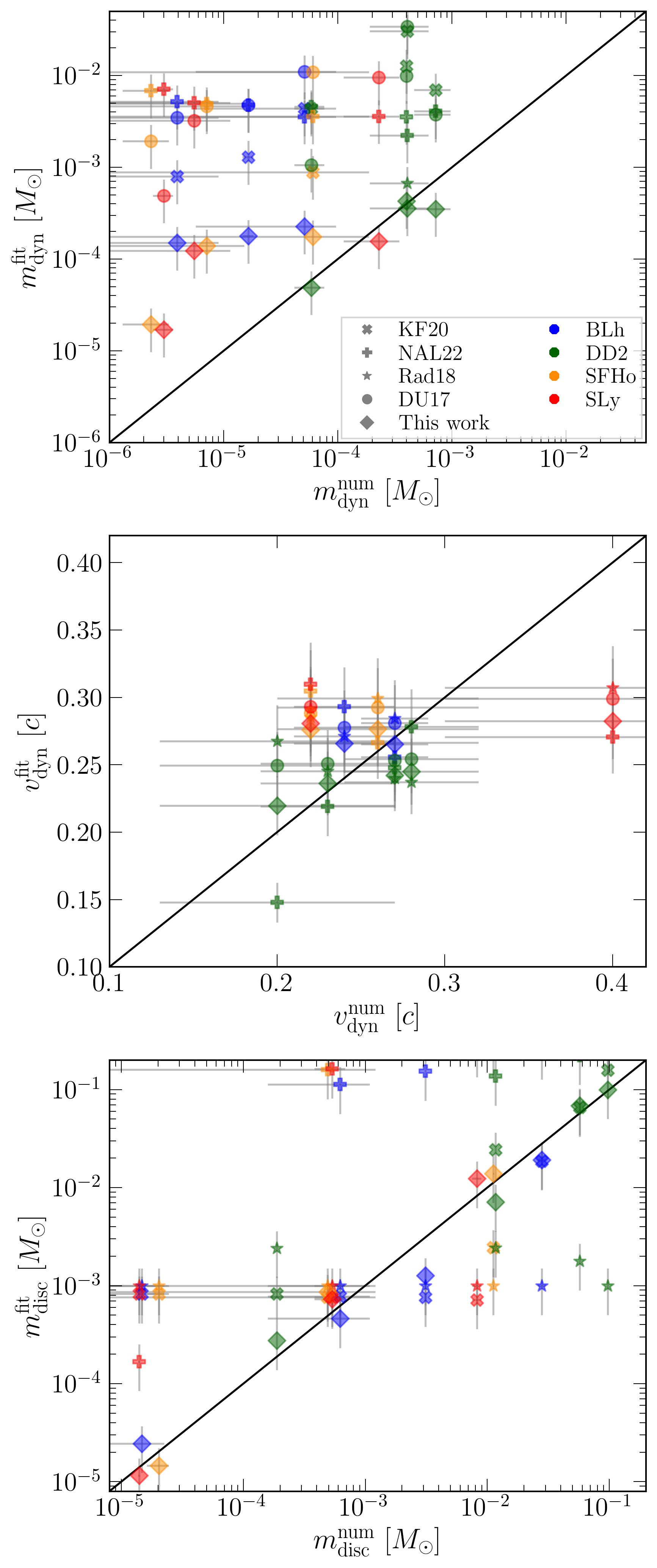}
    \caption{Mass and velocity of the dynamical ejecta and disc mass from GW190425-targeted numerical simulations in \cite{Camilletti2022MNRAS} ($m_{\rm dyn}^{\rm num}$, $v_{\rm dyn}^{\rm num}$, $m_{\rm disc}^{\rm num}$), and the corresponding values computed by means of fitting formulae ($m_{\rm dyn}^{\rm fit}$, $v_{\rm dyn}^{\rm fit}$, $m_{\rm disc}^{\rm fit}$). Different symbols correspond to the fitting formulae reported in KF20, NAL22, Rad18, DU17, and this work, while colours indicate the nuclear \ac{EOS}. Error bars are estimated as in \cite{Camilletti2022MNRAS}: y-axis errors represent 50\% ($m_{\rm dyn}$, $m_{\rm disc}$) and 10\% ($v_{\rm dyn}$) of the fit prediction, while x-axis errors represent the absolute difference between standard and low-resolution simulation results.}
    \label{fig:scatter_gw190425}
\end{figure}

\begin{figure*}[h!]
    \centering
    \includegraphics[scale = 0.41]{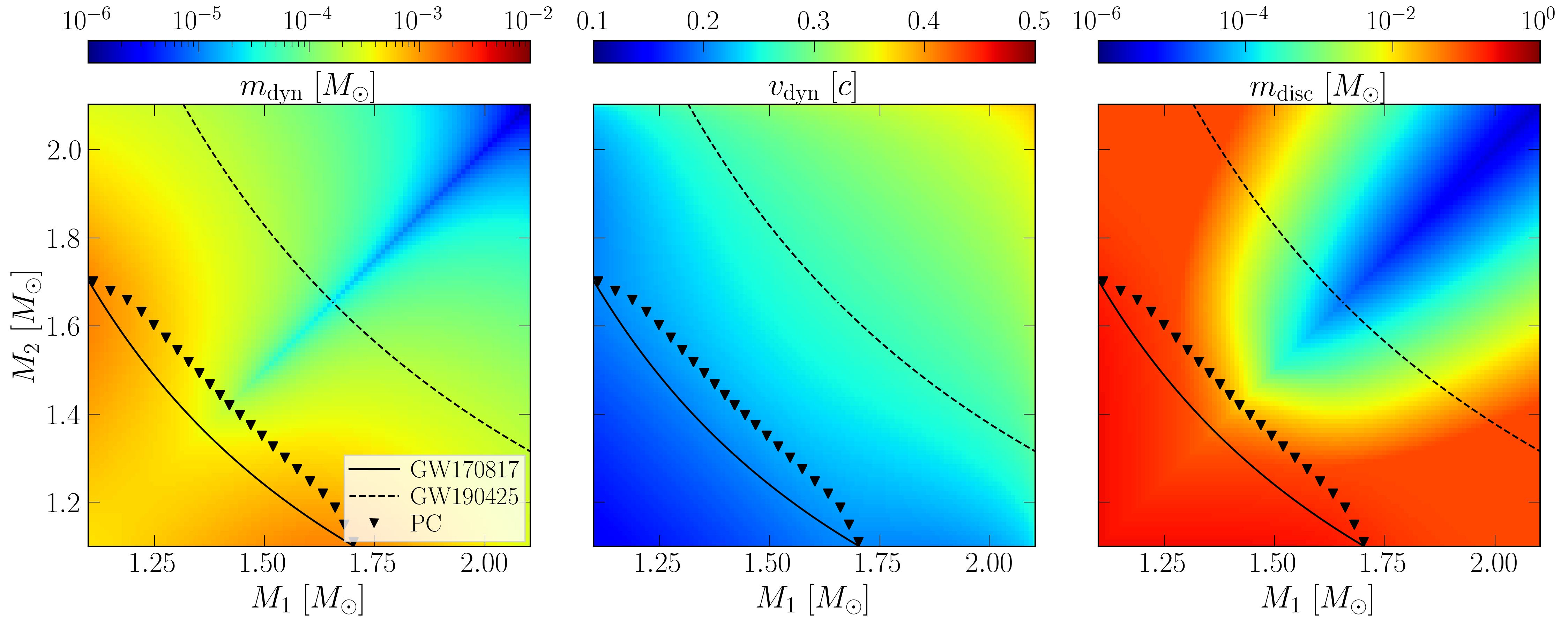}
    \caption{Mass of the dynamical ejecta (left), velocity of the dynamical ejecta (centre), and mass of the disc (right) for the \ac{EOS} BLh, as defined in Eq. (\ref{eq:mdyn})-(\ref{eq:mdisc}). $x$ and $y$ axes represent the masses $M_1$ and $M_2$ of the BNS. The solid and the dashed black lines mark the chirp mass of GW170817 and GW190425, respectively. Triangular markers show the prompt-collapse mass threshold $M_{\rm PC}$.}
    \label{fig:BLh_all_fits}
\end{figure*}

To model the PC scenario, the extrapolation of fitting formulas available in the literature is problematic, as shown by \cite{Henkel2023PhRvD}. The reliability of these formulas is heavily contingent on the sample of merger simulations they are calibrated against \citep{Nedora2021ApJ}, and the parameter space covered by current BNS merger simulations remains relatively narrow, typically restricted to the case of GW170817. Therefore, in order to compute the properties of the ejecta in the case of prompt collapse, we developed new fitting formulas calibrated on prompt-collapse simulations.
Our fits were specifically calibrated on the results of GW190425-targeted simulations published in \cite{Camilletti2022MNRAS}. Accordingly, we obtained the following fits for the dynamically ejected mass:
\begin{equation}\label{eq:mdyn_fit}
    m_{\rm dyn,~PC} = a \tilde{\Lambda} (q^{-1} - b) e^{c/q} \, , 
\end{equation}
where $a = 1.25 \times 10^{-4}, ~ b = 9.82 \times 10^{-1}, ~ c = -2.44$, while $q\leq 1$ and $\tilde{\Lambda}$ are the binary mass ratio and the reduced dimensionless tidal deformability, respectively; for the velocity of the dynamical ejecta:
\begin{equation}\label{eq:vdyn_fit}
     v_{\rm dyn,~PC} = \left[a \dfrac{M_1}{M_2} (1+ c~\mathcal{C}_1) \right] + (1 \leftrightarrow 2) + b \, ,
\end{equation}
where $\mathcal{C}$ is the NS compactness defined in \refeq{eq:compactness}, $a = -0.395,~ b = 0.798,~ c = -1.627$; and for the mass of the disc:
\begin{equation}\label{eq:mdisc_fit}
    \log_{10} \left( m_{{\rm disc,~PC}} \right) = \min \left( -1, ~a + b q + c \tilde{\Lambda}q^2 \right) \, , 
\end{equation}
where $a = 7.70,~ b = -1.34 \times 10,~ c = 8.16 \times 10^{-3}$. 
Our fit for $v_{\rm dyn,~PC}$ maintains the same functional shape as in \cite{Dietrich2017CQGra} and \cite{Radice2018ApJ}, but with different coefficients $a,~b,~c$. Our formulas reflect the ejecta's dependence on the EOS and the mass ratio as observed in \cite{Camilletti2022MNRAS}. Notably, both the mass of the dynamical ejecta and of the disc increase with $\tilde{\Lambda}$ (stiffer \acp{EOS}) and with the system's asymmetry (lower $q$), as evident from Figures 4 and 8 in \cite{Camilletti2022MNRAS}. Similarly, the velocity of the dynamical ejecta tends to increase with the NSs' compactness, a finding consistent with \cite{Dietrich2017CQGra, Radice2018ApJ}. In \reffig{fig:scatter_gw190425},  we compare the mass and velocity of the dynamical ejecta and the disc mass from the numerical simulations in \cite{Camilletti2022MNRAS} to the values obtained using our fitting formulas and those available in the literature for various \acp{EOS}.
Specifically, we make a comparison with the fitting formulas reported in \cite{Kruger2020PhRvD} (KF20), \cite{Nedora2022CQGra} (NAL22), \cite{Radice2018ApJ} (Rad18), \cite{Dietrich2017CQGra} (DU17), and this work (Eqs. (\ref{eq:mdyn_fit})-(\ref{eq:mdisc_fit})), considering four nuclear \acp{EOS}, specifically BLh \citep{Bombaci2018AAP, Logoteta2021AAP}, DD2 \citep{Hempel2010NuPhA, Typel2010PhRvC}, SFHo \citep{Steiner2013ApJ}, and SLy \citep{Douchin2001A&A, Schneider2017PhRvC}.
Our fits for the dynamically ejected mass and the disc mass outperform those in DU17, Rad18, KF20, and NAL22. As shown in \reffig{fig:scatter_gw190425}, all considered fitting formulas predict similar values for the velocity of the dynamical ejecta, which are fully compatible with the errors from the numerical simulations. However, our fit more accurately captures the general trend of the simulation data.

In summary, we used \refeq{eq:mdyn}-(\ref{eq:mdisc}) to smoothly interpolate between fits calibrated in the two different regimes, corresponding to the PC and non-PC scenarios. Other BNS merger observations would be instrumental in improving these fits.
In \reffig{fig:BLh_all_fits} and \reffig{fig:APR4_all_fits} (in \refapp{app:KNmodeling}), we display $m_{\rm dyn},~v_{\rm dyn}$, and $m_{\rm disc}$, as defined in \refeq{eq:mdyn}-(\ref{eq:mdisc}), in the $M_1 - M_2$ space for the BLh and APR4 \ac{EOS}, respectively. BLh consistently predicts a larger disc mass than APR4 for fixed $(M_1,~M_2)$ values due to its lower compactness and greater tidal deformability. Comparing the amount of dynamically ejected mass predicted by the two \acp{EOS} is less immediate. If the merger remnant does not promptly collapse to a \ac{BH}, then $m_{\rm dyn}$ is dominated by the mass ejected at the remnant's rebounce, which is larger for more compact \acp{EOS} \citep{Bauswein2013ApJ,Radice2018ApJ}. In contrast, for prompt collapse scenarios, the dynamical ejecta comes primarily from the tidal matter tails formed before the coalescence, leading to less compact \acp{EOS} with larger tidal deformability ejecting more mass \citep[e.g.][]{Bernuzzi2020MNRAS}.
The distinction between these two regimes is marked by the prompt collapse mass threshold, which also varies with the \ac{EOS}, being smaller for more compact \acp{EOS}. Therefore, given the binary masses, APR4 predicts a larger $m_{\rm dyn}$ than BLh only if the total binary mass is below the prompt-collapse mass threshold for APR4, as illustrated by comparing \reffig{fig:BLh_all_fits} and \reffig{fig:APR4_all_fits}.

\begin{figure}[h!]
    \centering
    \includegraphics[scale = 0.46]{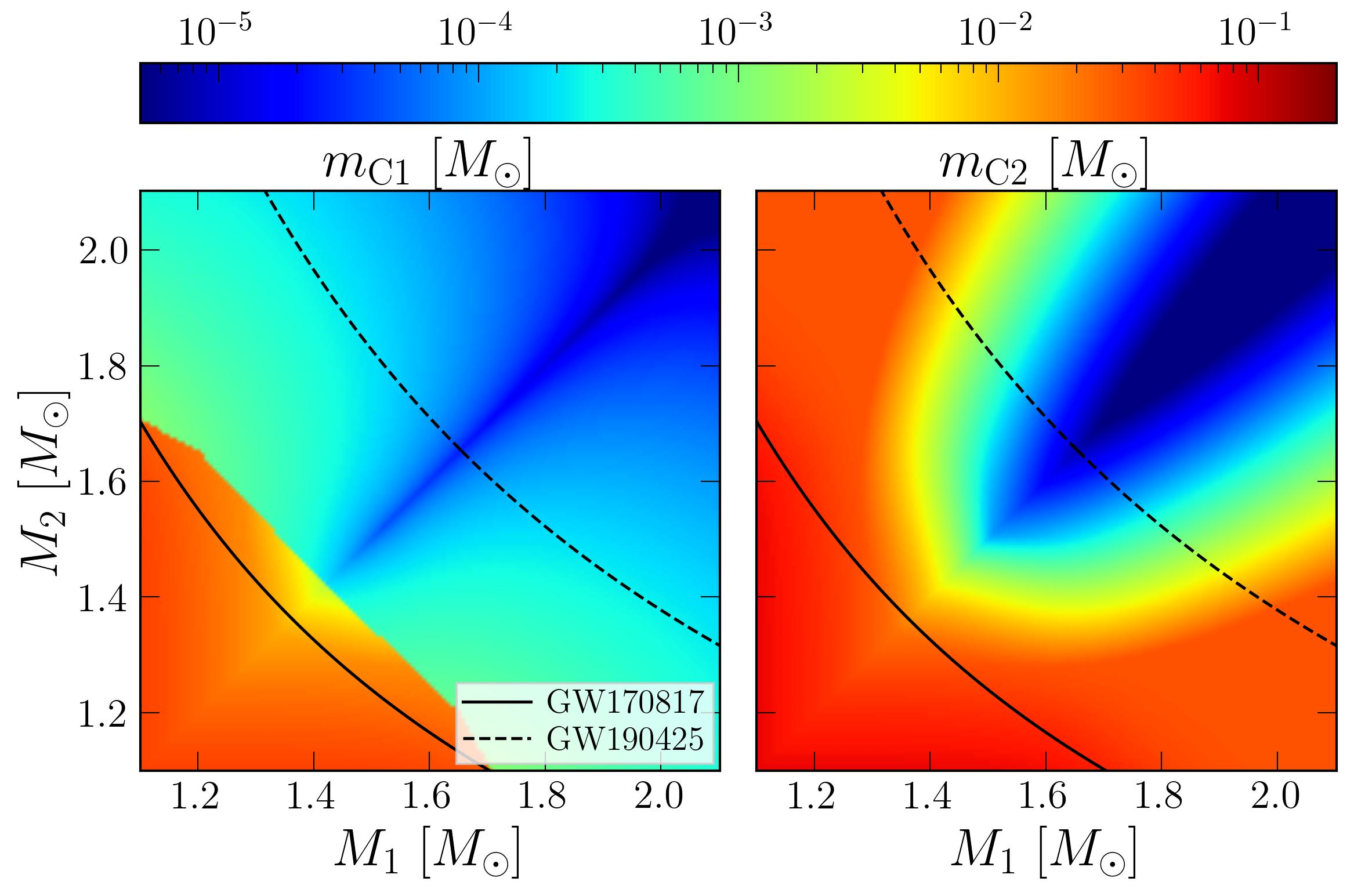}
    \caption{Mass of the first component (left) and the second component (right) of the ejecta, as defined in \refeq{eq:mass_C1} and \refeq{eq:mass_C2}, respectively, for the \ac{EOS} BLh. The solid and the dashed black lines mark the chirp mass of GW170817 and GW190425, respectively.}
    \label{fig:BLh_ej_comp}
\end{figure}

We now detail the computation of mass, velocity, opacity, and angular distribution for the two ejecta components, C1 (anisotropic, including both the dynamical ejecta and ejecta unbound from the disc via spiral wave winds) and C2 (isotropic, accounting for the secular ejecta unbound from the disc on secular timescales), which were the input of the \texttt{xkn} KN model.
We modelled the amount of mass in C1 as:
\begin{equation}\label{eq:mass_C1}
    m_{\rm C1} = m_{\rm dyn} + m_{\rm sw} \, ,
\end{equation}
where $m_{\rm dyn}$, defined in \refeq{eq:mdyn}, represents the mass ejected dynamically, and $m_{\rm sw}$ denotes the mass ejected through spiral wave winds. Spiral wave winds are generated by \ac{NS} remnants before the collapse to \ac{BH}. Therefore, we set $m_{\rm sw} = 0$ if $M_{\rm tot} \geq M_{\rm PC}$; otherwise, $m_{\rm sw} = 0.15~m_{\rm disc}$, where $m_{\rm disc}$ is defined in \refeq{eq:mdisc} and we assumed that spiral wave winds acting on timescales of several tens of ms are effective in unbind a significant fraction of the disc \citep{Radice:2023zlw}. We distributed $m_{\rm C1}$ as $F(\theta) = \sin(\theta)$, where $\theta$ is the polar angle \citep[cf.][]{Perego2017ApJ, Radice2018ApJ}. In general, matter ejected on a dynamical timescale and through spiral wave wind have different velocities. Hence, we define the central velocity of the first component as a weighted average:
\begin{equation}
    v_{\rm C1} = \dfrac{m_{\rm dyn} v_{\rm dyn} + m_{\rm sw} v_{\rm sw}} {m_{\rm dyn} + m_{\rm sw}}~,
\end{equation}
with $v_{\rm dyn}$ from \refeq{eq:vdyn} and $v_{\rm sw} = 0.17 c$ \citep[cf. Figure 2 in][]{Nedora2019ApJ}. The opacity for dynamical or spiral wave wind ejecta, $k_{\rm C1}$, varies with the polar angle. In general, the opacity is $\sim 10 - 20 \rm{~cm}^2~\rm{g}^{-1}$ in the equatorial region, while it is expected to reduce to $\sim 0.5 - 5 \rm{~cm}^2~\rm{g}^{-1}$ close to the poles, if strong enough shocks develop at the merger contact interface and if the merger remnant lives long enough to irradiate the polar ejecta with neutrinos \citep[e.g.][]{Kasen2013ApJ,Perego:2014fma,Metzger:2014ila,Martin2015ApJ,Sekiguchi:2016bjd,Perego2017ApJ}. For asymmetric binaries ($q \lesssim 0.7 - 0.75$), the lighter \ac{NS} is largely destroyed during the inspiral phase. Moreover, asymmetric binaries have a smaller prompt collapse mass threshold. Therefore, we applied the following prescription:
\begin{itemize}
    \item If $q \geq \tilde{q} = 0.725$, then we set $k_{\rm C1} = 1 \rm{cm}^2\rm{g}^{-1}$ in the polar region, $\theta \leq 50$ deg and $\theta \geq 130$ deg, and $k_{\rm C1} = 15 \rm{cm}^2\rm{g}^{-1}$ in the equatorial region, $ 50 <\theta < 130$ deg.
    \item If $q < \tilde{q} = 0.725$, then we set $k_{\rm C1} = 15 \rm{cm}^2\rm{g}^{-1}$ at all latitudes.
\end{itemize}
We assigned an entropy of $s_{\rm C1} = 10 k_{\rm B}/$baryon and an ejecta expansion timescale $\tau_{\rm C1} = 5$ ms for C1 \citep{Radice2018ApJ}.\\

We modelled the amount of mass in the second component as
\begin{equation}\label{eq:mass_C2}
    m_{\rm C2} = 0.3~m_{\rm disc}~,
\end{equation}
where $m_{\rm disc}$ is defined in \refeq{eq:mdisc} \citep[c.f. the estimated amount of mass from AT2017gfo fitting in][]{Perego2017ApJ}. We assumed a uniformly distributed mass density and set the central velocity to $v_{\rm C2} = 0.06 c$. We assumed the opacity to be uniformly distributed and we set $k_{\rm C2} = 10~\rm{cm}^2\rm{g}^{-1}$. Finally, we set the entropy to $s_{\rm C2} = 20 k_{\rm B}/$baryon, and the ejecta expansion timescale to $\tau_{\rm C2} = 33$ ms 
\citep[see e.g.][]{Wu2016MNRAS,Villar:2017wcc,Perego2017ApJ}.

In \reffig{fig:BLh_ej_comp} and \reffig{fig:APR4_ej_comp} (in \refapp{app:KNmodeling}), we show the ejecta component masses $m_{\rm C1}$ and $m_{\rm C2}$ in the $M_1 - M_2$ space for BLh and APR4 \acp{EOS}, respectively. We notice that $m_{\rm C1}$ and $m_{\rm C2}$ are always larger for BLh than APR4, implying in general brighter \ac{KN} light curves for BLh compared to APR4.

\subsubsection{Kilonova light curves}\label{sec:xkn}
We computed the \ac{KN} light curves for the previously described ejecta components using the \texttt{xkn} framework, a semi-analytic model designed for predicting and interpreting both the bolometric luminosity and broadband light curves of KNe \citep{Ricigliano2024MNRAS}. This tool is particularly effective for extensive parameter estimation and population studies since it is computationally extremely efficient. Specifically, we adopted the \texttt{xkn-diff} model, which relies on a semi-analytical solution of the radiative transfer equation for homologously expanding material, in line with approaches developed in \cite{Pinto2000ApJ} and in the appendix of \citet{Wollaeger2018MNRAS}. The \texttt{xkn-diff} model simultaneously accounts for emission from both optically thick and thin ejecta, marking a significant advancement over previous semi-analytic methods. It also incorporates cosmological and K-corrections \citep{Ricigliano2024MNRAS}.

For modelling the recombination effects in both lanthanide-free and lanthanide-rich ejecta, we established two floor temperatures: $T_{\rm Ni} = 3000$ K and $T_{\rm La} = 1500$ K \citep[see][]{Kasen2017Natur, Kasen2019ApJ,Villar:2017wcc, Breschi2021MNRAS}. 
Given the current uncertainties on the heating rate \citep[e.g.][]{Mendoza-Temis2015PhRvC, Rosswog2017CQGra,Zhu2021}, stemming from variances in the nuclear mass model, we multiplied the nuclear heating rate, $\dot{\epsilon}_{\rm nucl}$, computed as per \cite{Ricigliano2024MNRAS} using the FRDM nuclear mass model \citep{Moller2016ADNDT}, by a factor $f_{\dot{\epsilon}}$. Its value was determined by ensuring the reproduction of the AT2017gfo case within our ejecta and \ac{KN} light curve modelling (see \refapp{app:KNtest}).
For our two different EOS models, we find $f_{\dot{\epsilon}} = 1.50$ for BLh and $f_{\dot{\epsilon}} = 2.75$ for APR4. We notice that the heating rates $\dot{\epsilon}_{\rm nucl}$ resulting from these values are within their uncertainty range \citep[][]{Mendoza-Temis2015PhRvC, Rosswog2017CQGra,Zhu2021}.
In \refapp{app:KNtest}, we test our choice, by comparing AT2017gfo to the \ac{KN}e produced within our model for binaries with the same chirp mass as GW170817 and binary mass ratio $q > 0.725$. 

As discussed in \cite{Ricigliano2024MNRAS}, the simplified treatment of the opacity used in \texttt{xkn} introduces significant deviations in the KN light curves at early times after the merger, when compared to the results of radiative transfer calculations. For this reason, when using our light curves to develop an observational strategy with Rubin, we considered the computed emission to be reliable only starting from 1.5 hours after the merger.

\subsection{Kilonova light curves distribution and properties} 

\begin{figure}[h!]
    \includegraphics[scale=0.51]{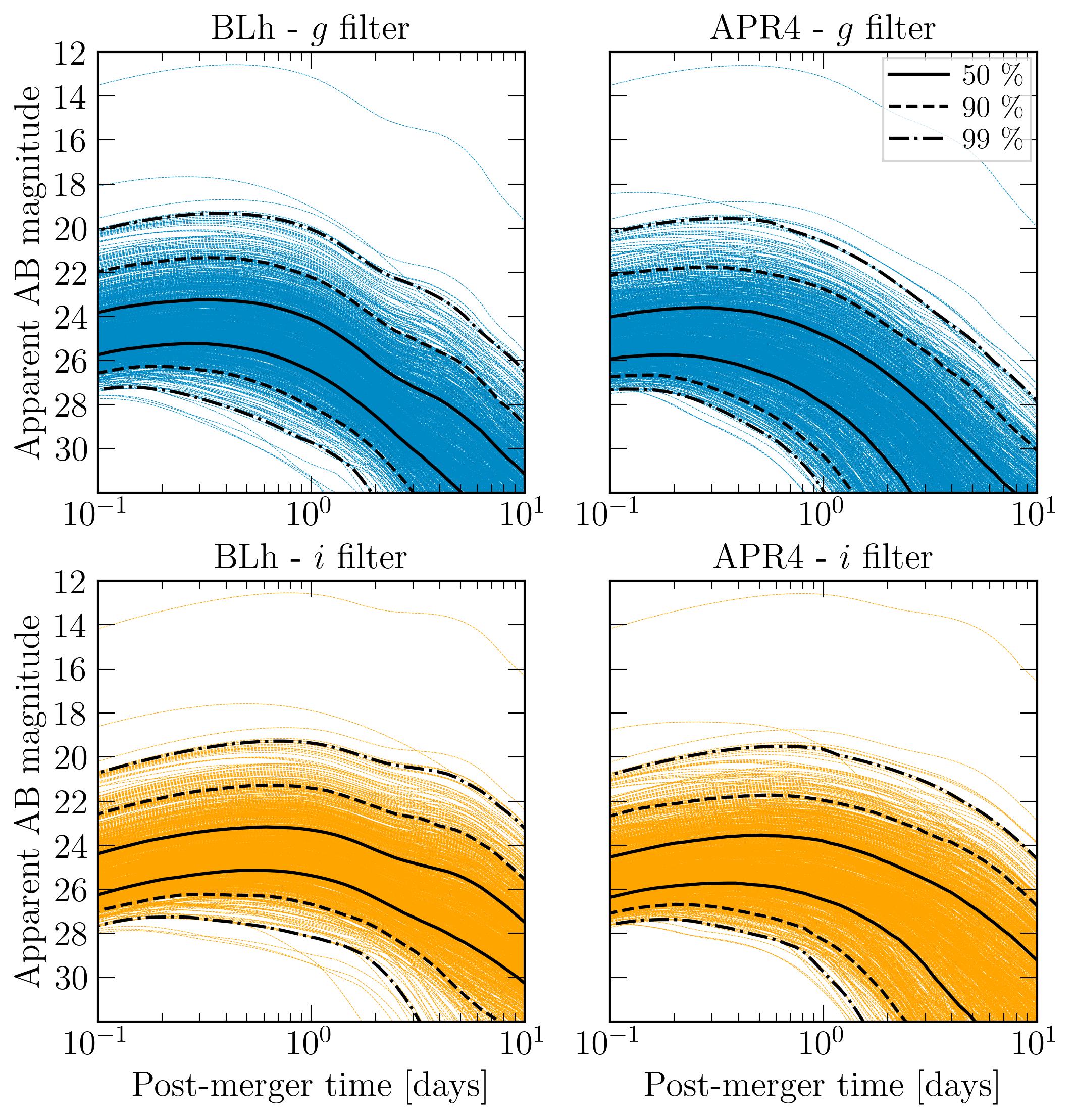}
    \caption{\ac{KN} light curves for the $g$ (top) and $i$ (bottom) filters. These light curves were generated from BNS merger fiducial populations,
    Gaussian NS mass distribution, and the BLh (left) and APR4 (right) \acp{EOS}. The plotted light curves are associated with BNS mergers detectable by \ac{ET} reference design with a sky localisation accuracy better than 100 deg$^2$. The solid, dashed, and dash-dot black lines mark the $50\%$, $90\%$, and $99\%$ probability distributions of the light curves.}
 \label{fig:lc_gaussian}
\end{figure}

In this section, we examine the typical magnitude range of our KN light curve populations, focusing on KNe associated with BNS mergers detectable by ET-triangle. We select sources with a sky localisation uncertainty smaller than 100 $\rm deg^2$, which are suitable to be followed up with wide field-of-view optical telescopes, such as the Vera Rubin Observatory.

\reffig{fig:lc_gaussian} presents the KN light curves and their probability distributions in the \textit{g} and \textit{i} filters of Rubin. These light curves are calculated for the BNS mergers detected by ET-triangle within 100 deg$^2$ from populations characterised by a common envelope efficiency $\alpha = 1.0$, Gaussian NS mass distribution, and BLh (left) and APR4 (right) EOSs.
In the case of BLh EOS, the peak apparent AB magnitude for 90\% of the KN distribution ranges between 26 mag and 21 mag in both filters. However, there's a noticeable variation in the peak times. For the 90\% distribution range, this spans approximately between 5 and 12 hours for the \textit{g} filter and between 9 and 24 hours for the \textit{i} filter. For both $g$ and $i$, the broad time range of the peak can be attributed to the differences in redshift among the light curves and time dilation in the observer frame. The delay observed going from redder to bluer emission is instead mainly due to the different opacities of the ejecta components.
Of particular interest are two exceptionally bright light curves, with one reaching a magnitude of about 13 mag in the \textit{g} filter, and the other around 18 mag. These correspond to relatively nearby and rare events, occurring at about 10 and 40 Mpc, respectively.
These findings are slightly different for the APR4 EOS, where the light curves generally appear fainter by half to one magnitude and evolve more rapidly. This difference is attributed to the BLh EOS producing more massive ejecta in both components, C$_1$ (dynamical and spiral wave wind ejecta) and C$_2$ (secular ejecta), than APR4.
This is evident when comparing \reffig{fig:BLh_ej_comp} and \reffig{fig:APR4_ej_comp} within the mass range of [1.1, 1.6] $M_\odot$ for both $M_1$ and $M_2$. Having a smaller mass threshold for prompt collapse and predicting smaller ejecta, the APR4 EOS consistently exhibits more light curves with a very rapid decay compared to BLh.

The distribution properties of the light curves exhibit more pronounced variations in the case of uniform NS mass distribution, as illustrated in \reffig{fig:lc_uniform} for the BLh (left) and APR4 (right) EOSs, respectively. While the peak magnitude range is similar to that of the Gaussian counterparts, with the light curves generally appearing slightly fainter, there is a notable spread in both the peak time and the duration of the light curve fading. The uniform mass distribution allows us to explore parts of the $M_1$ - $M_2$ parameter space which are not accessible with Gaussian distributions. Specifically, systems with high chirp mass and near-unity mass ratios (visible in the upper right corners of \reffig{fig:BLh_ej_comp} and \reffig{fig:APR4_ej_comp}) have minimal ejecta, resulting in extremely faint KNe. Conversely, binaries with high chirp masses and highly asymmetric mass ratios ($q \lesssim 0.7$, as seen in the upper left and lower right corners of \reffig{fig:BLh_ej_comp} and \reffig{fig:APR4_ej_comp}) produce a small amount of ejecta (below $10^{-3} M_\odot$) in the first component, while the second ejecta component, originating from the tidally disrupted disc, is significantly more massive (exceeding $0.02 M_\odot$). These conditions lead to brighter KNe, particularly with the APR4 EOS. This is because, although the ejected mass is comparable for both EOSs, the larger nuclear factor $f_{\dot{\epsilon}}$ adopted for APR4 (necessary to reproduce AT2017gfo) results in brighter light curves. A comparison between our modelling of the KN light curves and AT2017gfo is provided in \refapp{app:KNmodeling} and Figures \ref{fig:kn_comparison_at2017gfo}-\ref{fig:compare_gw170817}.

Finally, we compare our modelling of KN light curves with studies of short GRBs observed on-axis (some of them showing possible KN signatures in the afterglow emission) which evaluated the luminosity range for the KN emission \citep{Ascenzi2019MNRAS}.
To compare our results with the findings reported by \cite{Ascenzi2019MNRAS}, we computed
the AB absolute magnitudes of the KN light curves presented in \reffig{fig:agkn_g}. For the BLh EOS, the peak magnitudes in the \textit{g} filter span from  -13.9 mag to -16.6 mag, with the peak occurring between 1.1 and 10.2 hours. Similarly, the APR4 EOS shows a peak magnitude range from -14.2 mag to -16.5 mag, occurring between 1 hour and 10.2 hours. Remarkably, our modelling aligns with the results of \cite{Ascenzi2019MNRAS}, who reports a peaking magnitude range of [-12.3,-16.8] mag in the \textit{g} filter, occurring within 18 hours after the GRB prompt emission.

\subsection{Short GRB afterglow modelling}\label{sec:afterglow_modeling}
\begin{figure}[h!]
    \centering
    \includegraphics[scale = 0.495]{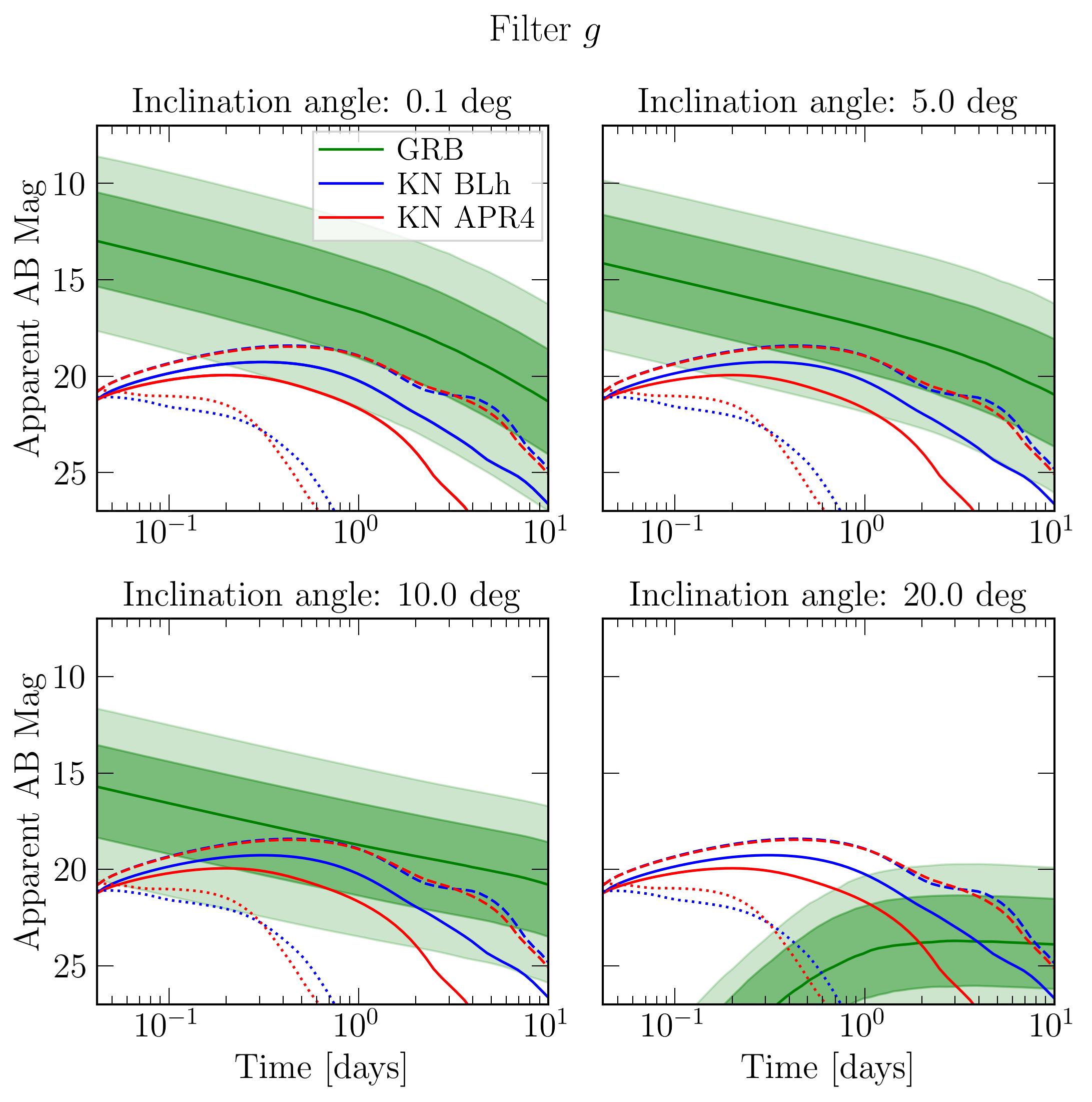}
    \caption{GRB afterglow distribution (green region) compared to \ac{KN}e light curves for BLh (blue lines) and APR4 (red lines) \acp{EOS} computed in the $g$ filter of Rubin and at different inclination angles, assuming luminosity distance $d_{\rm L} = 100$ Mpc. The green line, dark green region and light green region correspond to the median, the 16-84 percentiles, and the 2.5-97.5 percentiles of the afterglow distribution, respectively. The \ac{KN}e light curves correspond to three different BNS systems sampled from the BNS population with $\alpha = 1.0$ and Gaussian mass distribution. Solid lines correspond to $M_1 = M_2 = 1.33 M_\odot$, that is the peak of the \ac{NS} mass distribution; dashed lines correspond to $M_1 = 1.18 M_\odot, ~ M_2 = 1.10 M_\odot$, that is the BNS producing the most massive disc; dotted lines correspond to $M_1 = M_2 = 1.57 M_\odot$, that is the BNS producing the lightest disc.}
    \label{fig:agkn_g}
\end{figure}

\begin{figure}[h!]
    \centering
    \includegraphics[scale = 0.495]{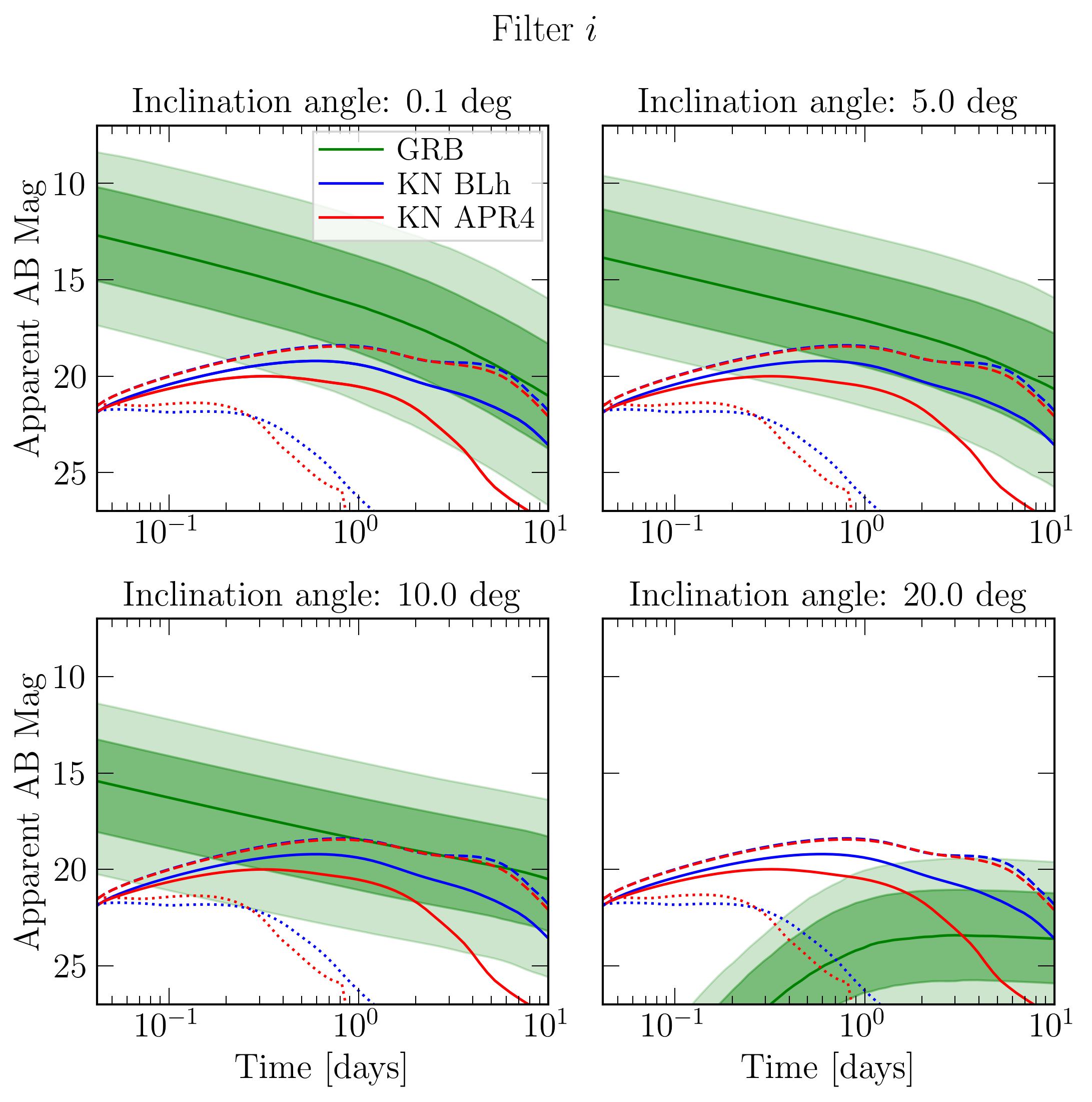}
    \caption{Same as in \reffig{fig:agkn_g} for the $i$ filter of Rubin.}
    \label{fig:agkn_i}
\end{figure}

In addition to the emission due to the KNe, we included the contribution of the optical afterglows associated with the production of relativistic jets.  One of the still uncertain parameters is the fraction $f_j$ of BNS mergers capable of successfully producing a jet. As detailed in \cite{Ron2022A&A...665A..97R}, the quantity $f_j$ can be deduced starting from a BNS merger population, knowing the local rate and its redshift evolution, and a phenomenological model of \ac{GRB} prompt emission.  Adopting a structured jet with an opening angle of 3.4$^{\circ}$ and an off-core power-law profile with a slope $s=4$, \cite{Ron2022A&A...665A..97R} obtain a median value $f_j=0.26$.  However, $f_j$ must be adjusted accordingly for different BNS population synthesis tracks to maintain the same local short GRB rate.
Therefore, denoting with a tilde the quantities adopted in \cite{Ron2022A&A...665A..97R}, we computed
\begin{equation}
    f_j = \min \left( \tilde{f_j} \frac{\int_0^1\tilde{\mathcal{R}}_{BNS}~dz}{\int_0^1 \mathcal{R}_{BNS} dz},1 \right)~.
\end{equation}
For our BNS population with $\alpha=1.0$, we obtain an $f_j$ value of 0.7, and for $\alpha=0.5$, $f_j$ reaches 1.0.

To evaluate the afterglow emission, we used the Python package \texttt{afterglowpy} \citep{Ryan2020ApJ}. The jet structure followed the functional form derived in \cite{ghi2019Sci...363..968G}. The isotropic-equivalent kinetic energy of the jet was derived starting from the isotropic prompt energy extracted from the posterior distribution in \cite{Ron2022A&A...665A..97R}. Adjusted for the jet's aperture angle, the distribution of the jet kinetic energy is compatible with the one reported in \cite{fon2015ApJ...815..102F}. Our micro-physical parameters include $\epsilon_e=0.1$, $p=2.2$, $\log_{10}{\epsilon_B}$ uniformly distributed in the range $[-4,-2]$, and $n \in [0.25,15] \times 10^{-3}$ cm$^{-3}$. These parameters allowed us to replicate the observed range of short GRB afterglow luminosities in optical and X-ray bands, as documented in \cite{kan2011ApJ...734...96K} and \cite{fon2015ApJ...815..102F}.

Figures \ref{fig:agkn_g} and \ref{fig:agkn_i} compare jet afterglows and the \ac{KN} emissions for APR4 (red lines) and BLh (blue lines) in \textit{g} and \textit{i} filters at 100 Mpc luminosity distance, considering different inclination angles. We display the median and percentiles (2.5, 16, 84, and 97.5) of afterglow distributions and three relevant cases of \ac{KN} emissions:
\begin{itemize}
    \item average \ac{KN}e (solid lines): we consider an equal mass BNS with masses corresponding to the peak of the Gaussian distribution, that is $M_1 = M_2 = 1.33 M_\odot$;
    \item very bright \ac{KN}e (dashed lines): we select the BNS producing the disc with the largest mass ($M_1 = 1.18 M_\odot$, $M_2 = 1.10 M_\odot$) from the BNS population with $\alpha = 1.0$ and Gaussian mass distribution and compute the corresponding KN;
    \item very faint \ac{KN}e (dotted lines): we select the BNS producing the disc with the smallest mass ($M_1 = M_2 = 1.57 M_\odot$) from the BNS population with $\alpha = 1.0$ and Gaussian mass distribution and compute the corresponding KN.
\end{itemize}
Jet afterglows outshine the faintest \ac{KN}e in the entire time domain for inclination angles $\iota < 15 - 20$ deg. In contrast, \ac{KN}e dominate early emissions at $\iota \geq 20$ deg, as expected. The brightest \ac{KN}e overcome a non-negligible portion of the afterglow emission distribution at any angle a few hours after the merger.
The behaviour of the average \ac{KN}e strongly depends on the nuclear \ac{EOS}. In particular, the peak of the \ac{KN} computed assuming the BLh \ac{EOS} is on average significantly brighter than the one corresponding to the APR4 \ac{EOS}. Accordingly, the probability of observing the peak of the \ac{KN} emission above the afterglow emission at $\iota \sim 5 - 10$ deg is enhanced for the BLh \ac{EOS}.

We simulated our observational strategy with Rubin by considering 1) only the KN emission and 2) the sum of the \ac{KN} emission and the afterglow emission from the jet.

\section{Rubin observational strategy} 
\label{sec:Rubin}
This section outlines our observational strategy with the Rubin Observatory \citep{ivezic19} to detect the optical counterparts of GW signals. Rubin is a large and wide field-of-view (9.6 $\rm deg^2$) ground-based telescope, comprising an 8.4 m primary mirror and a 3.2 gigapixel camera combined in a stiff compact design enabling an extremely fast slewing time. It is currently the most innovative observatory providing capabilities for rapidly covering large sky regions (through a mosaic strategy) reaching deep sensitivity with short exposures in several filters. This makes it the optimal instrument for observing the relatively faint KN counterparts of BNS mergers detected by next-generation GW detectors.

We propose to use ToO observations as the most effective follow-up strategy. This enables us to rapidly point to the sky-localisation associated with GW signal, to use the appropriate exposure time and to repeat the observation with the right cadence for KN detection. 
Considering that the next generation of GW observatories is expected to be operative after 2035 when Rubin is planned to have completed its decadal surveys and achieved its associated objectives, we assumed the availability of a large amount of observational time for ToO programs.

We designed an observational strategy based on following up all the events with a sky-localisation uncertainty (given as 90\% credible region) below a certain threshold value.  
This choice makes it possible to maximize the number of events to follow while keeping Rubin's time within a reasonable amount.
We considered only events within the sky regions accessible to Rubin\footnote{We use as Rubin's visible sky the EuclidOverlapFootprint of the \texttt{rubin\_sim} module available on \href{https://github.com/lsst/rubin_sim}{GitHub}.}. This made our simulations account for the combination of the real sky-direction sensitivity of the GW network and the Rubin-accessible sky. 

We adopted a multi-filter and multi-epoch observational strategy. This strategy enabled us to have indications of the source spectra and colour evolution, which is particularly important to identify the candidate counterparts and remove contaminant events.
Multi-filter observations were scheduled in two epochs over 2 or 3 nights, starting from the first night after the merger. We took sunrise and sunset times into account and started the follow-up as soon as it was feasible after sunset. Given the merger time $t_{\rm c}$, we started the observations as soon as possible, but never earlier than $t_{\rm c}+ 1.5$ hours after the merger, because our KN modelling is less reliable at very early times. Sunrise and sunset times were taken to be 6:30 am and 7:30 pm, allowing for 11 hours of darkness per night, which is an average for Cerro Pachon throughout the seasons\footnote{Using sunrise maps available at \href{https://sunrise.maplogs.com/}{this link}.}. We used the $g$ and $i$ filters. The $i$ filter is preferred over $z$ for this study since it allows for deeper point-source photometry in a single exposure while it is still expected to enable contaminant removal following the $g-i$ colour evolution \citep{Cowperthwaite2019ApJ, Andreoni2019PASP, Bianco2019PASP}. For each epoch, the observations were performed starting with the $g$ filter, completing the mosaic, and then switching to the $i$ filter and observing the mosaic again. If the mosaic was completed in one night in the first epoch, the second epoch started after the next sunset. If the mosaic could not be completed during the first night, observations of the first epoch were resumed from the point where they were interrupted immediately after the next sunset. 
Thus, the first and second epoch observations were performed during the two or three consecutive nights after the merger. We used single exposures of 600 s for each pointing, but we also tested strategies which make use of deeper exposures of 1200 s (see \refsec{sec:deeperexp}). The overhead for filter change and the slewing time to cover the mosaic in each filter was considered to be 220 s for each epoch\footnote{The overhead associated with the slewing time to cover the mosaic was defined conservatively, calculated for a sky localisation of 100 $\rm deg^2$ and also applied for events detected with a better localisation.}. This time takes into account that Rubin's mirrors have the ability to slew extremely fast with a median slew time of $\sim$5 s \citep{bianco22} and that the filter change time is of the order of two minutes if the filter is already mounted on the wheel (five out of the six filters will be housed on the wheel at all times).

The total Rubin follow-up time to observe all the selected GW events was determined as:
\begin{equation}
\rm Total\ time(\Omega_{90,max})= \sum_{i=1}^{N(< \Omega_{90,max})} t_{obs,i}
\label{totime}
\end{equation}
where $\rm N(< \Omega_{90,max})$ is the number of events detected by a GW network with sky localisation smaller than $\rm \Omega_{90,max}$ in the Rubin footprint, and $\rm t_{obs,i}$ is the Rubin observational time spent to tile the whole localisation area for each of these events in two filters and two epochs. For each event, this time was given by
\begin{equation}
\rm t_{obs,i}= n_{pointings,i} \times t_{exp} \times n_{\rm filters} \times n_{\rm epochs}+440 s
\label{eacheventtime}
\end{equation}
where $n_{\rm pointings,i}$ is the total number of pointings of the mosaic obtained by dividing the sky-localisation uncertainty of each event for the Rubin field of view as
\begin{equation}
\rm n_{pointings,i} = \frac{\Omega_{90,i}}{9.6 \rm deg^2} + 1 .   
\end{equation}
$\rm t_{\rm exp}$ is the exposure time (in seconds) for each pointing, $\rm n_{filters} = 2$ is the number of filters, and $\rm n_{epochs} = 2$ is the number of epochs. The $\rm t_{obs,i}$ estimate includes 440 s (about seven minutes) of overhead for the mosaic slewing time and for changing filters as detailed above.

\cite{andreoni22_too} propose using more than two filters and shorter exposure times as an effective strategy to follow GW events detected by current GW detectors. In contrast, our two-filter strategy seeks to minimise the observational resources required for a single event in order to maximise the number of events to be followed. This is important to make the most of the larger number of events detected by next-generation GW observatories compared to the current ones. Here, we favour longer exposures to reach a larger distance and therefore to detect weaker KNe. Additionally, we use $g$ and $i$ filters since they are capable of capturing the unique features that can distinguish the evolution of a KN from other transients.

The criteria for detection were set based on the $5\sigma$ depth for a single exposure of 600 s (or of 1200 s), calculated at a conservative airmass at $45\deg$ from zenith\footnote{More details are provided in this \href{https://smtn-002.lsst.io/}{LSST technote}. In particular, we used the m5 values from the LSST Operations Simulator.}. Specifically, for the $g$ band, the $5 \sigma$ depth is 26.53 mag at 600 s exposure and 26.91 mag at 1200 s in units of ABmag. In the case of the $i$ band, $5 \sigma$ depth is 25.59 mag at 600 s exposure and 25.97 mag at 1200 s. For each detected GW event, we computed the total number of pointings needed to cover its sky localisation (90\% credible region) and we randomly assigned a position to the GW source in one of these mosaic tiles, assuming a uniform distribution within the localization area\footnote{This is a conservative assumption since having the real localization probability distribution provided by a full posteriors analysis would make the observational strategy more effective, starting the mosaic from the tiles having a higher probability to contain the source.}. We then evaluated the flux at the time when the observation arrived at the tile containing the source and determined if it was brighter than the $5\sigma$ threshold. We highlight that numbers and plots referring to the joint detections provided in \refsec{JointGWOPT} were obtained by injecting and recovering the signals within the targeted 90\% credible region, thus, the readers should take into account a 10\% reduction for sources located outside the targeted region.

Our light curves were pre-processed to take into account the effect of Galactic absorption. We included the extinction correction for each filter obtained from \citet{dustmaps}, based on \citet{sfd98} and \citet{sf11}, considering the precise sky location of each BNS merger.
 
We considered two detection strategies, designated as {\it one epoch detection (1ep)}, and {\it two epoch detection (2ep)}. The {\it 1ep} strategy requires the optical counterpart to be detected at least once in both filters during the same epoch observations. 
 The {\it 2ep} criterion is more conservative and requires the source to be detected in the two filters in the first epoch and at least in the $i$ filter in the second epoch (the KN is typically expected to show a slower decline in the redder filter compared to the bluer one). The {\it 2ep} events are a subset of the ones detected following the {\it 1ep} criterion, but multiple detections in the first and second epoch enable us to follow the colour evolution in the first days which can be crucial to distinguish the optical counterpart of GW event among many contaminants. In \reftab{tab:rubin_strategy}, we list the requirements to classify an event as detected according to the \textit{1ep} and \textit{2ep} strategies.  

\begin{table}[h]
\centering
\caption{Truth table of our observational strategy with Rubin, using filters $g$ and $i$.}
\label{tab:rubin_strategy}
\setlength\extrarowheight{-2pt}
\begin{tabular}{cc|cc|c}
\hline
\multicolumn{2}{c|}{\textbf{Observations in Ep. 1}} & \multicolumn{2}{c|}{\textbf{Observations in Ep. 2}} & \textbf{Flag} \\
\textit{g} & \textit{i} & \textit{g} & \textit{i} &  \\
\hline
\cmark & \cmark & \cmark & \cmark & 2 \\
\cmark & \cmark & \xmark & \cmark & 2 \\
\cmark & \cmark & \xmark & \xmark & 1 \\
\cmark & \cmark & \cmark & \xmark & 1 \\
\xmark & \cmark & \cmark & \cmark & 1 \\
\cmark & \xmark & \cmark & \cmark & 1 \\
\xmark & \xmark & \cmark & \cmark & 1 \\
\hline
\end{tabular}
\tablefoot{We followed events in two epochs with both filters, assigning \cmark each time an event is detected in one filter. In the one epoch detection (\textit{1ep}) strategy, an event was considered detected with flag $\geq 1$, while in the two epoch detection (\textit{2ep}) strategy, we required flag $\geq 2$ to claim a detection. The magnitude threshold for detection is based on the 5$\sigma$ depth for a single exposure. Specifically, for the \textit{g} band (\textit{i} band) the 5$\sigma$ depth is 26.53 mag (25.59 mag) at 600 s exposures and 26.91 mag (25.97 mag) at 1200 s in units of AB mag.}
\label{tab:strategies}
\end{table}

This work focuses on KN detection based on photometric data. However, all the modelling and simulations developed for the present work have been used to explore KN spectroscopy (Bisero et al. in prep) and in particular to investigate the multi-messenger science case for the WST project \citep{Mainieri2024arXiv240305398M}. WST is a proposed innovative 12-m class spectroscopic telescope with simultaneous operation of a large field-of-view (3 $\rm deg^2$) and a high multiplex (20,000) multi-object spectrograph facility.

\section{Results and discussion}\label{sec:ResDisc}

\subsection{GW detection and parameter estimation forecasting}\label{sec:GWresults}

\begin{figure}[h!]
\centering
\includegraphics[scale=0.6]{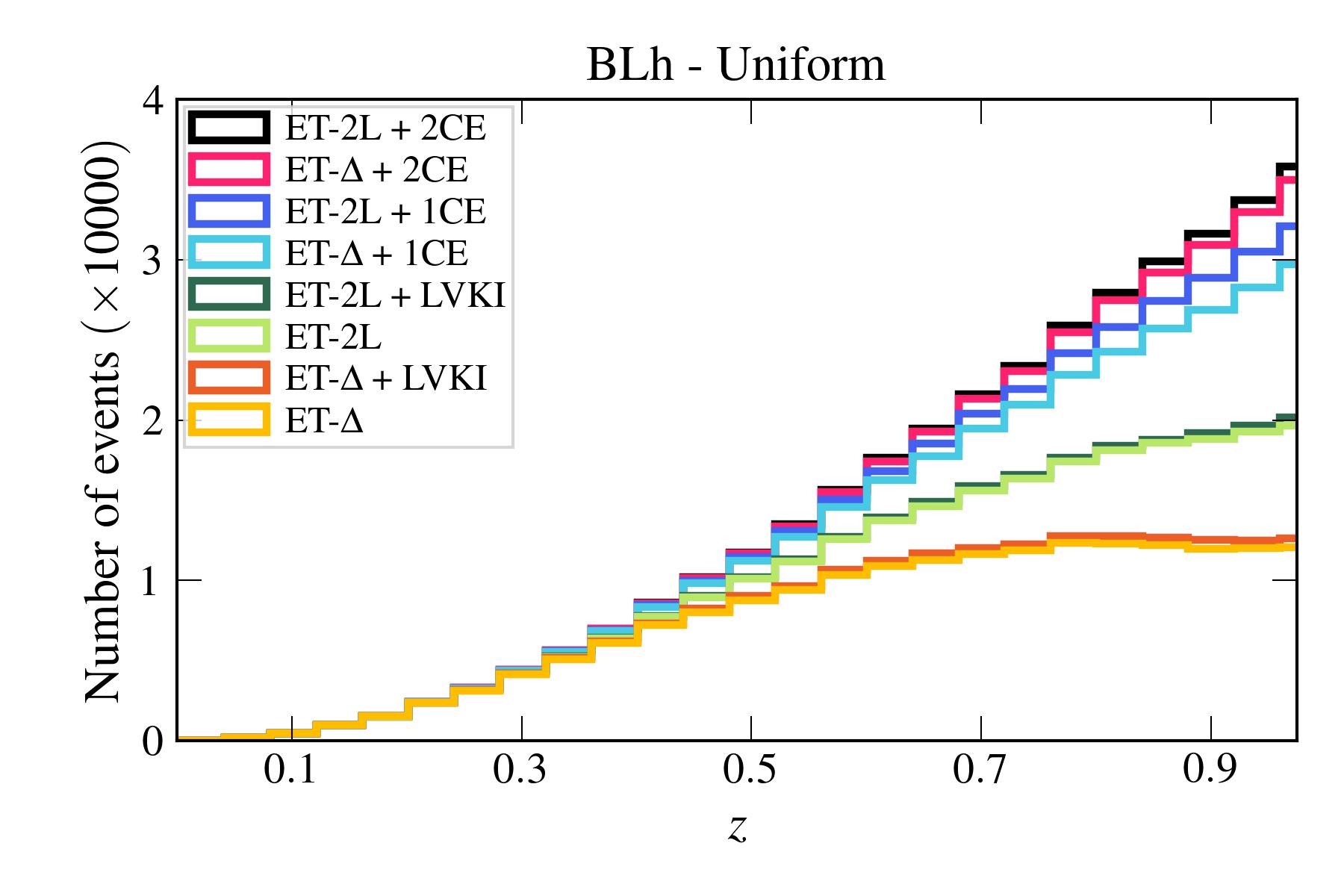}
\caption{Ten-year detection distribution as a function of redshift for the eight different networks using the fiducial population ($\alpha=1$) with BLh EOS and uniform mass distribution.
}
\label{fig:detections_z_blh_uniform_alpha1}
\end{figure}

\begin{figure}[h!]
\centering
\includegraphics[scale=0.6]{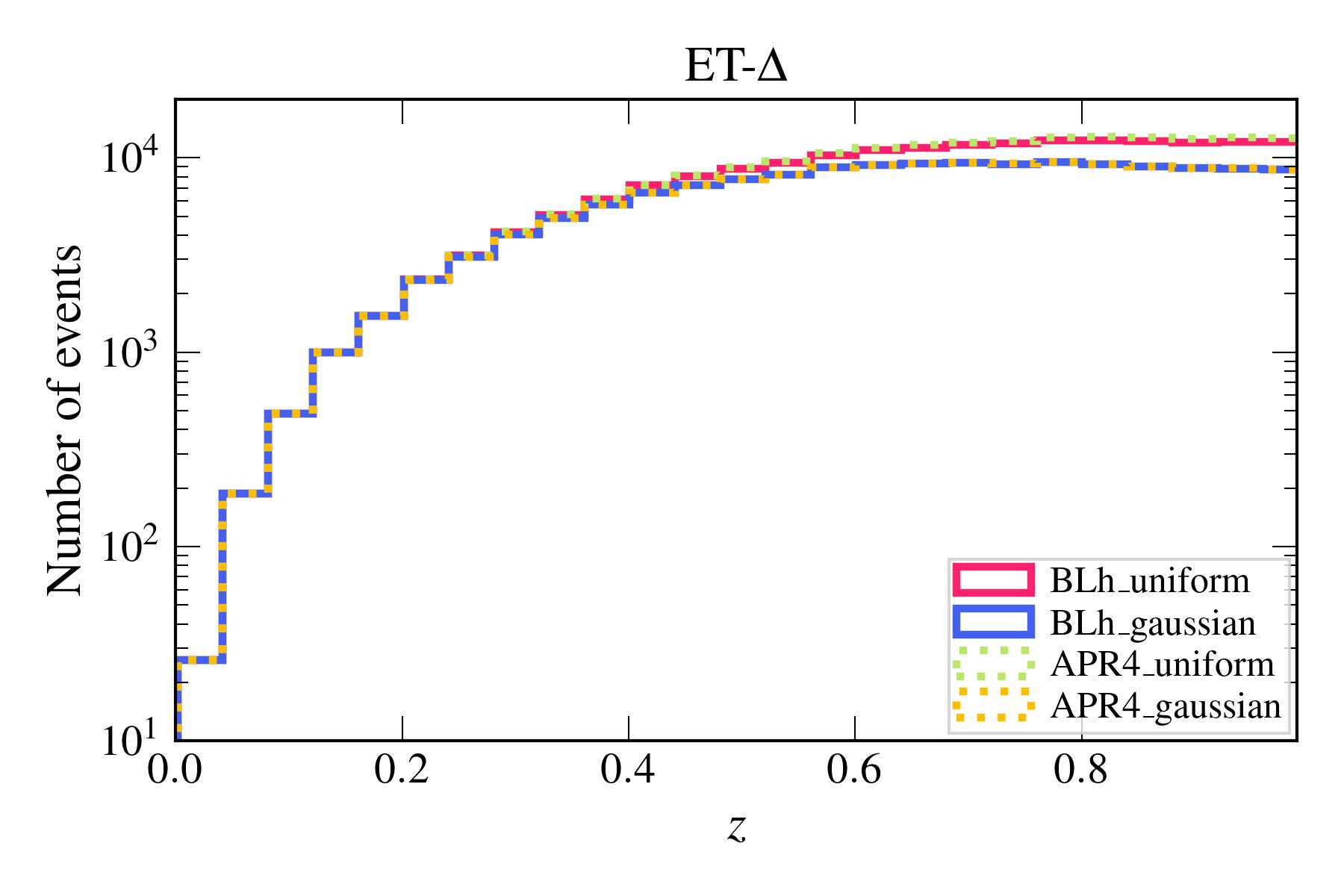}
\caption{Ten-year detection distribution as a function of redshift for the reference ET-triangle using the fiducial population ($\alpha=1$) and comparing different NS mass distributions and EOS models.}
\label{fig:detections_z_et_triangle_alpha1}
\end{figure}

This section summarises and discusses the outcomes of the 64 simulations described in \refsec{sec:GWsimul}, focusing on the detection capabilities and the associated parameter estimation of the considered GW detectors.

\reffig{fig:detections_z_blh_uniform_alpha1} shows the number of detections over ten years up to redshift $z=1$ for the fiducial population ($\alpha=1$), assuming a uniform NS mass distribution and the BLh EOS. As found in \cite{Branchesi2023}, the 2L ET performs better than the ET-triangle; the total number of detections up to $z=1$ increases by about 30\%. The detection capabilities of both the 2L and triangle configurations of ET are enhanced when operating within a network of detectors. In particular, the number of detections increases by about 3\% when ET operates with current detectors (LVKI O5 sensitivities). The number of detections increases by approximately 70\% (90\%) when ET-triangle is part of a network with next-generation detectors, including one CE (two CEs). For ET-2L, the number of detections increases by about 30\% (40\%). In \refapp{appD}, we detail the absolute number of detections for the different EOSs and NS mass distributions analysed in the present paper. In particular, \reffig{fig:detections_z_blh_gaussian_alpha1} shows the number of detections for the BLh EOS and Gaussian NS mass distribution, in parallel to \reffig{fig:detections_z_blh_uniform_alpha1}.  

\reffig{fig:detections_z_et_triangle_alpha1} compares ET's triangle detection numbers up to $z=1$ across different choices for the NS mass distribution and the EOS. For the uniform mass distribution, the number of detections increases by about 20-25\% compared to the Gaussian mass distribution, regardless of the EOS. This is expected due to the larger masses achieved by binary systems with a uniform NS mass distribution. Comparing the EOSs the differences are negligible. We only observe that APR4 yields a slightly higher number of detections (by 3\%) than BLh for the uniform mass distribution. This effect is mainly driven by the larger maximum NS mass supported by APR4, leading to louder GW signals.  
These findings agree with the analysis by \cite{Iacovelli2023PhRvD}, who observes an increased detection rate for EOSs with a higher maximum mass, specifically under a uniform NS mass distribution.

\begin{figure}[h!]
\centering
\includegraphics[scale=0.39]{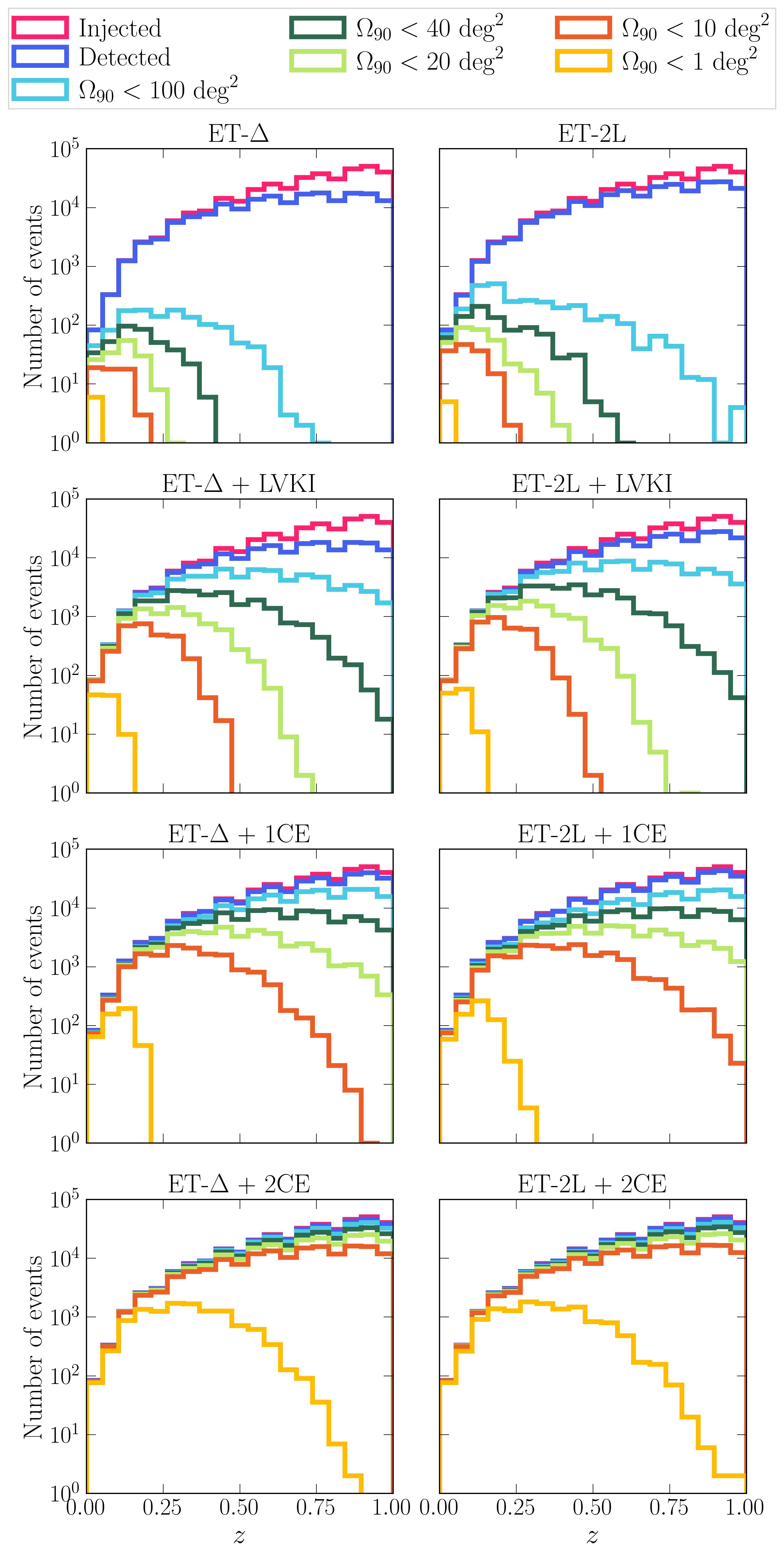}
\caption{Number of injected events and events localised better than 100 deg$^2$, 40 deg$^2$, 20 deg$^2$, 10 deg$^2$, and 1 deg$^2$ by the different detector networks considered in this work for the fiducial population ($\alpha = 1$) obtained assuming the BLh EOS and uniform NS mass distribution. The left plots show the ET-triangle and the right plots ET-2L shape. In \refapp{appD}, we show the same plots for BLh Gaussian NS mass distribution, and APR4 uniform and Gaussian NS mass distribution.}
\label{fig:skyloc_blh_uni}
\end{figure}

The sky-localisation capabilities of ET, operating either alone or in a network of detectors, are summarised in \reftab{tab:loc_events_triangle} and \reftab{tab:loc_events_2l} for ET-triangle and 2L, respectively. These tables list the number of detections over ten years by various detector networks with sky-localisation accuracy, $\Omega_{90}$, smaller than certain thresholds in the case of the pessimistic population ($\alpha=0.5$) and the fiducial one ($\alpha=1$), for the analysed NS EOSs (APR4 and BLh) and NS mass distributions (uniform and Gaussian). The results obtained for $\alpha=0.5$ (top of the tables) and $\alpha=1.0$ (bottom of the tables) show the large uncertainties in the absolute detection numbers due to the difference in the common envelope efficiency.  
The fiducial population provides a factor five higher number of detections with sky localisation smaller than 100 $\rm{deg}^2$ compared to the pessimistic one. As previously underlined, this is consistent with the poor observational constraints on the local BNS merger rate.

Comparing \reftab{tab:loc_events_triangle} and \reftab{tab:loc_events_2l}, we find that ET-2L operating as a single observatory localises better than ET-triangle by detecting a factor 2.4 more events within sky-localisation $100~\rm{deg}^2$. This is also found for ET operating together with current detectors, where ET-2L allows for 40-50\% more detections within $100~\rm{deg}^2$ compared to ET-triangle. The better performances of ET-2L in source localisation are still visible when ET operates in a network with one CE; in this case, ET-2L localises better than $10~\rm{deg}^2$ about 30-40\% more events compared to ET-triangle. When ET is operating with two CEs, the difference between 2L and triangle is negligible, except for the events localised better than $1~\rm{deg}^2$. These results are fully consistent with the findings of \cite{Branchesi2023} for ET alone and ET plus one CE in the USA.

The number of well-localised events dramatically increases when ET operates in a network of detectors, as shown in
\reftab{tab:loc_events_triangle} and \reftab{tab:loc_events_2l}. For the fiducial population, the number of events localised better than $100~\rm{deg}^2$ increases from a few hundred per year to several thousand per year when ET operates with current detectors (LVKI) rather than alone. ET will significantly benefit from the presence of current detectors (mainly from the distant ones, while Virgo gives a negligible contribution due to its vicinity), and the number of events detected better than $10~\rm{deg}^2$ will increase from a few tens to a few hundreds per year. \reffig{fig:skyloc_blh_uni} shows the benefits carried by detector networks in terms of larger redshifts reached by well-localised events. The presence of a network of next-generation detectors brings the number of events localised better than $10~\rm{deg}^2$ to $10^3$ ($10^4$) per year for ET+1CE (ET+2CE) up to $z \sim 0.9-1.0$.

The larger number of GW detections for the uniform mass distribution compared to the Gaussian one results in a systematically larger number of well-localised events in the case of uniform mass distribution. Moreover, for a given luminosity distance, the accuracy of sky localisation tends to be higher in the uniform distribution scenario due to the presence of more massive systems, which generate higher SNRs.

Finally, also the NS EOS affects the number of \ac{GW} detections at a given sky localisation. A comparison between the number of detected events for APR4 and BLh at $\Omega_{90} < 100,~40,~20, ~10 ~\rm{deg}^2$ shows that typically the number of detections increases for APR4 up to $4 \%$, in the case of Gaussian mass distribution, and $10 \%$, in the case of uniform mass distribution. This is because APR4 is more compact, therefore it has smaller tidal deformability compared to BLh (see \reffig{fig:mass_lambda_gaussian} and \reffig{fig:mass_lambda_uniform}), so the inspiral of the \ac{GW} signal lasts more time in the detector. Moreover, as mentioned above, in the case of uniform mass distribution, APR4 supports a larger maximum NS mass with respect to BLh (see \reftab{tab:eos_prop}) providing systems with larger mass, and thus larger SNR which can improve the sky localisation.

\begin{figure}[h!]
\centering
\includegraphics[scale=0.6]{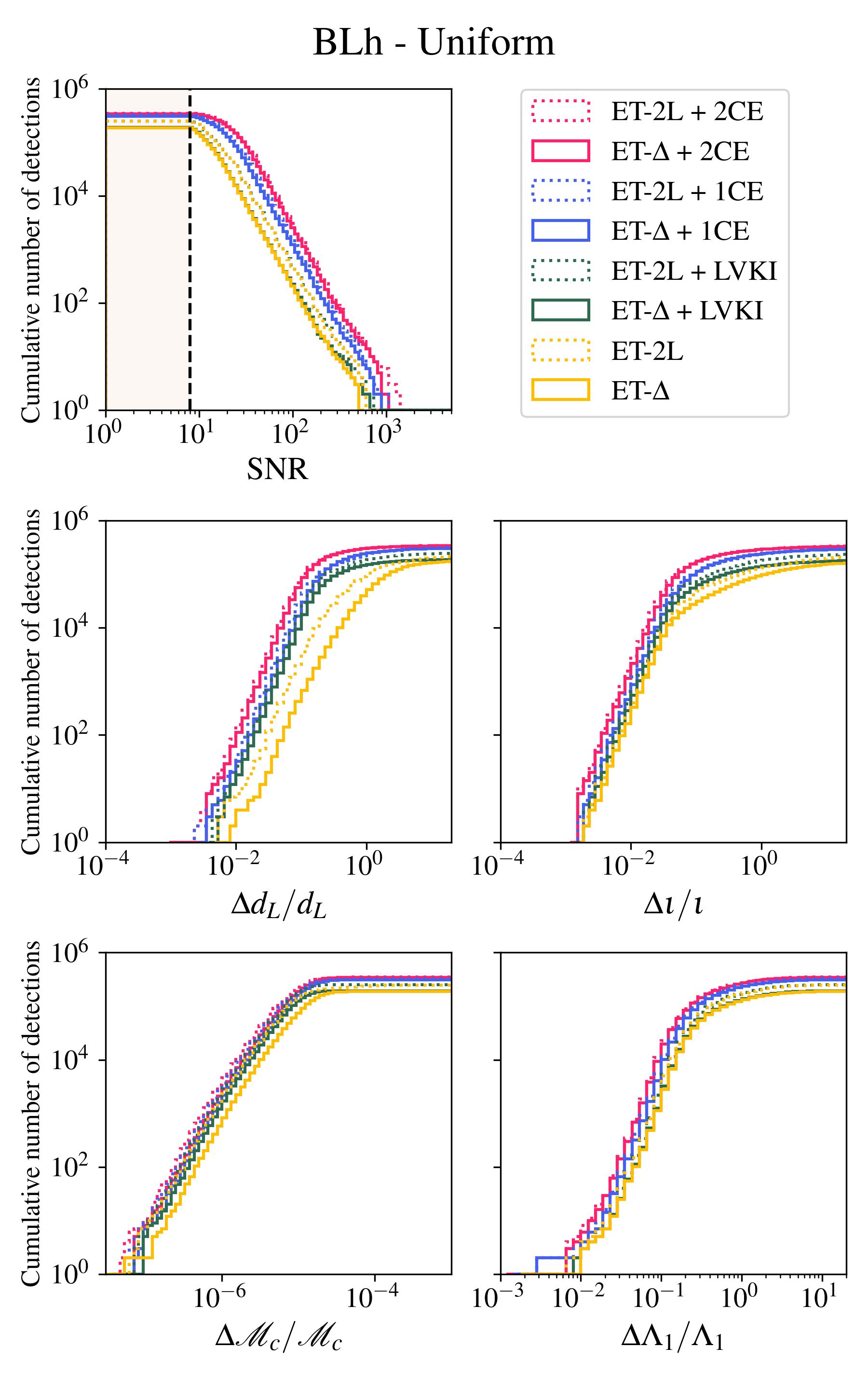}
\caption{Distribution of SNR and relative errors on detector frame chirp mass, $\mathcal{M}_c$, luminosity distance $d_{L}$, inclination angle $\iota$, and first component tidal deformability $\Lambda_{1}$. These results are obtained for the fiducial population ($\alpha=1.0$) and assuming a uniform NS mass distribution with BLh EOS.}
\label{fig:rel_errors_blh_uniform_alpha1}
\end{figure}

\reffig{fig:rel_errors_blh_uniform_alpha1} presents the SNR distributions for the injected signals and the parameter estimation for the detected sources (SNR $>8$), highlighting the uncertainties in chirp mass, luminosity distance, inclination angle, and tidal deformabilities for the fiducial population ($\alpha=1$) with a uniform NS mass distribution and BLh EOS. Notably, the parameter estimation accuracy is significantly enhanced with the 2L ET compared to the triangular configuration when each operates as a single detector. This enhancement, albeit less significant, persists when ET is operating in a network of detectors. It is noteworthy to highlight the improved estimation of parameters already reached when ET operates in synergy with current detectors. This is particularly relevant for the estimate of the luminosity distance, which is the key parameter for cosmology studies. This finding is pertinent to the redshift range $z = 0 - 1$ considered in this study, as extending to higher redshifts the presence of current detectors would not improve parameter estimation. The parameter estimation accuracy further increases with the inclusion of next-generation detectors, as evidenced by the distribution of well-localised events as a function of redshift (see \reffig{fig:skyloc_blh_uni} and Figures in \refapp{appD}). 

In \refapp{appD}, we show the SNR distributions and the relative uncertainties on chirp mass, luminosity distance, inclination angle, and tidal deformabilities for the fiducial population, but considering the Gaussian NS mass distribution with BLh and the uniform NS mass distribution with APR4. 
These plots show results consistent with those described above. 

\subsection{Joint GW and optical detections}
\label{JointGWOPT}

\begin{figure*}[ht!]
    \centering
    \includegraphics[scale=0.38]{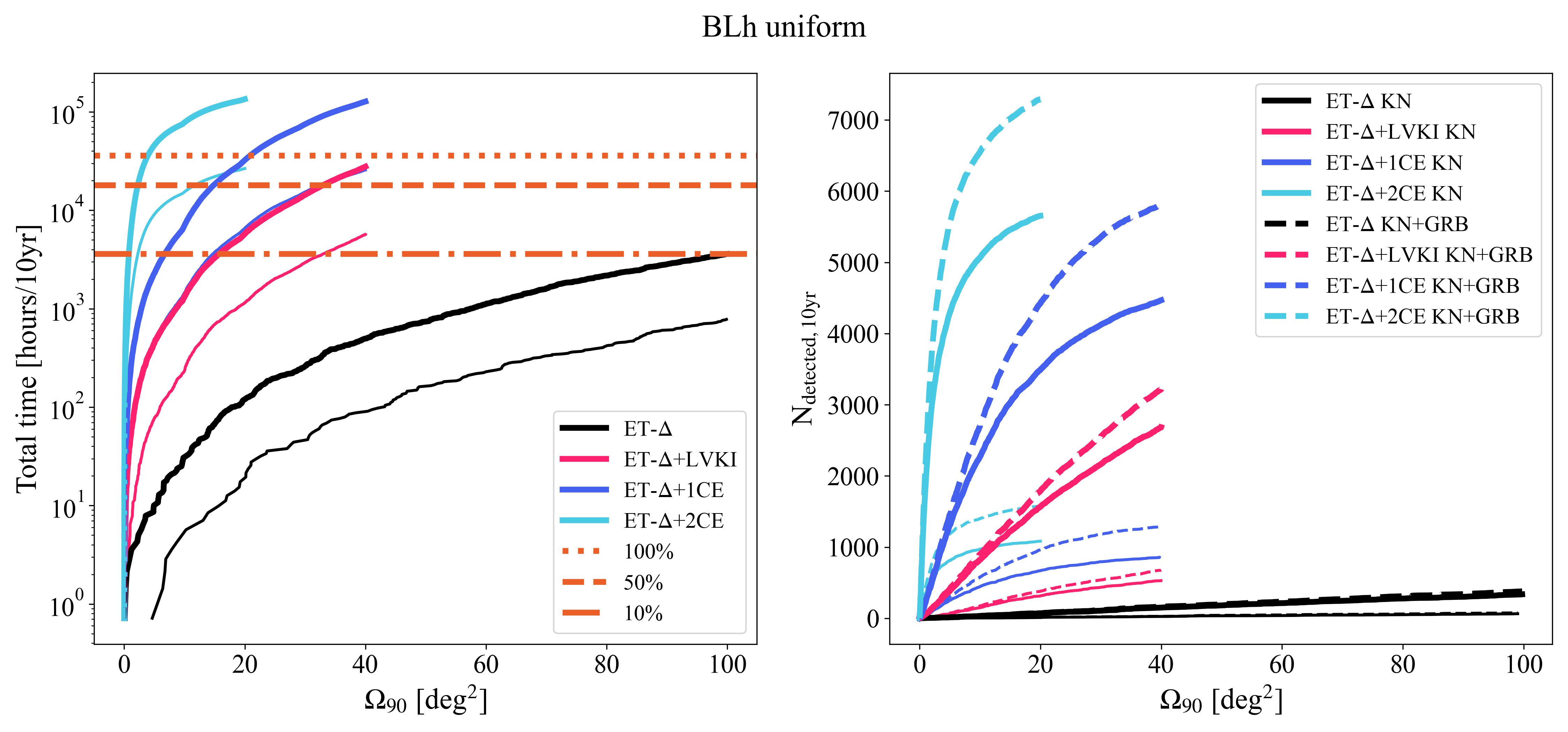}
    \includegraphics[scale=0.38]{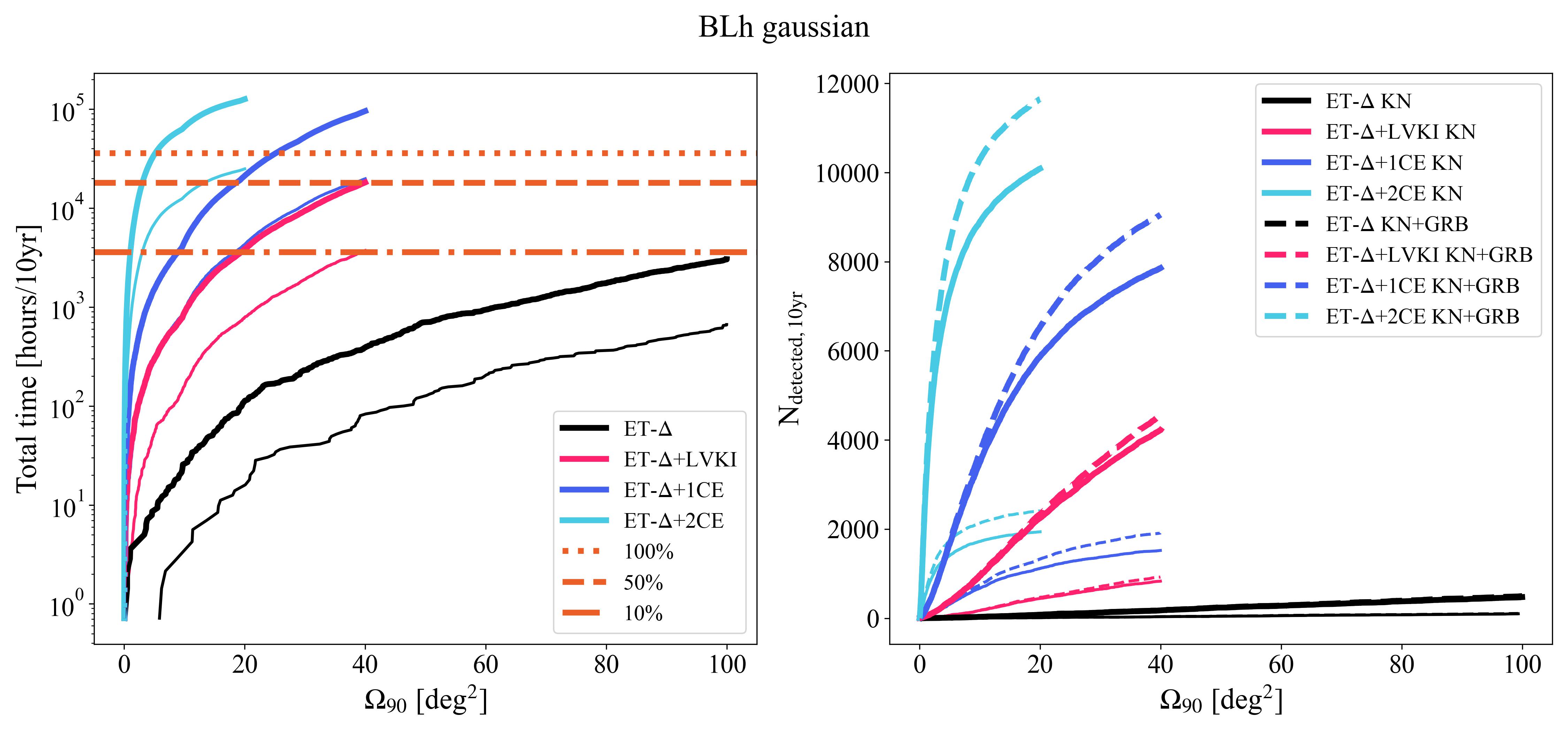}
    \caption{Cumulative time required and total number of joint detections for Rubin follow up. Left panels: cumulative time necessary to follow up all the events in the Rubin footprint with a sky localisation smaller than $\Omega_{90}$ (indicated in the x-axis) for ET-triangle as a single observatory, and included in a network of current and future GW detectors as indicated in the legend. The horizontal dashed lines indicate the time corresponding to 10\%, 50\%, and 100\% of Rubin's total observational time in 10 years. Right panels: the corresponding number of optical detections by Rubin. Solid lines are for the KN emission, while dashed lines are for the KN+GRB afterglow.
    All plots refer to the BLh EOS; the top row plots are obtained using the uniform NS mass distribution and the bottom row ones for the Gaussian NS mass distribution. Thick lines represent the fiducial population ($\alpha$=1.0) whereas the thin lines represent the pessimistic population ($\alpha$=0.5).}
    \label{fig:time_number_BLh}
    \end{figure*}

The results of our simulations reproducing different scenarios of GW detectors including ET-triangle operating in synergy with Rubin are summarised in \reffig{fig:time_number_BLh}, \reffig{fig:time_number_APR4}, and \reftab{tab:rubin_det_triangle}. Similarly, the results for the 2L ET are reported in Figures \ref{fig:2L_time_number_BLh}-\ref{fig:2L_time_number_APR4}  and \reftab{tab:rubin_det_2L}. 

Fig. \ref{fig:time_number_BLh} refers to the BLh EOS and shows: 1) the total observational time (see \refeq{totime}) required to follow up all the GW events visible by Rubin with sky-localisation uncertainty within threshold $\Omega_{90}$, and 2) the corresponding joint (KN and KN+GRB) detection numbers. In other words, if the observer decides to follow up all the events within a certain sky localisation ($\Omega_{90}$, x-axis), the required Rubin time is indicated on the y-axis, while the joint detection number is shown on the y-axis of the corresponding right plot. The total time and number of detections are given for ten years of observations. The left panels show horizontal lines corresponding to the percentages (10\%, 50\%, 100\%) of the Rubin's total observational time which is obtained by dividing the total observational time for the GW follow-up by the total 10-year observing time of Rubin. We assumed a total Rubin observational time of 3600 hours per year considering 11 hours of dark time each night in a year and a Rubin duty cycle of 90\%. Each plot shows the results for the fiducial (thick lines) and pessimistic (thin lines) populations. All the panels refer to the \textit{1ep} strategy. In \refapp{app:joint_GW_EM}, we provide the corresponding figures showing the results for the APR4 EOS and the 2L-shaped ET.

Tables \ref{tab:rubin_det_triangle} and \ref{tab:rubin_det_2L} list the number of events followed and detected over ten years by Rubin operating in synergy with \ac{ET} triangle-shaped and 2L-shaped, respectively. They show the results for ET alone and operating within the network of GW detectors, assuming for each event only \ac{KN} emission or \ac{KN} and \ac{GRB} afterglow emission, and considering the two NS EOSs (APR4 and BLh), and the two NS mass distributions (uniform and Gaussian). The results are given for both the pessimistic ($\alpha =0.5$) and fiducial ($\alpha=1.0$) populations of BNS mergers, and considering the two detection strategies (\textit{1ep} and \textit{2ep}) described in \refsec{sec:Rubin}. 
Our selection criteria for the events to be followed up are determined by the sky-localisation accuracy achievable through GW signal analysis, which varies by detector network configuration.
In \reffig{fig:time_number_BLh}, we provide readers with all the necessary information to define their own sky-localisation thresholds for selecting triggers while considering the time required for the follow-up. In contrast, Tables \ref{tab:rubin_det_triangle} and \ref{tab:rubin_det_2L}, present specific thresholds to illustrate a feasible example of selection.
In particular, we use $\Omega_{90} < 100~{\rm deg^2}$ for \ac{ET} alone (both triangular and 2L shapes), $\Omega_{90} < 20~{\rm deg^2}$ when ET operates with current detectors, $\Omega_{90} < 10~{\rm deg^2}$ for ET and one CE, and $\Omega_{90} < 5~{\rm deg^2}$ for ET and two CEs. Our threshold choice wants to maximize the number of joint detections while keeping the Rubin time allocated for the GW follow-up reasonable (within 30\% of the Rubin observational time). An exception is made for the ET + 2CEs network with $\alpha = 1.0$, since in this case the number of events localised better than $5~{\rm deg^2}$ corresponds to several thousands per year and the time exceeds the total Rubin observational time. In \refapp{app:joint_GW_EM}, \reftab{tab:rubin_2CE} shows the number of joint detections and corresponding Rubin's time allocation resulting from a narrower cut $\Omega_{90} < 1~{\rm deg^2}$ on the sky localisation of the events detected by ET + 2CEs. Selecting the events within $1~{\rm deg^2}$ reduces the required observational time while maintaining a high number of detections (about hundreds per year), similar to ET + 1CE. However, in this case, also smaller field-of-view telescopes could be used instead of Rubin.

\begin{table*}[ht!]
\centering
\caption{Number of joint detections by Rubin operating in synergy with \ac{ET} triangular shape operating as a single observatory, and as part of a global network of \ac{GW} detectors (indicated in the second column from the left).}
\scalebox{0.8}{

\begin{tabular}{l l l l c r c r c r c r}
\hline
& & &
&\multicolumn{4}{c}{\textbf{APR4}}
&\multicolumn{4}{c}{\textbf{BLh}}\\
\hline
& &$\Omega_{90}$&Transient 
&\multicolumn{2}{c}{Uniform} 
&\multicolumn{2}{c}{Gaussian}
&\multicolumn{2}{c}{Uniform}
&\multicolumn{2}{c}{Gaussian}\\
& &[deg$^2$] & &Followed&Detected &Followed&Detected &Followed&Detected &Followed&Detected\\
& & & &&1ep/2ep &&1ep/2ep &&1ep/2ep &&1ep/2ep\\
\hline
\hline
\multirow{8}{*}{\rotatebox[origin=c]{90}{\parbox[c]{1.2cm}{\centering $\alpha = 0.5$}}}
&\multirow{2}{*}{\rotatebox[origin=c]{90}{\parbox[c]{1.2cm}{\centering ET-$\Delta$}}}
&\multirow{2}{*}{\centering 100}
&KN 
&\multirow{2}{*}{\centering 168}
&$72/47$ 
&\multirow{2}{*}{\centering 144}
&$76/57$ 
&\multirow{2}{*}{\centering 172}
&$64/49$ 
&\multirow{2}{*}{\centering 144}
&$100/79$\\
& & & KN+GRB 
&(2.1\%)&$82/59$ &(1.8\%)&$86/70$ &(2.2\%)&$77/62$ &(1.9\%)&$108/91$\\
\cline{2-12}
&\multirow{2}{*}{\rotatebox[origin=c]{90}{\parbox[c]{1.2cm}{\centering ET-$\Delta$ \\\textbf{+}\\ LVKI}}}
&\multirow{2}{*}{\centering 20}
&KN 
&\multirow{2}{*}{\centering 1043}&$383/228$ &\multirow{2}{*}{\centering 655}&$326/204$ &\multirow{2}{*}{\centering 956}&$313/188$ &\multirow{2}{*}{\centering 646}&$444/296$\\
& & &KN+GRB &(3.5\%)&$447/279$ &(2.2\%)&$366/246$ &(3.2\%)&$378/237$ &(2.2\%)&$469/337$\\
\cline{2-12}
&\multirow{2}{*}{\rotatebox[origin=c]{90}{\parbox[c]{1.2cm}{\centering ET-$\Delta$ \\\textbf{+}\\ 1CE}}}
&\multirow{2}{*}{\centering 10}
&KN 
&\multirow{2}{*}{\centering 2097 }&$546/285$ 
&\multirow{2}{*}{\centering 1242}&$470/252$ 
&\multirow{2}{*}{\centering 1930}&$445/241$ 
&\multirow{2}{*}{\centering 1227}&$673/380$\\
& & &KN+GRB &(4.4\%)&$680/381$ &(2.6\%)&$552/327$ &(4.0\%)&$579/328$ &(2.6\%)&$747/456$\\
\cline{2-12}
&\multirow{2}{*}{\rotatebox[origin=c]{90}{\parbox[c]{1.2cm}{\centering ET-$\Delta$ \\\textbf{+}\\ 2CE}}}
&\multirow{2}{*}{\centering 5}
&KN 
&\multirow{2}{*}{\centering 13435 }&$978/442$ 
&\multirow{2}{*}{\centering 10033 }&$900/387$ 
&\multirow{2}{*}{\centering 12701}&$812/368$ 
&\multirow{2}{*}{\centering 9671}&$1408/613$\\
& & &KN+GRB &(26.6\%)&$1358/628$ &(19.9\%)&$1238/564$ &(25.2\%)&$1186/541$ &(19.2\%)&$1744/799$\\
\hline
\hline
\multirow{8}{*}{\rotatebox[origin=c]{90}{\parbox[c]{1.2cm}{\centering $\alpha = 1.0$}}}
&\multirow{2}{*}{\rotatebox[origin=c]{90}{\parbox[c]{1.2cm}{\centering ET-$\Delta$}}}
&\multirow{2}{*}{\centering 100}
&KN 
&\multirow{2}{*}{\centering 846}&$346/251$ 
&\multirow{2}{*}{\centering 695}&$392/277$ 
&\multirow{2}{*}{\centering 825}&$337/246$ 
&\multirow{2}{*}{\centering 695}&$475/381$\\
& & & KN+GRB &(10.2\%)&$383/282$ &(8.4\%)&$416/301$ &(10.0\%)&$379/273$ &(8.4\%)&$497/403$\\
\cline{2-12}
&\multirow{2}{*}{\rotatebox[origin=c]{90}{\parbox[c]{1.2cm}{\centering ET-$\Delta$ \\\textbf{+}\\ LVKI}}}
&\multirow{2}{*}{\centering 20}
&KN 
&\multirow{2}{*}{\centering 5231}&$1896/1148$ &\multirow{2}{*}{\centering 3287}&$1740/1132$ &\multirow{2}{*}{\centering 4825}&$1566/971$ &\multirow{2}{*}{\centering 3224}&$2248/1545$\\
& & &KN+GRB &(17.4\%)&$2108/1306$ &(10.9\%)&$1859/1229$ &(16.0\%)&$1787/1127$ &(10.7\%)&$2335/1646$\\
\cline{2-12}
&\multirow{2}{*}{\rotatebox[origin=c]{90}{\parbox[c]{1.2cm}{\centering ET-$\Delta$ \\\textbf{+}\\ 1CE}}}
&\multirow{2}{*}{\centering 10}
&KN 
&\multirow{2}{*}{\centering 10188 }&$2772/1533$ &\multirow{2}{*}{\centering 6072 }&$2505/1436$ &\multirow{2}{*}{\centering 9320 }&$2269/1263$ &\multirow{2}{*}{\centering 6002 }&$3496/2023$\\
& & &KN+GRB &(21.4\%)&$3201/1806$ &(12.7\%)&$2793/1654$ &(19.5\%)&$2702/1538$ &(12.6\%)&$3726/2255$\\
\cline{2-12}
&\multirow{2}{*}{\rotatebox[origin=c]{90}{\parbox[c]{1.2cm}{\centering ET-$\Delta$ \\\textbf{+}\\ 2CE}}}
&\multirow{2}{*}{\centering 5}
&KN 
&\multirow{2}{*}{\centering 67949 }&$5102/2276$ &\multirow{2}{*}{\centering 50440 }&$4666/2147$ &\multirow{2}{*}{\centering 64113 }&$4290/1911$ &\multirow{2}{*}{\centering 48557 }&$7301/3276$\\
& & &KN+GRB &(134.8\%)&$6356/2827$ &(100\%)&$5828/2680$ &(127.2\%)&$5548/2473$ &(96.3\%)&$8423/3857$\\
\hline
\end{tabular}
}
\tablefoot{The table reports the numbers of \ac{GW} events detected with a sky localisation smaller than the threshold, $\Omega_{90}$, and within Rubin's footprint. These events are the ones followed up by Rubin and indicated as {\em followed} in columns 5, 7, 9 and 11. The percentage of Rubin's observational time required to follow up all of them is given in parentheses. The corresponding numbers of KN and KN+GRB detections are given in columns 6, 8, 10, and 12 and indicated with {\em detected} both for the \textit{1ep} and \textit{2ep} detection strategies. The table lists the results for the APR4 and BLh EOSs, the uniform and Gaussian NS mass distributions both for the fiducial ($\alpha = 1.0$) and pessimistic ($\alpha = 0.5$) populations. All listed numbers refer to 10 years of observations and include the duty cycle for the GW detectors and Rubin as described in the text.}
\label{tab:rubin_det_triangle}
\end{table*}

\subsubsection{Joint detection efficiency}
\label{sec:efficiency}
We evaluate the joint GW/optical detection efficiency as a function of redshift by dividing the number of joint detections by the number of followed-up events (namely, events detected by the GW detector network within a certain sky-localisation threshold and in the Rubin footprint) in increasing redshift bins of 0.05. The sky-localisation thresholds correspond to the maximum values of  $\Omega_{90}$ used for Fig. \ref{fig:time_number_BLh}, namely $\Omega_{90} < 100~{\rm deg^2}$ for \ac{ET} alone, $\Omega_{90} < 40~{\rm deg^2}$ for ET+LVKI and for ET+1CE, and $\Omega_{90} < 20~{\rm deg^2}$ for ET+2CEs. As an example, \reffig{fig:eff_theta_blh_uniform} shows the detection efficiency for the fiducial population ($\alpha$=1.0), the BLh EOS, the uniform NS mass distribution, and the \textit{1ep} strategy.  
The right panels of the same figure show the distribution of the inclination angle, $\iota$, for the GW events detected within the aforementioned sky localisation thresholds (GW), the subset of these events within the Rubin footprint (fp), and the corresponding KN and the KN+GRB counterpart detections. Each row in the figure represents a different GW detector network, starting from the top with ET as a single observatory, ET+LVKI, ET+1CE, and ET+2CE. The same plots for the Gaussian NS mass distribution and APR4 EOS are given in \refapp{app:joint_GW_EM}.

Focusing on the inclination angle plots (right panels of \reffig{fig:eff_theta_blh_uniform}), the distribution of $\iota$ for the joint detections closely resembles that of the sources detected in GWs displaying fewer events for face-on systems ($\iota$ around $0$ and $180$ deg). This is fully consistent with the expectations for a population of randomly oriented systems in a sphere leading to fewer observable face-on systems than edge-on ones. There is also a lower density of GW detections around the region of the edge-on systems ($\iota = 90$ deg). This under-density is due to the SNR being lower for edge-on systems than for face-on ones with the same intrinsic parameters. 
Lower SNR values cause a smaller ratio of detected to injected signals for edge-on systems and the worsening of sky localisation.

Focusing on the GW/KN detection efficiency plots, we note that the efficiency drops to 50\% already at redshift $\sim 0.25-0.3$. \reffig{fig:skyloc_blh_uni} shows that the GW events selected for the follow-up based on sky-localisation thresholds extend to much higher redshifts. Thus, this reduction in efficiency at relatively low redshifts is due to the limited detection capability of the Rubin observations whose sensitivity is not enough to detect KN emission at larger redshifts. Adding the GRB afterglow emission from the relativistic jet, a tail of detections at higher redshifts appears (KN+GRB case). Indeed, the distribution of the $\iota$ for the KN+GRB detections shows that the overdensity of events largely comes from BNS systems with low inclination angles (on-axis and also some off-axis events, $0 < \iota < 15$ deg or  $165~{\rm deg} < \iota < 180$ deg). When the BNS merger is able to produce a jet and the system is oriented almost on-axis, the beamed GRB afterglow emission which is typically brighter than the KN emission enables us to detect the optical counterparts at higher redshifts. This is also aided by the fact that the GW signal of on-axis events is louder compared to the off-axis ones leading to on-axis events being well-localised even at higher $z$.

It is noteworthy that the efficiency never quite reaches 100\%, implying that not all KNe are detected even when the sources are at low redshifts. This is due to a combination of multiple factors, the first of which is the fact that we have considered realistic extinction correction for each merger based on their location on-sky. The large absorption for sources in the Galactic plane causes the light curves to become fainter than the Rubin detection limit. Analyzing sources within $z= 0.1$ (where we can limit the effect of distance making the source fainter) high extinction events make up $2-7\%$ of the events followed up by Rubin but account for $100\%$ of events not detected for ET-triangle and ET-2L for all the networks in the case of the NS Gaussian mass distribution, for both BLh and APR4 EOS. For the NS uniform mass distribution, high extinction events make up $40-75\%~(40-60\%)$ of the events not detected for the BLh (APR4) EOS. While the Gaussian mass distribution produces brighter light curves, and the only events not detected at low redshifts are due to high extinction close to the Galactic plane, the NS uniform mass distribution produces fainter light curves that cannot be detected by Rubin due to their intrinsic properties and not necessarily due to higher galactic extinction. Since BLh produces slightly brighter light curves than APR4, the percentage of events not detected due to being intrinsically faint is slightly lower for the case of BLh than for APR4.
 
\subsubsection{Network comparison: joint detection numbers versus observational time}
\label{sec:network}
The right panels of \reffig{fig:time_number_BLh} show that by setting a threshold on the sky localisation of events to be followed up by Rubin, a large increase in detections can potentially be achieved with networks of GW detectors. For example, selecting a threshold of 20 $\rm deg^2$, ET operating together with the network of current detectors (LVKI) enables us to detect an order of magnitude more events than ET operating as a single observatory. The joint GW/KN detection number further increases by a factor of two if ET operates together with one CE. On the other hand, taking as reference the fiducial population with uniform BNS mass distribution and the BLh EOS (see \reffig{fig:time_number_BLh}), if we set a threshold on the observational time allocated for GW follow-up to a value equal to 10\% (50\%) of Rubin's time, this corresponds to the sky location thresholds of 15 $\rm deg^2$ (30 $\rm deg^2$) and 5 $\rm deg^2$ (15 $\rm deg^2$) for ET+LVKI and ET+1CE, respectively. This selection results in ET+1CE detecting about 30\% (40\%) more KNe than ET+LVKI. These results show that while the GW detections within similar sky-localisation thresholds significantly increase (by about a factor of 5), we have a much lower increase in KNe detection. The detection efficiency of KNe with ET operating with new-generation detectors (CE) is not much better than that obtained with ET+LVKI. This is largely due to Rubin's detection efficiency being limited to relatively small redshifts (see \refsec{sec:efficiency}), where the network of current detectors is sensitive, and the large baselines created by distant detectors would already enable optimal sky-localisation capability with ET. In the scenario of ET+CE, there is a much larger number of GW triggers relatively well-localised to be followed up, and among them, a larger fraction is expected to not have a detectable optical counterpart due to the source being too distant and faint. In the case of ET in a network with next-generation detectors, it will be necessary to develop a more careful prioritization of the events to be followed up, which must be tailored for specific science goals and eventually increase the exposure time of each pointing. Such a prioritization would need to make use of other source parameters to identify the events to be followed (e.g. the distance and/or SNR). 

\subsubsection{ET-triangle versus 2L}
\label{sec:ETdesign}
Here, we compare the KN detection performance of Rubin operating with ET-triangle versus ET-2L, either as a single observatory or within a network of \ac{GW} detectors. As a reference example, we focus on the \textit{1ep} detection strategy with Rubin, the fiducial population, the BLh \ac{EOS}, and the uniform \ac{NS} mass distribution. Our findings reveal that the 2L configuration of ET significantly outperforms the triangular one in terms of KN detection capability when ET operates as a single observatory; the 2L-shaped configuration detects a factor of 2.2 more KNe compared to the triangular one. However, this increase is reduced to $20\%$ when ET operates in a network with LVKI and becomes negligible when ET operates together with one or two CEs. Similar results are obtained by varying the NS mass distribution and EOS. The observed difference in KN joint detection rates between ET-triangle and ET-2L for ET operating as a single observatory is closely linked to their source localisation capabilities. As discussed in \refsec{sec:GWresults}, the ET-2L has a superior ability to localise sources within an area smaller than 100 deg$^2$, identifying 2.4 times more sources than the ET-triangle. The reduced difference in the joint KN detection rates between the ET-triangle and the ET-2L when operating in a network of GW detectors is because localisation is provided by the much wider baseline provided by distant detectors of the network which allows for better triangulation based on the signal arrival time. Adding the GRB optical afterglow to the KN emission, we find that ET-2L outperforms ET-triangle, exhibiting the same percentual increase in the number of detections already observed for KN-only emission.

\begin{figure}[ht!]
\centering
\begin{subfigure}{0.49\columnwidth}
    \centering
    \includegraphics[width=\linewidth]{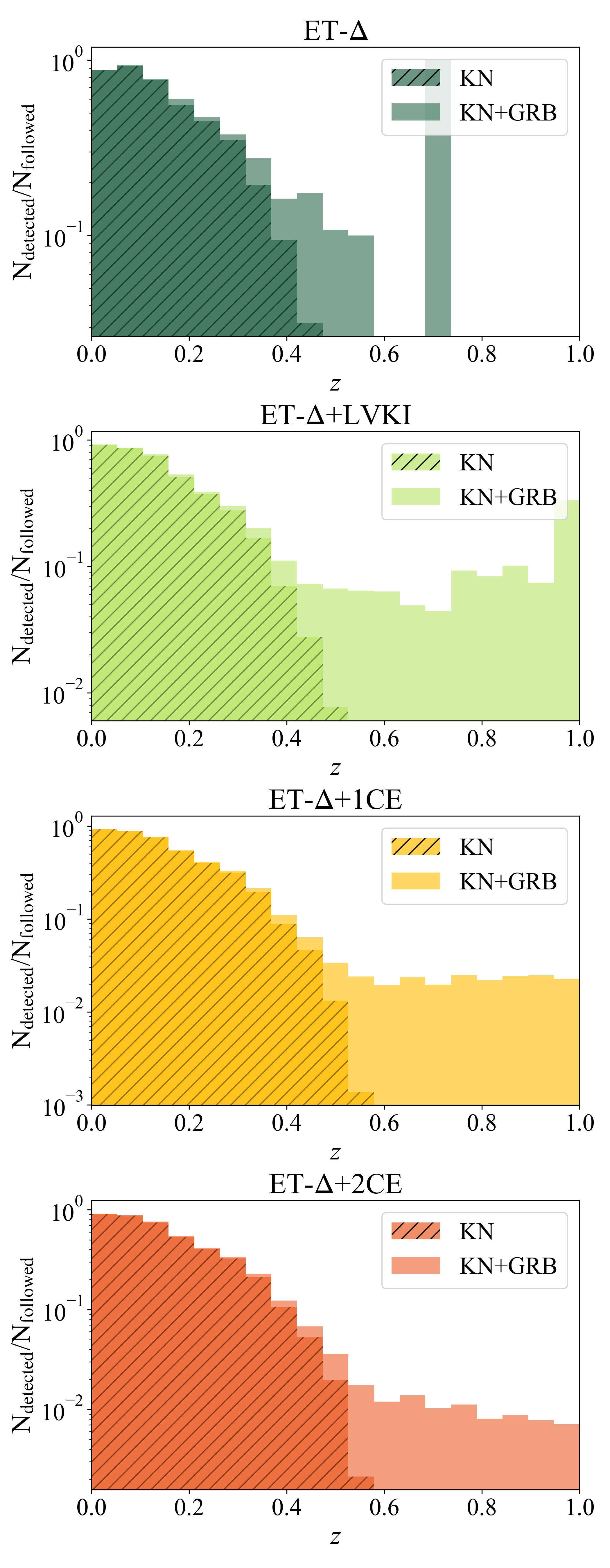}
\end{subfigure}
\hfill
\begin{subfigure}{0.49\columnwidth}
    \centering
    \includegraphics[width=\linewidth]{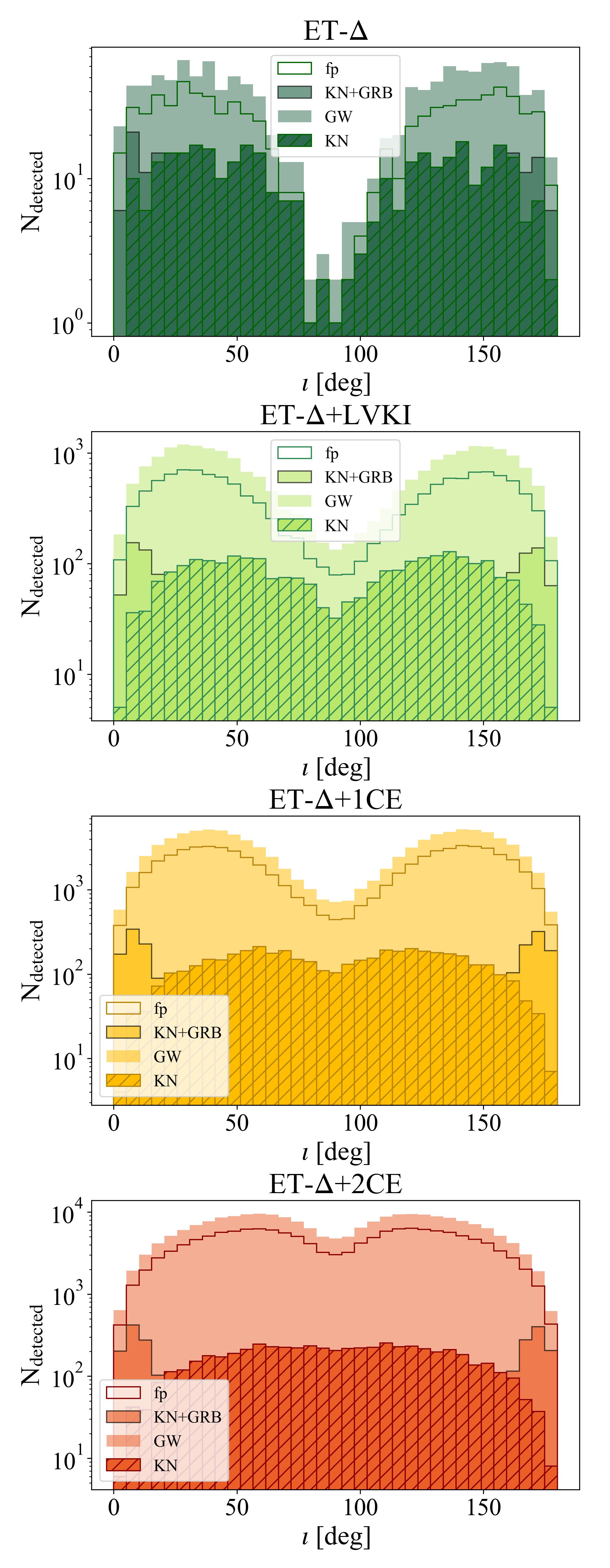}
\end{subfigure}
\caption{Efficiency of joint detections and distribution of inclination angles of BNS mergers. Left panels: efficiency of joint detections as a function of redshift for ET-$\Delta$ in each network configuration for both KN (hashed) and KN+GRB (solid) histograms. Right panels: distribution of the inclination angle, $\iota$, 
    for each network configuration indicated in the legends of the left panels. The plots show all the detected (GW) events with a sky localisation within  $\Omega_{90} < 100~{\rm deg^2}$ for \ac{ET} alone, $\Omega_{90} < 40~{\rm deg^2}$ for ET+LVKI and ET+1CE, and $\Omega_{90} < 20~{\rm deg^2}$ for ET+2CE (as used for \reffig{fig:time_number_BLh}), the events followed-up as within the Rubin footprint (fp), the corresponding KN detections, and the KN+GRB detections. The plots show the results obtained for the BLh EoS and the uniform NS mass distribution using the \textit{1ep} strategy.}
\label{fig:eff_theta_blh_uniform}
\end{figure}

\subsubsection{KN detections by Rubin}
\label{sec:impacts}
\subsubsection*{Impact of the BNS populations}
The analysis of \ac{KN} detection numbers with $\alpha = 0.5$ and $\alpha = 1.0$, as detailed in Tables \ref{tab:rubin_det_triangle} and \ref{tab:rubin_det_2L}, reveals that $\alpha = 1.0$ yields approximately a factor of 5 more detections across all configurations of \ac{GW} networks, \ac{NS} \acp{EOS} (APR4 and BLh), and \ac{NS} mass distributions (uniform and Gaussian). The same increase is found in the number of events located within a certain threshold, as reported in \reftab{tab:loc_events_triangle} and \reftab{tab:loc_events_2l}. 
Such findings are consistent with the different normalization in the local counts of BNS mergers for the two populations considered in this work.

This uncertainty is within the range of astrophysical BNS merger rates derived from the first three observation runs of LIGO and Virgo (10 to 1700 $\rm Gpc^{-3} yr ^{-1}$) \citep{Abbott2023popGWTC3}. Our fiducial population is consistent with the fourth run of LVK observations, which has been ongoing for more than a year without any BNS detection, disfavouring the most optimistic merger rate.

\subsubsection*{Impact of NS EOSs and NS mass distributions}
We explore the influence of the KN modelling on the detection rates of KNe by Rubin using the \textit{1ep} observational strategy. We focus on the results obtained with the fiducial population. To provide a clear analysis, we organise the comparison by holding either the NS mass distribution or the EOS constant. 

When the mass distribution is Gaussian, the detection rates of KNe favour the BLh EOS over APR4. Specifically, the BLh EOS results in 20\% more detections for ET operating as a single observatory. This advantage increases for ET in a network, yielding 30\% more detections with ET + LVKI, 40\% more detections with ET + 1CE, and 60\% more detections with ET + 2CEs, regardless of the ET's geometry. Conversely, when the mass distribution is uniform, the APR4 EOS leads to higher KNe detection rates than BLh; by less than 10\% for an ET operating alone and approximately 20\% for ET in networks, consistently across different ET geometries. 

With the APR4 EOS, uniform mass distributions are generally associated with higher KNe detection rates than Gaussian distributions, with the difference around 10\% across all networks\footnote{An exception is noted for the ET operating independently, where a Gaussian mass distribution results in more detections than a uniform distribution within the APR4 framework.}. For the BLh EOS, Gaussian mass distributions significantly outperform uniform distributions in KNe detections, with differences ranging between 40\% to 70\%, depending on the network configuration.

\begin{figure}[]
\centering
\includegraphics[scale = 0.465]{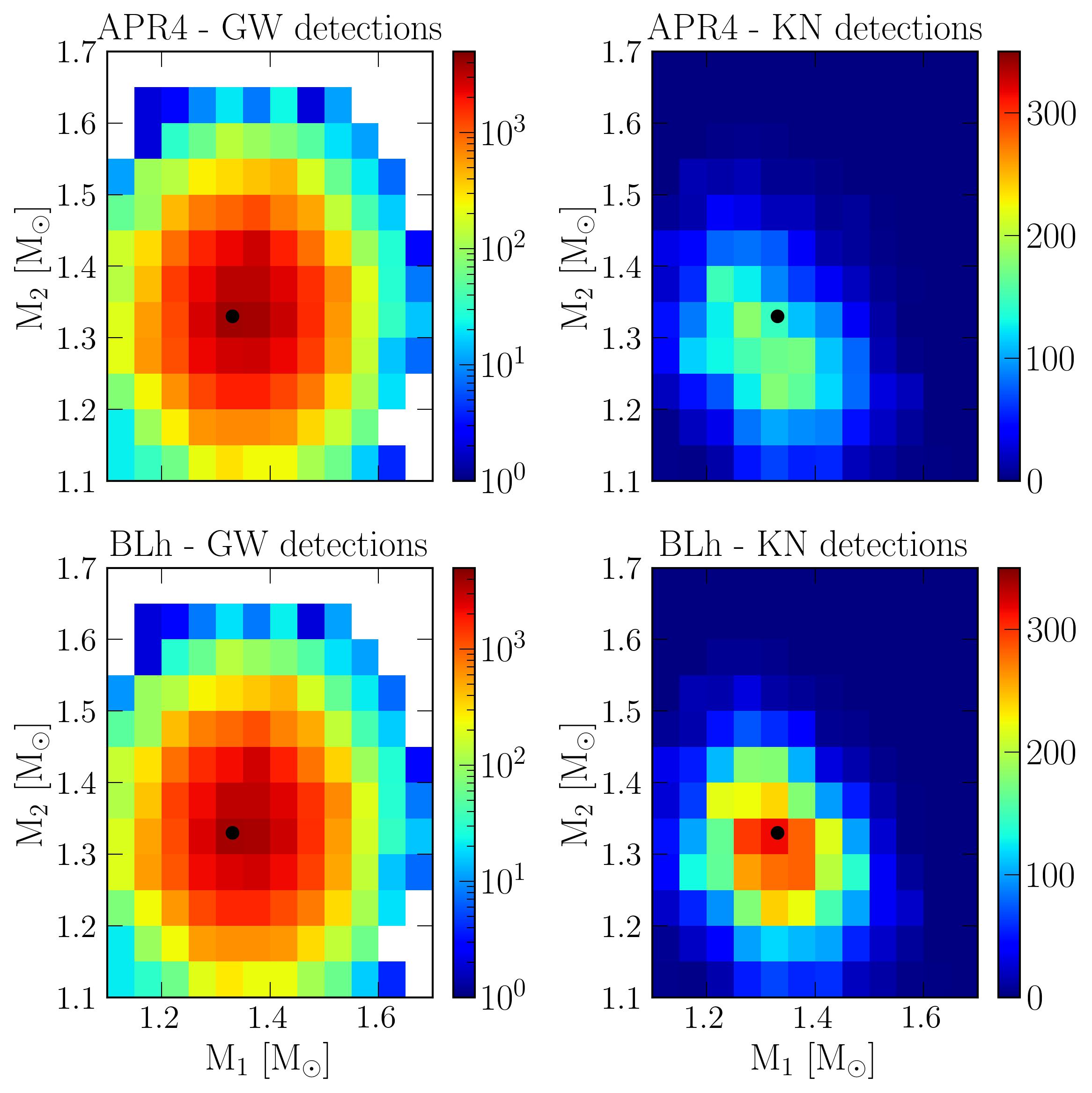}
\caption{Number of \ac{GW} and KN detections in the $M_1 - M_2$ space. The colorbar indicates the number of GW detections with sky localisation $\Omega_{90} < 5 \rm{deg}^2$ (left) and the number of corresponding KN detections for the \textit{1ep} detection strategy (right) obtained assuming APR4 (top) and BLh (bottom) EOSs, Gaussian NS mass distribution, $\alpha = 1.0$, and \ac{GW} network ET-$\Delta$ + 2 CE. The black dot shows the peak of the Gaussian mass distribution, that is $M_1 = M_2 = 1.33 \rm{M}_\odot$.}
\label{fig:hist2d_gaussian_rubin}
\end{figure}

\begin{figure}[]
\centering
\includegraphics[scale = 0.465]{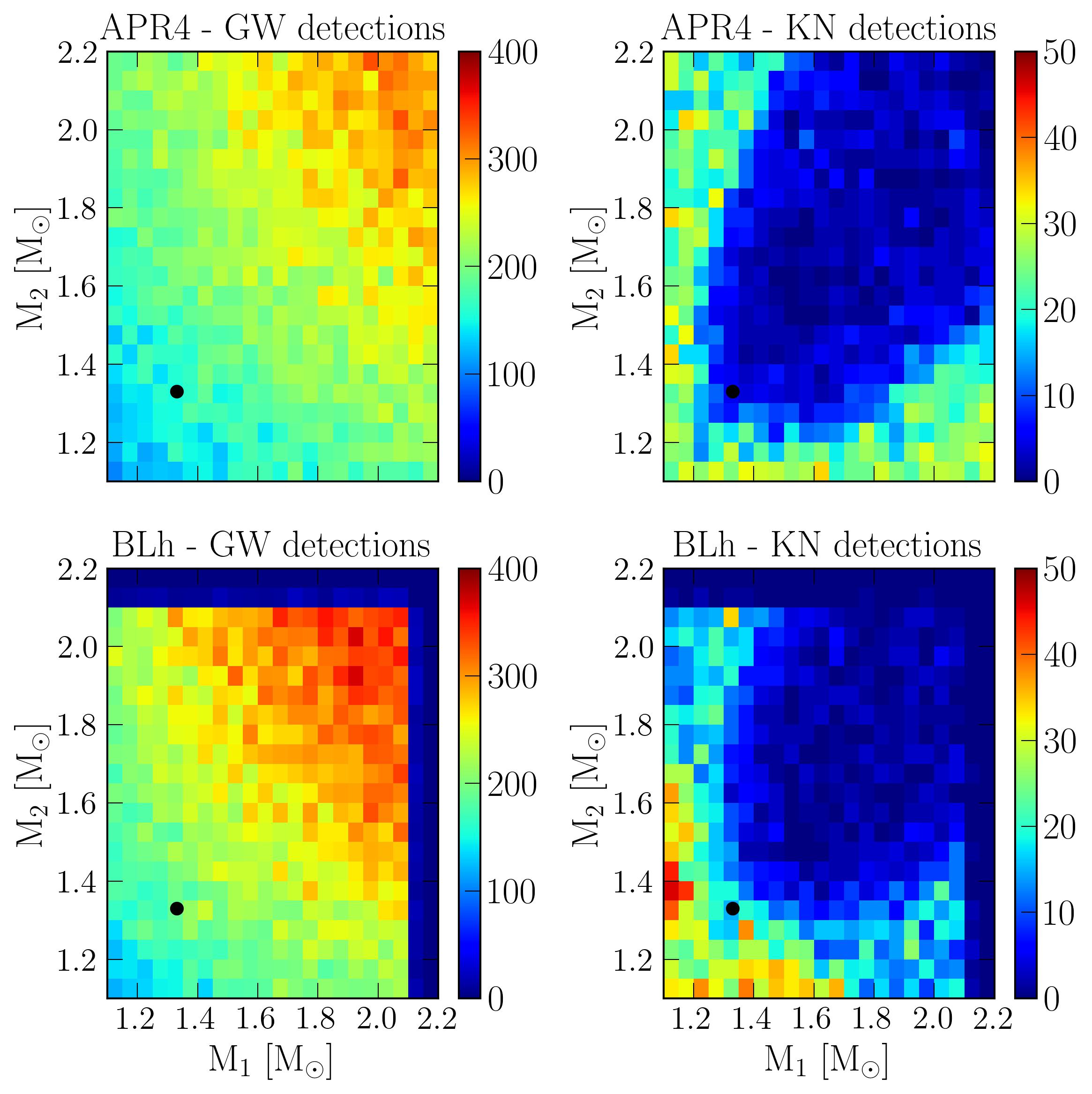}
\caption{Same as \reffig{fig:hist2d_gaussian_rubin} but for the uniform NS mass distribution. 
 The black dot shows the peak of the Gaussian mass distribution, that is $M_1 = M_2 = 1.33 \rm{M}_\odot$.}
\label{fig:hist2d_uniform_rubin}
\end{figure}

To understand these trends, we analyse and compare the number of well-localised \ac{GW} events and corresponding KN detections in the space defined by the binary masses $M_1~{\rm and}~M_2$. Figs. \ref{fig:hist2d_gaussian_rubin} and \ref{fig:hist2d_uniform_rubin} display the number of \ac{GW} detections localised by ET-triangle + 2CEs within $\Omega_{90} < 5~\rm{deg}^2$ and the corresponding KN detections, considering Gaussian and uniform mass distributions, respectively, for both APR4 and BLh \acp{EOS}. In the case of Gaussian NS mass distribution, \ac{NS} masses span from $1.1 M_\odot$ to approximately $1.7 M_\odot$, maintaining an identical component mass distribution for both the \acp{EOS} under consideration (also refer to \reffig{fig:mass_lambda_gaussian}). 
This reflects on the distribution of \ac{GW} well-localised events in the $M_1 - M_2$ space, which does not depend on the \ac{EOS}, consistently peaking near the peak of the Gaussian mass distribution $M_{1,2} = 1.33 M_\odot$, as shown in the left plots of \reffig{fig:hist2d_gaussian_rubin}. 
Conversely, the distribution of \ac{KN}e detections within the $M_1 - M_2$ plane exhibits a pronounced \ac{EOS} dependence. For both the EOSs, the KN detections lie in the region corresponding to binaries with total mass below the prompt collapse mass threshold (lower left quadrant). However, the BLh 
presents a peak around $M_{1,2} \simeq 1.33 M_\odot$, which is instead suppressed in the case of APR4 \ac{EOS}, and more \ac{KN} detections across the entire $M_1 - M_2$ space. Using the Gaussian mass distribution, we explore a portion of the $M_1 - M_2$ space where BLh EOS predicts ejecta masses, $m_{\rm C1}$ and $m_{\rm{C2}}$, %, to reach their maximum, 
largely exceeding those predicted by APR4, as shown by comparing \reffig{fig:BLh_ej_comp} and \reffig{fig:APR4_ej_comp}.
Therefore, BLh generates brighter KNe enhancing their detectability by Rubin.

Unlike the Gaussian distribution, the uniform mass distribution scenario shows a distinct EOS-dependent variation in the distribution of well-localised GW events within the $M_1 - M_2$ space (see \reffig{fig:hist2d_uniform_rubin}). Although in the case of uniform NS mass distribution there is a trend of having more well-localised GW events with the APR4 EOS compared to BLh reaching up to a 10\% increase, as detailed in \refsec{sec:GWresults}, \reffig{fig:hist2d_uniform_rubin} illustrates how for binaries with component masses $M_{1,2} \leq 2.1 M_\odot$ the BLh EOS is associated with a higher rate of GW detections per mass bin compared to APR4. This discrepancy arises because all mergers of the population (injected events) are spread across a broader $M_1 - M_2$ space for APR4 than for BLh since APR4 accommodates a larger maximum NS mass (see \reftab{tab:eos_prop}). To better understand the pattern observed in KNe detections, it is instructive to divide the binary mass space ($M_1 - M_2$) into three distinct regions, each characterised by a different distribution of KNe detections. These regions are delineated based on the masses of the BNSs and their mass ratios ($q$): 
\begin{enumerate}
    \item low-mass binaries, including binaries with individual masses ranging from $1.1 M_\odot$ to $1.6 M_\odot$ (lower left corner of KN detections plots in \reffig{fig:hist2d_uniform_rubin}). The total mass of the majority of these binaries is below the prompt collapse mass threshold.
    \item high-mass and symmetric binaries, including binaries with individual masses ranging approximately from $1.6 M_\odot$ to $2.20 M_\odot$, with mass ratios close to unity $q \sim 0.8 - 1.0$ (upper right corner of KN detections plots in \reffig{fig:hist2d_uniform_rubin}).
    \item highly asymmetric binaries, including binaries where one NS is significantly heavier than its companion, with masses for the heavier component in the  $1.6 M_\odot - M_{\rm max}$ range and the lighter one in the $1.1 M_\odot - 1.6 M_\odot$ range, resulting in mass ratios $q \sim 0.5 - 0.7$. The total mass of the majority of these systems surpasses the prompt collapse mass threshold.
\end{enumerate}
\reffig{fig:hist2d_uniform_rubin}, specifically the plots on the right, reveals that BLh yields a larger number of \ac{KN}e detections compared to APR4 within the low-mass binary region. This outcome aligns with expectations, as this portion of the $M_1 - M_2$ space is characterised by a significantly higher ejected mass for BLh than for APR4 (see \reffig{fig:BLh_ej_comp} and \reffig{fig:APR4_ej_comp}), in analogy with the case of Gaussian mass distribution. 
The region of the $M_1 - M_2$ space populated by high-mass and symmetric binaries, instead, is characterised by a minimal amount of ejected mass across both the \acp{EOS}, leading to uniformly dimmer KNe. This outcome results in KNe emissions that frequently fall beneath Rubin's sensitivity threshold, as set in our strategy, and are not detected.
Finally, in the region of highly asymmetric binaries, the dominant contribution to the ejected mass comes from material unbound from the accretion disc. Interestingly, the amount of ejected mass is relatively similar for both the APR4 and BLh EOSs, as depicted in Figures \ref{fig:BLh_ej_comp} and \ref{fig:APR4_ej_comp} and discussed in \refsec{sec:ejecta_properties}. Within our modelling, when the ejected masses are comparable across different EOSs, the differences in the brightness and temporal evolution of the KN emission become primarily influenced by the choice of nuclear factor, $f_{\dot{\epsilon}}$, which multiplies the nuclear heating rate. Notably, in order to correctly reproduce AT2017gfo, we associate APR4 with a higher value of $f_{\dot{\epsilon}}$. Given that the ejected mass and its velocity remain essentially the same across both EOSs, the APR4 EOS leads to KNe which are not only brighter but also exhibit a slower fading pattern in their light curves. According to our analysis, approximately 20\% of the well-localised binaries fall within the low-mass binary region, where BLh results in brighter KNe. Conversely, about 50\% of the well-localised binaries are situated in the highly asymmetric binary region, which favours APR4 in terms of producing brighter KNe. Therefore, the total count of KNe detections is higher for the APR4 EOS.

This analysis also explains why, if we fix the EOS and vary the mass distribution, we observe an opposite trend in KNe detections for APR4 and BLh. For the APR4 EOS, a uniform mass distribution enhances KNe detection rates, as it samples regions of the $M_1 - M_2$ space where APR4 generates KN light curves brighter than Rubin's sensitivity threshold. These regions are instead excluded from the Gaussian mass distribution. In contrast, for the BLh EOS, the Gaussian mass distribution proves more advantageous in maximizing detection rates. This is because it concentrates binary systems within the zone of peak ejected mass specific to the BLh parameter space, thereby optimizing the conditions for KNe detections. Interestingly, among the different combinations of EOSs and NS mass distributions, the BLh EOS with Gaussian (uniform) mass distribution is the one yielding the highest (smallest) number of KN detections for all the analysed configurations of GW networks.

\subsubsection*{Comparison among \textit{1ep} and \textit{2ep} strategies}
Here, we analyse the perspectives for KNe detections using the more stringent two-epoch Rubin's detection strategy (\textit{2ep}). As summarised in Table \ref{tab:rubin_strategy}, the events detected according to the \textit{2ep} criterion are a sub-sample of the \textit{1ep} ones. We recall here that the \textit{2ep} strategy is intended to provide the possibility of monitoring the colour evolution of the KN from the first to the second epoch, in order to be able to better identify the KN counterpart among many contaminant transients. We thus require a detection in $i$ and $g$ bands during the first epoch of observations and a detection in at least the $i$ band in the second epoch of observations, accounting for the possibility that the bluer component which is expected to have a faster decline might go under the detection limits in the second epoch. The numbers of light curves detected in the \textit{1ep} strategy and not detected in the \textit{2ep} one give us an idea of how many real KNe we might miss with this more stringent strategy.

The reduction in the number of detections going from \textit{1ep} to \textit{2ep} is stable against variations of the EOS or the NS mass distribution, while it is quite sensitive to the GW detector network. In the case of ET-triangle, the number of detections from \textit{1ep} to \textit{2ep} decreases by $20-35\%$ for ET operating alone, by $30-40\%$ for ET+LVKI, by $40-45\%$ by ET+1CE, and by $55\%$ for ET+2CEs. The decrease from \textit{1ep} to \textit{2ep} is very similar for ET-2L.

The KN light curves not detected in the \textit{2ep} for ET operating as a single observatory are mainly the ones that evolve too rapidly and become fainter than the detection limit in the $i$ and $g$ filters on the second night of observation. For ET operating in a network of current and next-generation detectors, about half of the light curves are missed for the same reason but the other half are missed because they are too faint to be detected in the $i$ filter on the first night of observations. These latter undetected events represent those for which the red KN component is slowly rising on the first night, and at the time of the first epoch of observation is not yet bright enough to be detected in the $i$ filter. They are prominent especially for detector networks because networks localise BNS better, hence the time to cover the entire mosaic of sky localisation is smaller (making the observations closer to the merger time). This factor could be mitigated by performing multiple multi-filter observations during the first night. The remaining events coming from non-detections in both $g$ and $i$ filters on the second night represent fast-evolving KNe which become fainter than the detection limit on the second night irrespective of the efficiency of sky localisation. In this case, a network of optical telescopes around the world could help to cover with shorter cadence observations during the Rubin daytime. 

\subsubsection{KN and GRB afterglow detections by Rubin}
Taking the fiducial population and the one epoch detection strategy (\textit{1ep}) as a reference, we analyse the detection rate when, in addition to the KN emission, we include the GRB afterglow. Compared to the case of KN-only emission, the rate of detections increases by 5-10\% for ET-triangle operating alone or with LVKI, 10-20\% for ET-triangle operating with one CE, and by 15-30\% for ET-triangle operating in synergy with two CEs. The probability of detecting the GRB afterglow is enhanced by GW detector networks providing better sky localisation at larger redshifts, such as ET with two CEs. Making it possible to access larger distances keeping a good sky-localisation capability, increases the probability to detect face-on BNS systems whose optical GRB afterglow emission is bright enough to be detected by Rubin at those distances.

The increase in the number of detections going from the KN-only scenario to the KN+GRB is more significant for the pessimistic population than the fiducial one. Specifically, for the pessimistic population the detection rate increases by 10-20\% for ET-triangle alone or operating with LVKI, 10-30\% for ET-triangle operating with one CE, and by 20-50\% for ET-triangle operating in synergy with two CEs. This difference between the populations with $\alpha = 0.5$ and $\alpha = 1.0$ can be explained by the fraction of BNS mergers powering a successful jet, $f_j$. As discussed in \refsec{sec:afterglow_modeling}, we calibrate $f_j$ on the observed local GRB rate, obtaining a larger $f_j$ for populations with $\alpha = 0.5$, where we find $f_j = 1$, compared to populations with $\alpha = 1.0$, where $f_j = 0.7$. Therefore, the stronger enhancement in the detection rate observed for $\alpha = 0.5$ follows from the higher percentage of BNS powering a relativistic jet, and thus a short GRB.

\subsection{Comparison with Rubin operating with current GW detectors and their upgrades}
\label{sec:current}
Considering that joint detections are limited by Rubin's efficiency in detecting KNe, this section compares the performance of ET with networks of current GW detectors and their upgrades. In particular, we evaluate the performance of LVKI and three A\# (3A\#) operating without ET. A\# is the upgrade project of LIGO, expected to be implemented after the end of O5 and to start operations around 2030 \citep{Fritschel2022}. For LVKI, we used the same configurations used for ET+LVKI, that is based on the best sensitivities expected for the fifth observing run \citep{Abbott2020LRR_senscurves} and considering an operating LIGO detector in India. For the upgrade of current detectors, we considered a scenario with three A\# two located in the USA and one in India. The LVKI and 3A\# represent possible scenarios operating around 2030 and before the next-generation detectors come into operation. 

\begin{figure}
\includegraphics[scale=0.4]{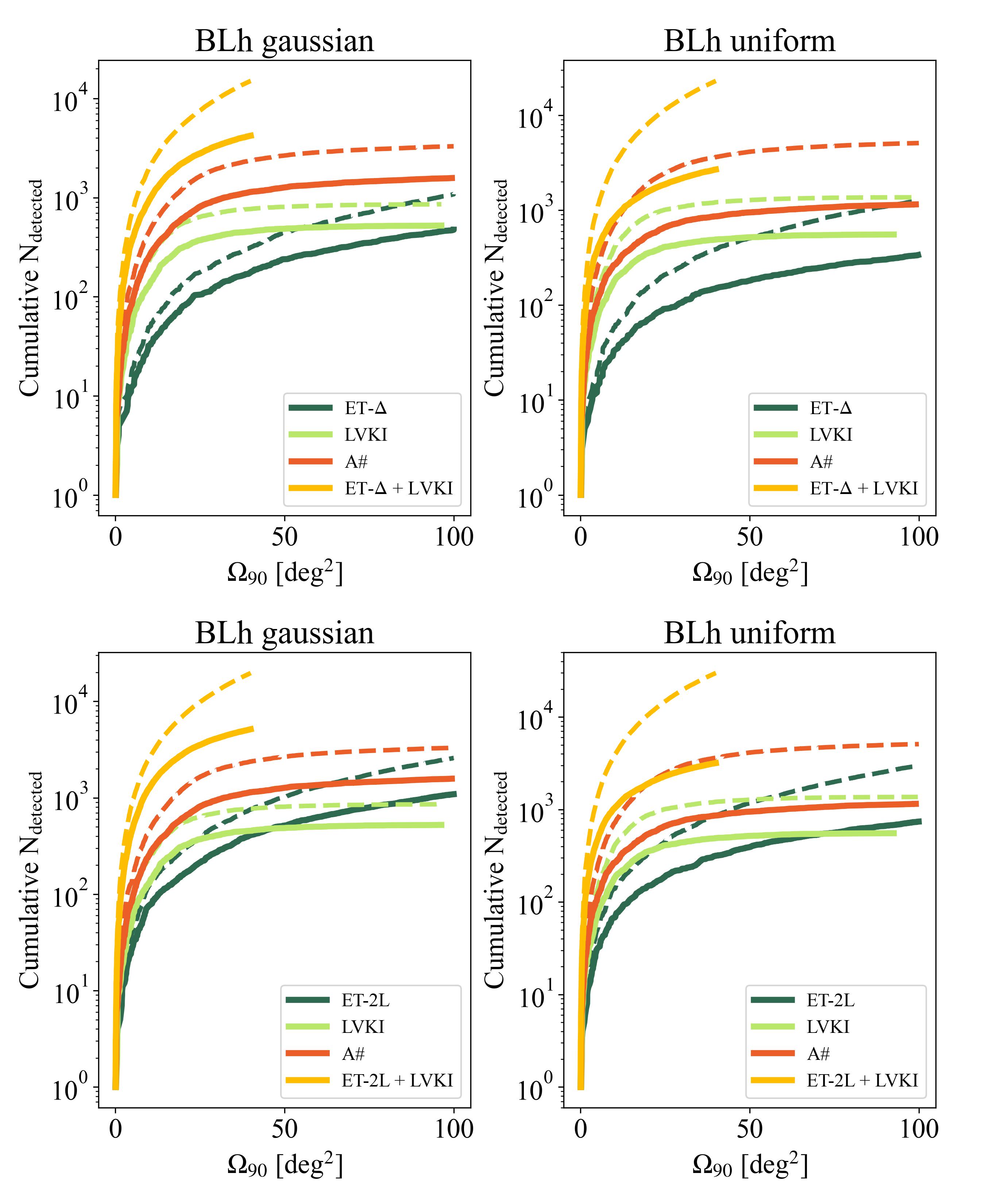} 
\caption{Cumulative number of GW detections (dashed lines) and corresponding KN counterpart detections by Rubin (solid lines) as a function of the sky localisation for upgrade configurations of the current detectors; LVKI and 3A\#. As a comparison, we show ET alone and in a network with LVKI. The top panels refer to ET-$\Delta$, while the bottom panels to ET-2L. The shown results are obtained using the fiducial population, the BLh EOS, both the uniform and Gaussian NS mass distributions, and \textit{1ep} strategy.}

\label{fig:skyloc_lvki_blh}
\end{figure}

\reffig{fig:skyloc_lvki_blh} shows the number of GW detections in ten years and the corresponding GW/KN joint detections for Rubin operating with the LVKI and 3A\# networks. We consider the fiducial population and BLh EOS. For comparison, both the ET-triangle (top panels) and 2L (bottom panles) configurations are shown. The same figure for APR4 EOS is provided in \refapp{app:joint_GW_EM}. Table \ref{tab2gen} summarises the results for both KN-only and KN plus GRB afterglow emissions.
\begin{table*}[]
\caption{Number of joint detections by Rubin operating in synergy with current generation detectors (LVKI) and their upgrade (3A\#).}

\makebox[\textwidth][c]{
\scalebox{0.88}{
\begin{tabular}{l l l l c r c r c r c r}
\hline
& & 
&\multicolumn{4}{c}{\textbf{APR4}}
&\multicolumn{4}{c}{\textbf{BLh}}\\
\hline
& $\Omega_{90}$ [deg$^2$] &Transient 
&\multicolumn{2}{c}{Uniform} 
&\multicolumn{2}{c}{Gaussian}
&\multicolumn{2}{c}{Uniform}
&\multicolumn{2}{c}{Gaussian}\\
& & &Followed&Detected &Followed&Detected &Followed&Detected &Followed&Detected\\
& & & &1ep/2ep &&1ep/2ep &&1ep/2ep &&1ep/2ep\\
\hline
\hline
\multirow{2}{*}{LVKI} 
&\multirow{2}{*}{100} 
&KN 
&\multirow{2}{*}{981 (5.2\%)} &$635/425$ &\multirow{2}{*}{562 (2.9\%)} &$497/409$ &\multirow{2}{*}{907 (4.8\%)} &$554/399$ &\multirow{2}{*}{562 (2.9\%)} &$523/496$\\
&& KN+GRB &&$652/434$ &&$503/415$ &&$575/411$ &&$527/501$\\
\hline
\multirow{2}{*}{3A\#} 
&\multirow{2}{*}{100} 
&KN 
&\multirow{2}{*}{3735 (27.9\%)} &$1353/852$ &\multirow{2}{*}{2211 (16.2\%)} &$1230/853$ &\multirow{2}{*}{3421 (25.5\%)} &$1151/735$ &\multirow{2}{*}{2201 (16.3\%)} &$1578/1157$\\
&& KN+GRB &&$1443/906$ &&$1281/885$ &&$1247/776$ &&$1608/1198$\\
\hline
\end{tabular}
\label{tab2gen}
}
}\tablefoot{The table lists the results obtained for the fiducial BNS population and corresponds to ten years of observations.}
\end{table*}

\begin{figure}
\includegraphics[scale=0.4]{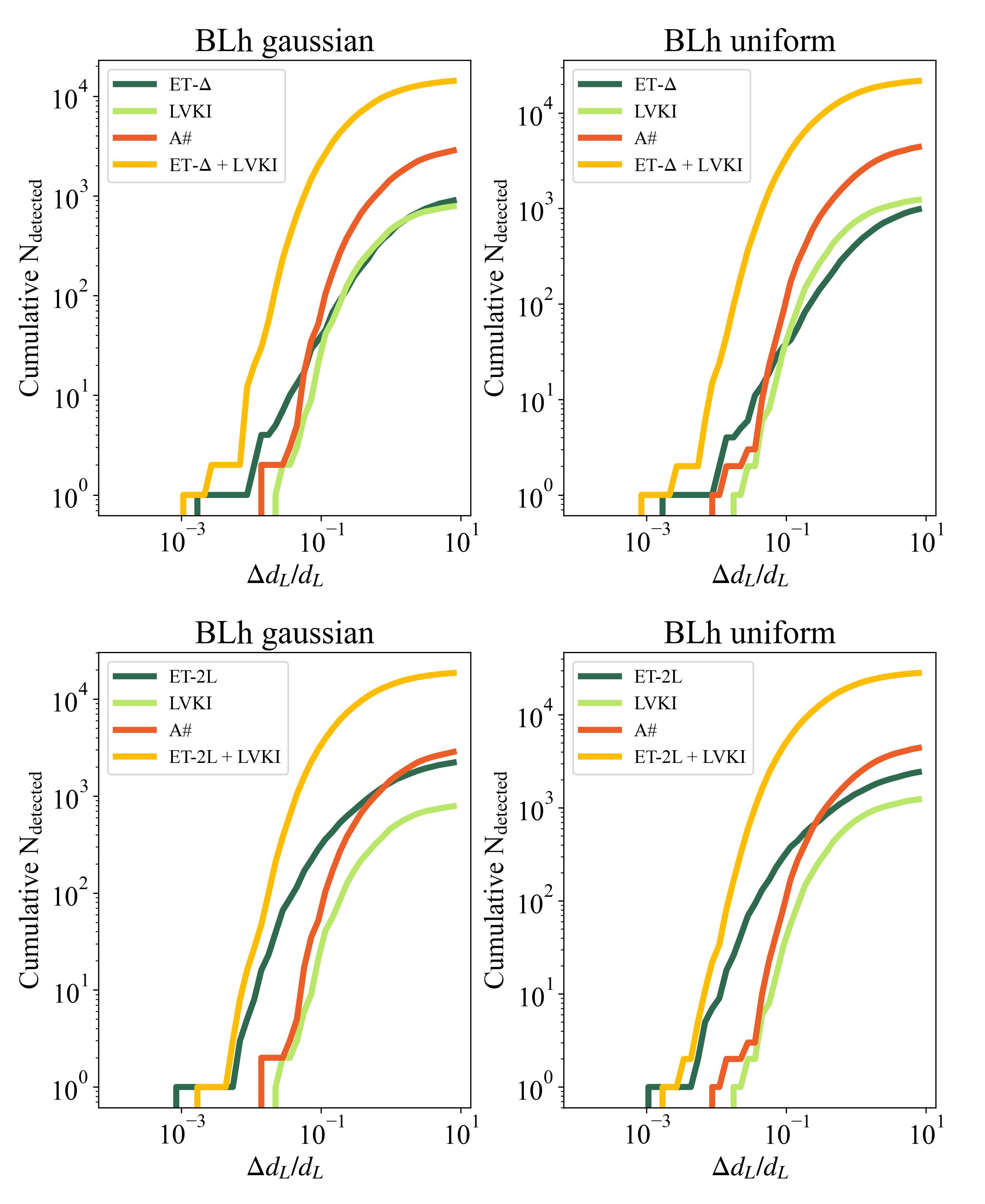} 

\caption{Parameter estimation capability of different GW detectors (ET alone, LVKI, 3A\#, ET+LVKI) in terms of relative error in the luminosity distance ($\Delta d_L/d_L$) corresponding to the GW detections of Figure \ref{fig:skyloc_lvki_blh}. The top panels refer to ET-$\Delta$, while the bottom panels to ET-2L. Each curve represents a different GW detector network, and the y-axis shows the cumulative number of GW detections with a relative uncertainty on $d_L$ smaller than the values in the x-axis. The maximum $\Omega_{90}$ used in the analysis is set at 100 $\rm deg^2$ for each GW network, except for ET+LVKI for which it is set at 40 $\rm deg^2$. The shown results are obtained using the fiducial population, the BLh EOS, both the uniform and Gaussian NS mass distributions, and \textit{1ep} strategy.}
\label{fig:ddlvdl_blh}
\end{figure}

The upper panels of \reffig{fig:skyloc_lvki_blh} show that for both NS mass distributions, the network of current detectors (LVKI) and the network of LIGO upgrade (3A\#) outperforms ET-triangle operating as a single observatory in terms of sky-localisation capabilities for events localised better than 100 $\rm deg^2$.  This is reflected in the joint GW/KN detection numbers. Comparing \reftab{tab2gen} and \reftab{tab:rubin_det_triangle}, we find that Rubin operating with LVKI can detect 60\%-80\% (10\%-30\%) more KNe than ET-triangle as a single observatory for the uniform (Gaussian) NS mass distribution, with a larger increment in the case of APR4 than the BLh EOS. Rubin, operating together with 3A\#, is able to detect a factor  3-4 more KNe compared to Rubin with ET-triangle. Interestingly, the lower panels of \reffig{fig:skyloc_lvki_blh} show that ET-2L alone outperforms LVKI and approaches the performance of 3A\# in sky localisation. In the case of both the triangle and 2L, including ET in a network of current detectors (ET+LVKI) vastly outperforms LVKI and 3A\#. Rubin operating with ET+LVKI detects a factor 3-4 more events than 3A\#, following up events localised within $\Delta \Omega=40\ {\rm deg}^2$. Consistent results are obtained for the APR4 EOS (see \reffig{fig:skyloc_lvki_apr4}). 

\reffig{fig:ddlvdl_blh} shows the comparative performance of the ET compared to the LVKI, 3A\#, and ET+LVKI networks in estimating the relative uncertainty on the luminosity distance. Focusing on the events with relative errors smaller than 10\%, ET-triangle (top panels) operating alone performs better than LVKI and 3A\# by showing a tail of detections with tiny uncertainties where neither LVKI nor 3A\# are capable of the same performance. In the case of ET-2L (bottom panels), ET alone largely outperforms LVKI and 3A\#. ET, LVKI and 3A\# are vastly outperformed by ET operating in a network with current detectors. ET+LVKI provides almost an order of magnitude more detections with relative errors smaller than 10\% than ET alone, LVKI and 3A\#. It provides several dozen detections with relative errors of less than 1\%. Consistent results are obtained for the APR4 EOS (see \reffig{fig:ddlvdl_apr4}).

\subsection{Rubin observations with longer exposures}
\label{sec:deeperexp}
\begin{figure}
\centering
\includegraphics[width=0.5\textwidth]{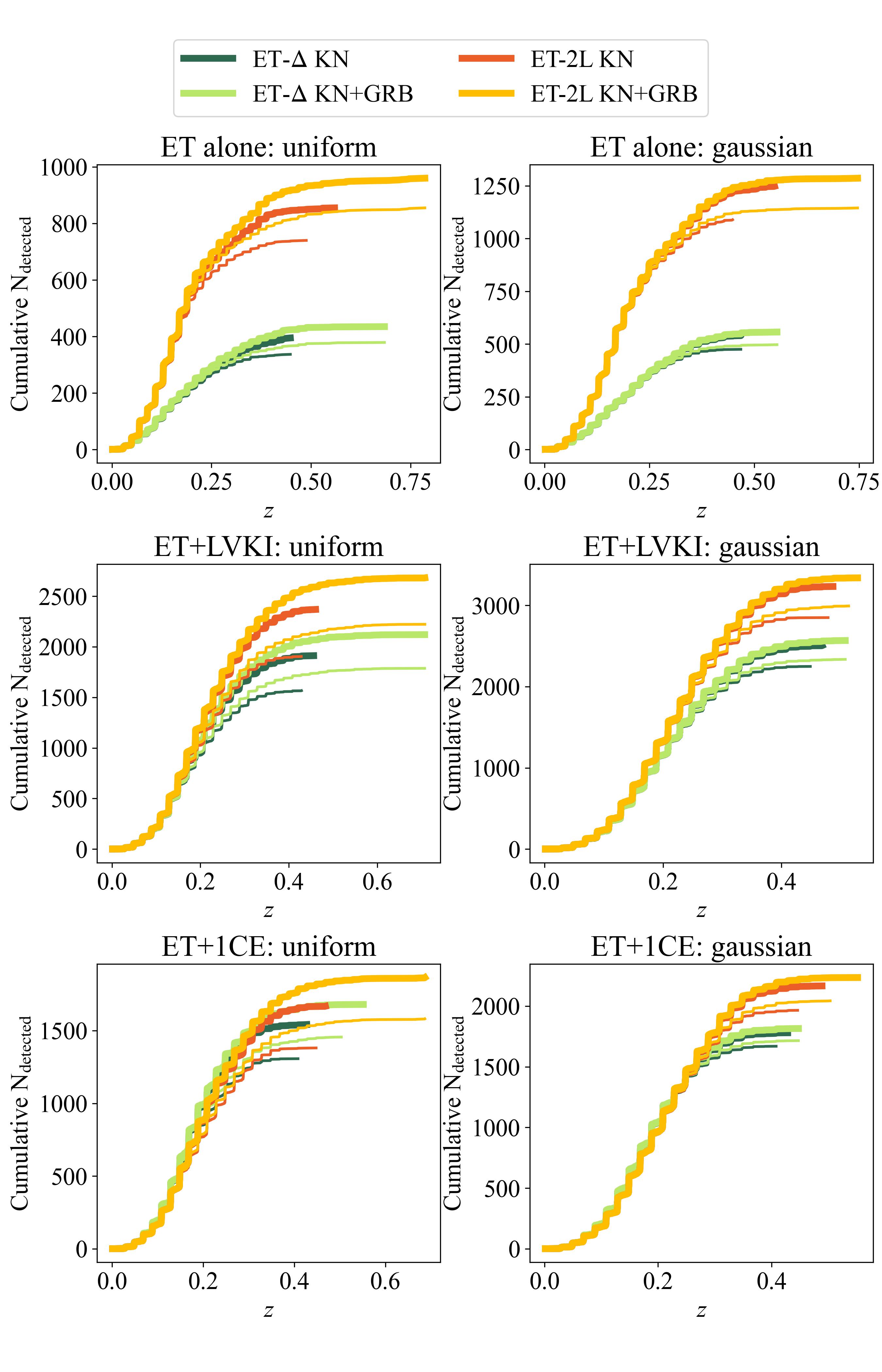}

\caption{Cumulative number of KN and KN+GRB counterpart detections obtained as a function of the redshift for ET for both ET-triangle and 2L. The thick lines indicate the 1200 s exposure strategy and the thin lines the 600 s exposure. The detections were obtained following up all the events with a sky localisation smaller than 100 $\rm deg^2$ for ET alone (top panels), 20 $\rm deg^2$ for ET+LVKI (middle panels) and 5 $\rm deg^2$ for ET+1CE (bottom panels). The left panels show the results obtained for the uniform NS mass distribution, while the right panels show the ones obtained for the NS Gaussian distribution. All plots refer to the BLh EOS. 
}
\label{fig:deeper_blh}
\end{figure}

To explore possible improvements in Rubin's efficiency for detecting KNe, we evaluated the perspectives with an alternative strategy that increases the exposure time for each pointing from 600 s of our nominal strategy to 1200 s. Using the same mosaic strategy with observations in two epochs and in the $g$ and $i$ filters, the time needed for Rubin to follow up all selected events in the footprint using deeper exposures of 1200 s doubles the budget of our nominal strategy. \reffig{fig:deeper_blh} shows the cumulative number of KN and KN+GRB counterpart detections obtained with 1200 s exposures as a function of the redshift for ET alone, ET+LVKI, and ET+CE for both the triangle and 2L geometries. The plots show the results obtained for the fiducial population, BLh EOS, the NS uniform and Gaussian mass distributions. For comparison, the plots also show the cumulative numbers obtained for the nominal observational strategy (600s) using thinner lines. We use a sky-localisation threshold of 100 $\rm deg^2$ to select the events to be followed-up for ET alone, of 20 $\rm deg^2$ for ET+LVKI, and 5 $\rm deg^2$ for ET+CE. Equivalent plots for APR4 EOS are shown in \reffig{fig:deeper_apr4} in \refapp{app:joint_GW_EM}.

For ET operating as a single observatory (selecting GW alerts with sky localisation < 100 deg$^2$), there is an increase of about 15-20\% for the KN detections and 10-15\% for the KN+GRB detections both for the triangle and 2L. The observational time increases to 20\% of the total Rubin time budget for ET-triangle and 50\% for ET-2L. Interestingly, we observe (more prominently for the 2L configuration) that the major increase of KN detections happens at redshifts larger than 0.2, showing an increase in Rubin's detection efficiency at larger $z$.

The increase at larger redshift is also observed for ET+LVKI and ET+CE. In absolute numbers, an exposure of 1200 s makes the detections increase by 11-23\% (13-24\%) for the KN (KN+GRB) in the case of ET-triangle (ET-2L) observing in a network with LVKI using 21-34\% (28\%-44\%) of the Rubin total observational time. For ET+CE, the increase in the number of detections is 7-20\% (10-24\%) for the KN (KN+GRB) for the ET-$\Delta$ (ET-2L). The required observational time is limited to 8-14\% for the ET-triangle and 10-18\% for the ET-2L. 

For all cases analysed, the longer exposure increases the number of joint optical/GW detections. Although the increase is limited in absolute number to less than 25\%, it concentrates at relatively high redshifts allowing a good number of detection at redshifts otherwise inaccessible or with low detection numbers. This is crucial for studies requiring access to high \textit{z}, such as cosmological parameter estimation.

\section{Summary and conclusions}
\label{sec:summary}

In this work, we explored the prospects of multi-messenger astronomy in the \ac{ET} era by evaluating the number of detections of \ac{KN}e and \ac{GRB} optical afterglows associated with BNS mergers. We considered \ac{ET} operating as a single observatory or within a network of current or next-generation \ac{GW} detectors, and observing in synergy with Rubin.

Starting from populations of BNS mergers up to $z=1$ generated by population synthesis codes, we conducted 64 simulations over ten years of observations to compute the detection capabilities and parameter estimation by ET, ET + LVKI, ET + 1CE, and ET + 2CE, considering two possible configurations for ET (triangular with 10 km arms or 2L with 15 km arms). To account for present uncertainties in the local merger rate, NS mass distribution, NS EOS, and KN modelling, we considered a pessimistic and a fiducial BNS merger population with local merger rates $\mathcal{R}_{\rm BNS} = [23, 107]~\rm{Gpc^{-3}~yr^{-1}}$, two NS mass distributions (uniform and Gaussian), and two \acp{EOS} (APR4 and BLh). The APR4 (BLh) \ac{EOS} reproduces more (less) compact \acp{NS}. 

We evaluated the properties of the ejecta for each BNS merger based on the system progenitor properties and we associated each merger with an optical light curve. We calibrated the \ac{KN} ejecta components based on the results of numerical-relativity simulations targeted to GW170817 and GW190425, introducing new fitting formulas, and modelled the impact of prompt-collapse on the properties of the ejecta. Taking into account the KN ejecta from high-mass BNS mergers, based on the results of GW190425-targeted numerical relativity simulations, makes our population of KNe including fainter and faster-evolving light curves compared to models calibrated using only GW170817-targeted numerical relativity simulations or AT2017gfo data (\reffig{fig:kn_comparison_at2017gfo} in \refapp{app:KNmodeling}). We also highlight that in the case of a uniform NS mass distribution, we account for the possibility of having highly asymmetric binaries ($q \sim 0.5 - 0.7$). These systems eject a large amount of matter ($\gtrsim 0.02 M_\odot$) from the tidally disrupted disc, leading to the formation of bright KNe.

To reproduce a realistic scenario, we also added the afterglow emission from the relativistic jet of typical short GRBs. The light curves were corrected for the Galactic extinction based on the source position. We then implemented a realistic follow-up strategy with Rubin, including Rubin's accessible sky, average seeing, slewing and filter change times, and night-day duty cycle.

Our findings indicate that \ac{ET} already operating as a single observatory will enable detections of a large number of \ac{KN}e in the range of $10-100$ per year, with uncertainties firstly dominated by the local BNS merger rate, then by the \ac{ET} configuration (triangle versus 2L), and finally dependent on the modelling of \ac{NS} mass distribution, \ac{EOS}, and \ac{KN} microphysics. The number of \ac{KN} multi-messenger detections significantly increases when \ac{ET} is operating in a network with current or next-generation detectors up to several hundreds of \ac{KN}e per year.\\

\noindent
In the following, we provide a summary of major results from our study:\\

\begin{itemize}
\item \ac{ET} as a single detector will increase the number of BNS merger detections of two orders of magnitude with respect to the current detectors (LVKI) for mergers happening within $z < 1$. Our analysis shows the rate of \ac{GW} detections is enhanced by $70-90\%$ when ET operates with one or two \acp{CE}, whereas the inclusion of \ac{LVKI} results in a modest 3\% increase. 

\item \ac{ET} as a single observatory is able to localise long signals, such as the BNS ones, thanks to the imprint of the Earth rotation in the inspiral signal lasting long in the ET observable band, which starts at relatively low frequencies. However, the number of well-localised events dramatically increases when ET operates in a network of GW detectors. For the fiducial population, the number of events localised better than 100 deg$^2$ up to $z=1$ increases from a few hundred per year to several thousand when ET operates with LVKI. Thousands of events per year localised better than 10 deg$^2$ are detected by ET observing in a network with CE.

\item  The increase of well-localised events of ET+LVKI makes the detected optical counterparts by Rubin increase by an order of magnitude than ET operating as a single observatory. However, the further increase of detections when Rubin operates with ET+CE does not correspond to the increase of well-localised events but it is limited to a factor of a few. This is mainly due to the Rubin detection efficiency associated with our strategy which already drops significantly at redshifts larger than 0.3. This makes it impossible to gain from the improved localisation accuracy at higher $z$ provided by ET+CE because Rubin's sensitivity is not sufficient to detect KN emission (whereas systems producing an optical afterglow from a jet are observable). 

\item The use of deeper Rubin observations (and thus spending more Rubin's time for the follow-up) does not dramatically increase the absolute numbers of detections but interestingly significantly increases the probability of detecting KNe at larger $z$. To limit the amount of Rubin time required when ET will operate with CE, it will be crucial to select the events to be followed up by prioritising them on the basis of specific scientific targets to achieve. The selection will need to use source parameters beyond the sky-localisation accuracy  (e.g. distance, SNR, masses).

\item The combination of sky-localisation capabilities by the GW network
and the KN detection efficiency by Rubin makes the networks of current GW detectors (LVKI) and a possible scenario of their upgrades (3A\#) outperforming ET-triangle as a single observatory; operating with 3A\#, Rubin can detect a factor 3-4 more KNe than ET-triangle. The better sky localisation of ET-2L with respect to ET-triangle makes ET-2L outperform LVKI and approach the performance of 3A\#. However, it is important to underline that ET (both geometries) enables us to better estimate crucial parameters such as the source distances. ET+LVKI significantly outperforms LVKI and 3A\# in terms of number of joint detections and parameter estimation. 

\item The dominant source of uncertainty in the GW detection rate is the still uncertain local BNS merger rate and its evolution with redshift.  Since we limit our analysis to events with $z < 1$, the shape of the evolution is expected to be similar for different populations, as shown in \cite{Santoliquido2021MNRAS}, and the main uncertainty thus remains associated with the normalization. The difference in the normalization of our fiducial and pessimistic populations is consistent with the local merger rate range as constrained by current \ac{GW} detectors, and gives the magnitude of this uncertainty. Another source of uncertainty on the detection rate is given by the poor knowledge of the NS mass distribution. Here, we find that a uniform distribution leads to $20-25 \%$ more detections than a Gaussian distribution. On the contrary, the EOS has a small impact (less than 5\%) on the number of detections.

\item The uncertainty in the absolute number of KNe detections is also dominated by the poorly constrained local rate of BNS mergers. The NS mass distribution, EOS, and KNe modelling impact the detection rates, albeit to a lesser extent. These factors interplay in complex ways, making their individual effects challenging to disentangle. In the case of a Gaussian NS mass distribution, the BLh EOS (with nuclear factor $f_{\dot{\epsilon}} = 1.5$) yields more KN detections than APR4 ($f_{\dot{\epsilon}} = 2.75$) with an increase ranging between 20\% and 60\% depending on the selected GW network of detectors. Conversely, when adopting a uniform NS mass distribution, the scenario reverses. In this case, KNe detections under the APR4 framework ($f_{\dot{\epsilon}} = 2.75$) exceed those predicted under the BLh framework ($f_{\dot{\epsilon}} = 1.5$) by 10\% to 20\%, varying with the GW network configuration.

\item Once the jet's optical afterglow emission is included, the rate of multi-messenger detections further increases, especially for the GW network ET + 2CE. This network can localise the sources at larger redshifts where the afterglow (intrinsically brighter than the KN emission) is detectable. Such an increase varies with the GW detector networks, the NS mass distribution and EOS, and the KN modelling, ranging between 5\% and 30\% for the fiducial BNS merger rate, and between 10\% and 50\% for the pessimistic one.
\item ET, especially in networks with next-generation GW observatories, will generate an extremely large number of alerts, making it prohibitive to follow up on every one of them. Here, we propose a simple observational strategy based on sky-localization selection, however, searches can be further optimised to be effective based on specific science cases. The selection and prioritization of alerts to be followed will require incorporating additional information about the source parameters. For instance, nuclear physics studies could focus on golden SNR events, which offer the best parameter estimates; cosmological studies could prioritize alerts based on source distance estimates; and studies of ejecta physics and nucleosynthesis could be guided by component masses. In the latter case, our study provides predictions of the expected ejecta and light curves based on the component masses. To explore science-case-dependent observational strategies based on source parameter estimates, we make available to the community the catalogues containing the uncertainty estimate for all the source parameters for the BNS merger populations and networks presented in this paper\footnote{The parameter estimation catalogues are publicly available at \href{https://zenodo.org/records/13850416}{this link}.}.
\end{itemize}

Our study provides a comprehensive overview of KN multi-messenger prospects for the next-generation \ac{GW} observatories operating with innovative optical wide field-of-view observatories, such as the Vera Rubin Observatory. 
It makes a significant step forwards in quantifying theoretical uncertainties affecting predictions on KNe multi-messenger detections, which need to be taken into account to devise robust observational strategies. As highlighted in the previous paragraphs, the largest source of uncertainty is given by the rate of BNS mergers. In addition to that, the mass distribution within BNS systems is still poorly constrained on both the sides of observational data and population synthesis simulations. By using either a Gaussian or a uniform NS mass distribution, our analysis explores the two extreme scenarios, including very asymmetric BNS systems in the case of uniform mass distribution. Both the total mass and the binary mass ratio of the two NSs have a profound impact on the brightness and time evolution of the KN light curves. Further theoretical studies are needed to understand how frequent these asymmetric systems ($q \lesssim 0.7$) are in the BNS mergers population. The time that bridges current detectors to the ET era is expected to improve our knowledge of BNS and the associated KN emission which are currently mainly constrained by only two BNS merger detections in GWs and a single KN associated with GW170817. Upcoming observations will provide important information (on BNS merger rate, NS properties, KN emission) to overcome current analysis limitations. 

\small{
\textit{Acknowledgements} M.B. and E.L. acknowledge financial support from the Italian Ministry of University and Research (MUR) for the PRIN grant METE under contract no. 2020KB33TP. M.B. and F.S. acknowledge financial support from the AHEAD2020 project (grant agreement n. 871158). E.L. acknowledges funding by the European Union – NextGenerationEU RFF M4C2 1.1 PRIN 2022 project 2022RJLWHN URKA. This work has received funding from the European Research Council (ERC) under the European Union’s Horizon 2020 research and innovation programme, Grant agreement No.\ 770017. M.M. acknowledges support from the Deutsche Forschungsgemeinschaft (DFG) under Germany’s Excellence Strategy EXC 2181/1-390900948.
We acknowledge useful discussion with M. Arca Sedda, S. Borhanian, A. Colombo, M. Drago, 
 G. Ghirlanda, F. Gulminelli, D. Logoteta, A. Maselli, J. Tissino.
We acknowledge Stefano Bagnasco, Federica Legger, Sara Vallero, and the INFN Computing Center of Turin for providing support and computational resources.
}

\section*{Data availability}
The output data of the GW simulations presented in this work (see \refsec{sec:GWsimul} and \refsec{sec:GWresults}) are publicly available on \href{https://zenodo.org/records/13850416}{Zenodo}
at \cite{dupletsa_2024_13850416}. \texttt{GWFish} is publicly available on \href{https://github.com/janosch314/GWFish}{Github}: the version used in this work is the commit \textsc{cf26d3c} in the main branch. 

\bibliographystyle{aa}
\bibliography{biblio}

\begin{appendix}

\section{Properties of the BNS populations}\label{app:A}
\begin{figure}[h!]
    \includegraphics[scale = 0.52]{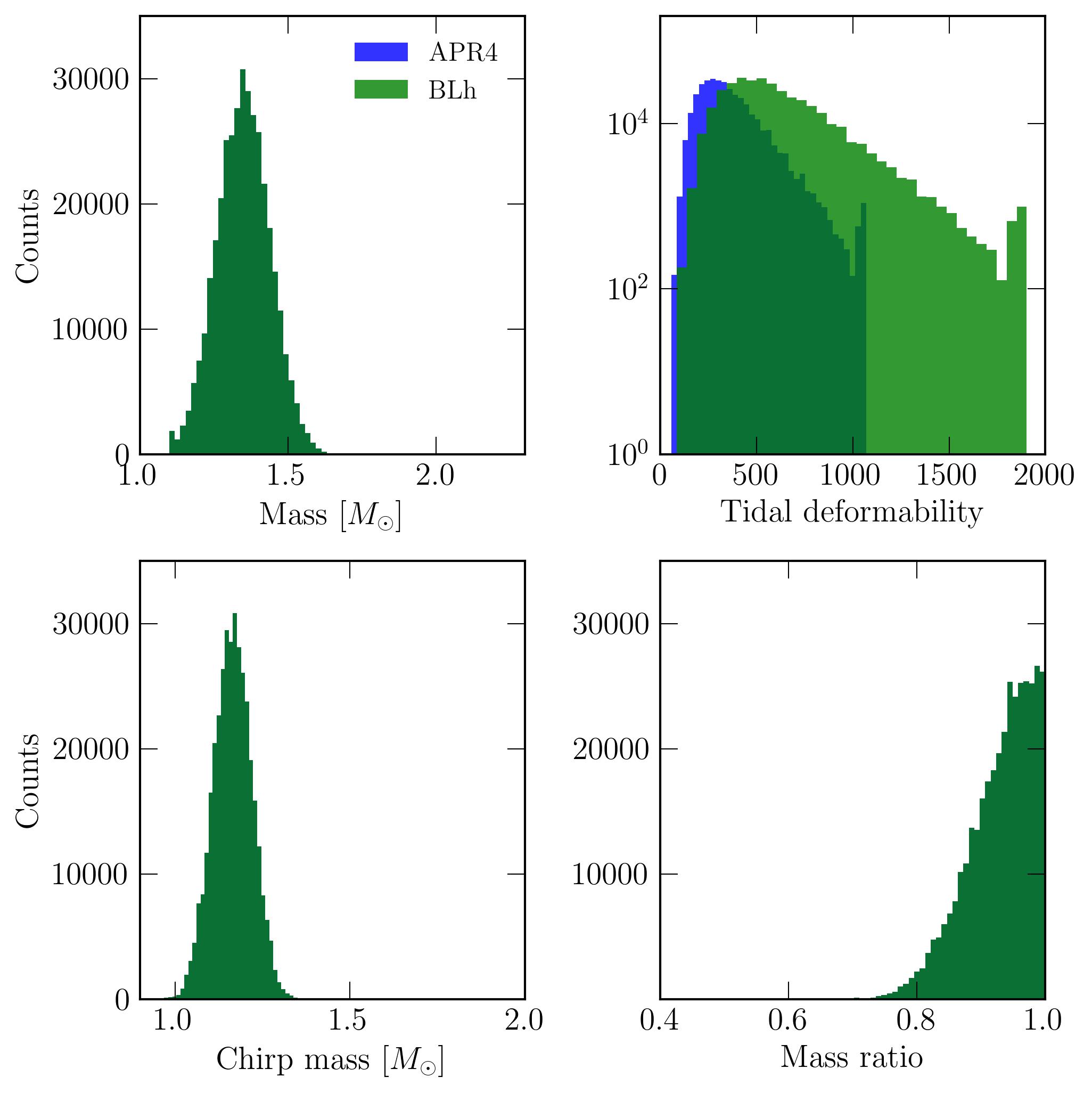}
    \caption{ Distribution of NS masses and tidal deformability defined in \refeq{eq:NS_tidal_def} (top) and BNS chirp masses and mass ratios (bottom) corresponding to the population with $\alpha = 1.0$ and Gaussian mass distribution, for APR4 and BLh \acp{EOS}. }
    \label{fig:mass_lambda_gaussian}
    \end{figure}

\begin{figure}[h!]
    \includegraphics[scale = 0.52]{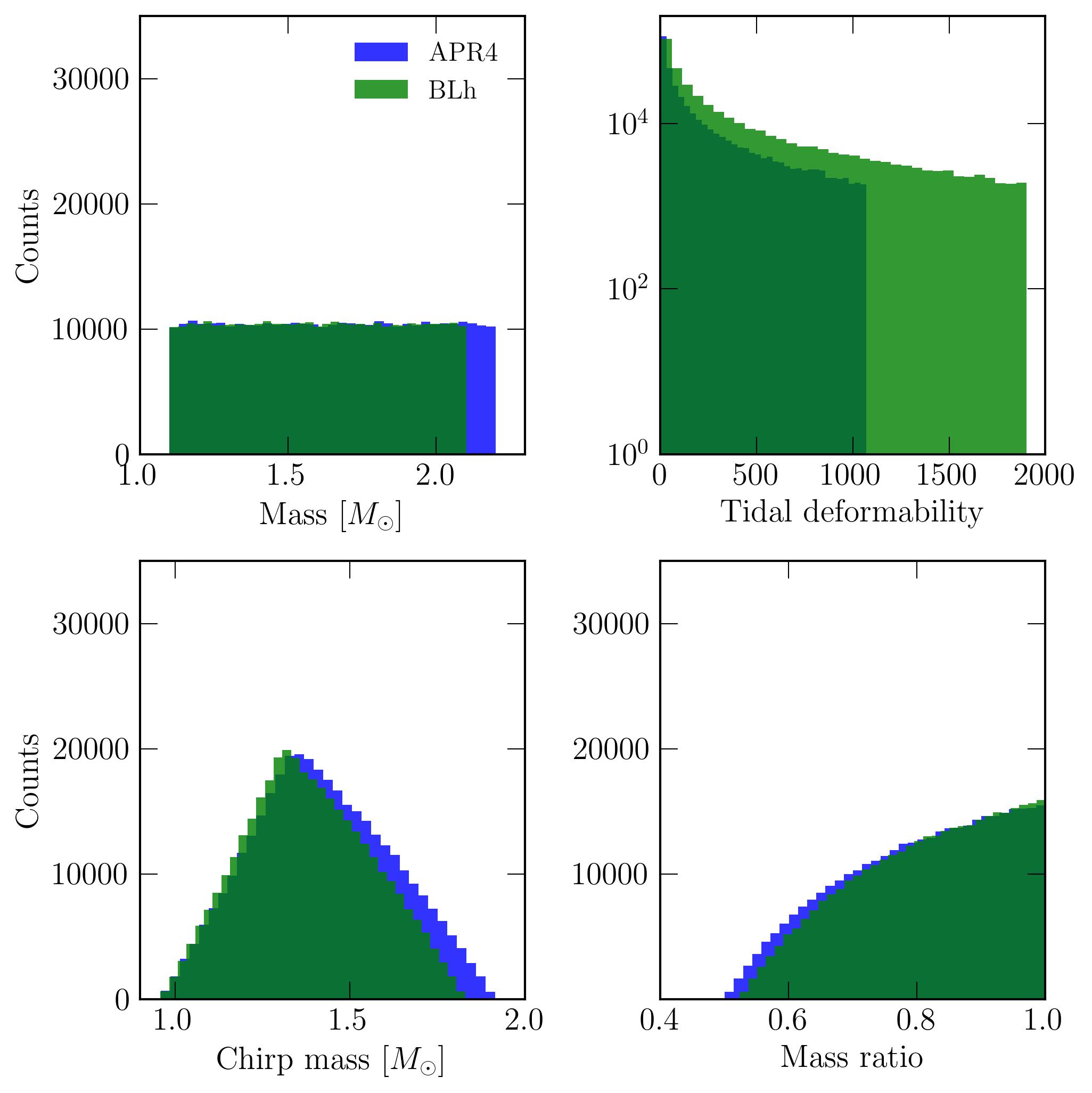}
    \caption{Distribution of NS masses and tidal deformability defined in \refeq{eq:NS_tidal_def} (top) and BNS chirp masses and mass ratios (bottom) corresponding to the population with $\alpha = 1.0$ and uniform mass distribution, for APR4 and BLh \acp{EOS}. }
    \label{fig:mass_lambda_uniform}
    \end{figure}

Here we present the properties of the BNS populations analysed in our work. 
Figs. \ref{fig:mass_lambda_gaussian} and \ref{fig:mass_lambda_uniform} illustrate the distributions of NS mass and tidal deformability (as defined in \refeq{eq:NS_tidal_def}), as well as BNS mass ratios and chirp masses for the populations with Gaussian and uniform mass distribution, respectively.\\
Notably, when comparing the two \acp{EOS}, the most significant differences are observed in the tidal deformability distributions. The BLh EOS shows generally larger tidal deformability values compared to APR4. For the uniform mass distribution (see \reffig{fig:mass_lambda_uniform}), the maximum masses attained are larger for APR4, leading to subtle variations in both the mass ratio and chirp mass distributions between APR4 and BLh.\\
Furthermore, a comparison between the Gaussian and uniform mass distribution populations (\reffig{fig:mass_lambda_gaussian} and \reffig{fig:mass_lambda_uniform}) reveals marked differences in the distributions of tidal deformability, mass ratio, and chirp mass. Particularly in the population with a uniform mass distribution, we see a broader distribution of chirp masses, extending to almost 2, and mass ratios descending to values as low as 0.5. This indicates a significant presence of asymmetric systems within the uniformly distributed mass population. This has a significant impact on the KN light curves. However, the asymmetric mass ratio of our systems is constructed a posteriori and does not come from the real evolution of the binaries.

\FloatBarrier

\section{GW detectors characteristics}\label{app:GW_detectors}

\begin{figure}[hb!]
\centering
\includegraphics[scale=0.6]{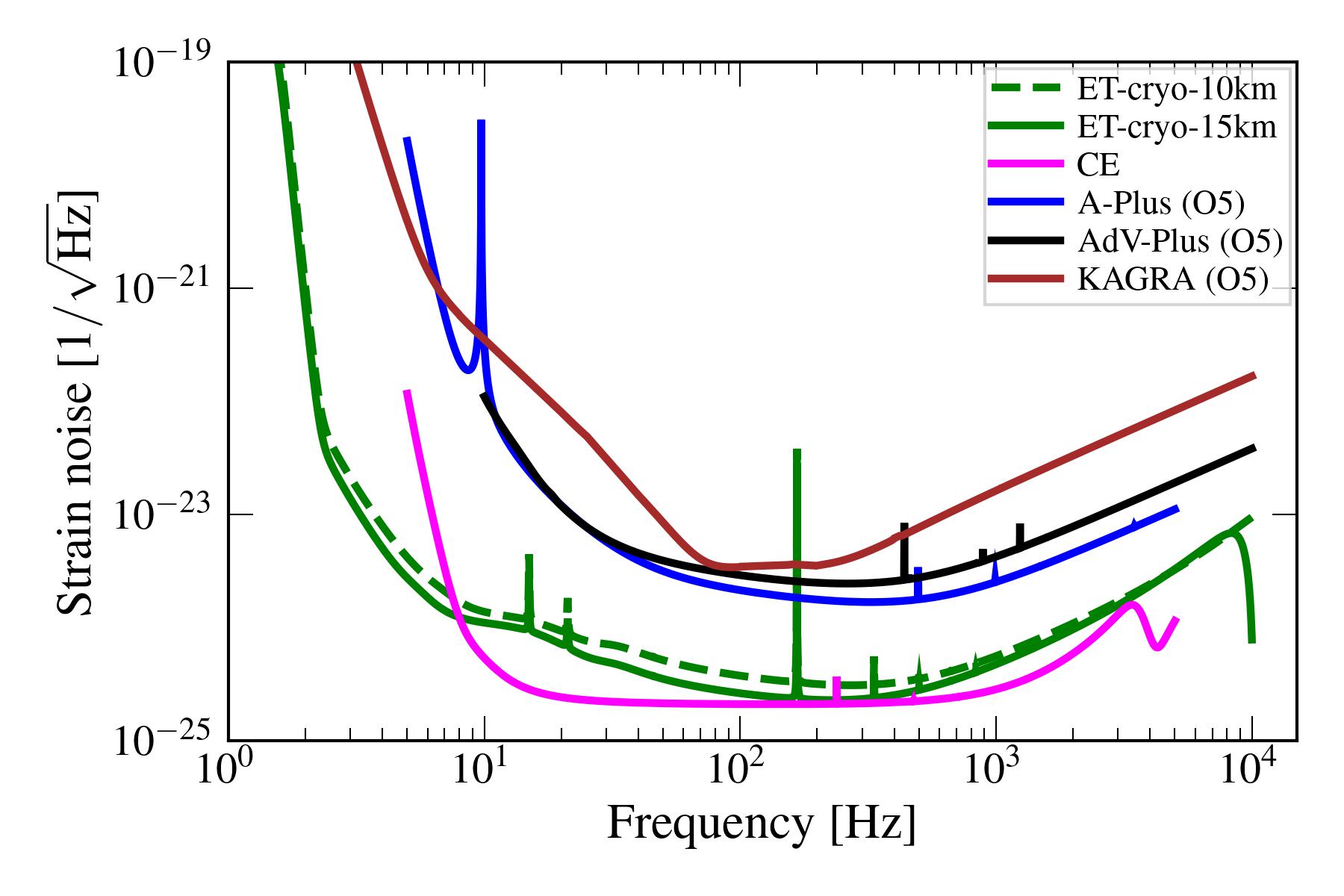}
\caption{Sensitivity curves used in the analysis. The ET strain noise is plotted considering a single nested detector, composed of a low-frequency and a high-frequency interferometer. Our analysis takes into account that the triangular-shaped ET consists of three nested detectors, each of them with the same sensitivity curve.}
\label{fig:sens_curves}
\end{figure}

In \reftab{tab:ifos} we summarise the main characteristics of the GW detectors in our simulation set-up. Specifically, we show the arm length, the location (country, latitude and longitude), the minimum frequency used for the analysis, and the adopted sensitivity curve. The sensitivity curves are then shown in \reffig{fig:sens_curves}.

\begin{table*}[ht!]
\centering
\caption{Detectors list and their main characteristics.}
{\small
\begin{tabular}{l r r r r r r}
\hline
\textbf{IFO name} &\textbf{Length} [km] &\textbf{Country} &\textbf{Latitude} [deg] &\textbf{Longitude}[deg] &\textbf{f$\_$min} [Hz] &\textbf{Sensitivity curve}\\
\hline
\rowcolor{Gray}
ET-$\Delta$ &10 &Italy &40+31/60 &9+25/60 &2 &ET-cryo-10\,km\\
\hline
\rowcolor{Gray}
ET-L1 &15 &Italy &40+31/60 &9+25/60 &2 &ET-cryo-15\,km\\
\hline
\rowcolor{Gray}
ET-L2 &15 &Netherlands &50+43/60 &5+55/60 &2 &ET-cryo-15\,km\\
\hline
\rowcolor{Gray}
CE1 &40 &USA &46+30/60 &-119+24/60 &8 &CE\\
\hline
\rowcolor{Gray}
CE2 &40 &Australia &-25+30/60 &152+4/60 &8 &CE\\
\hline
LIGO-Hanford &4 &USA &46+28/60 &-119-24/60 &8 &A-Plus (O5)\\
\hline
LIGO-Livingston &4 &USA &30+34/60 &-90-46/60 &8 &A-Plus (O5)\\
\hline
Virgo &3 &Italy &43+36/60 &10+30/60 &8 &  AdV-Plus (O5)\\
\hline
LIGO-India &4 &India &19+37/60 &77+2/60 &8 &A-Plus (O5)\\
\hline
KAGRA &3 &Japan &36+25/60 &137+18/60 &8 &KAGRA (O5)\\
\hline
\end{tabular}
}
\tablefoot{We have highlighted in grey the future third-generation ground-based detectors.}
\label{tab:ifos}
\end{table*}

\FloatBarrier

\section{Complements to kilonova modelling}\label{app:KNmodeling}

In this section, we provide additional figures and tests complementing the description of our KN modelling from \refsec{sec:kn_modeling}.

\begin{figure}[h!]
    \centering
    \includegraphics[scale = 0.363]{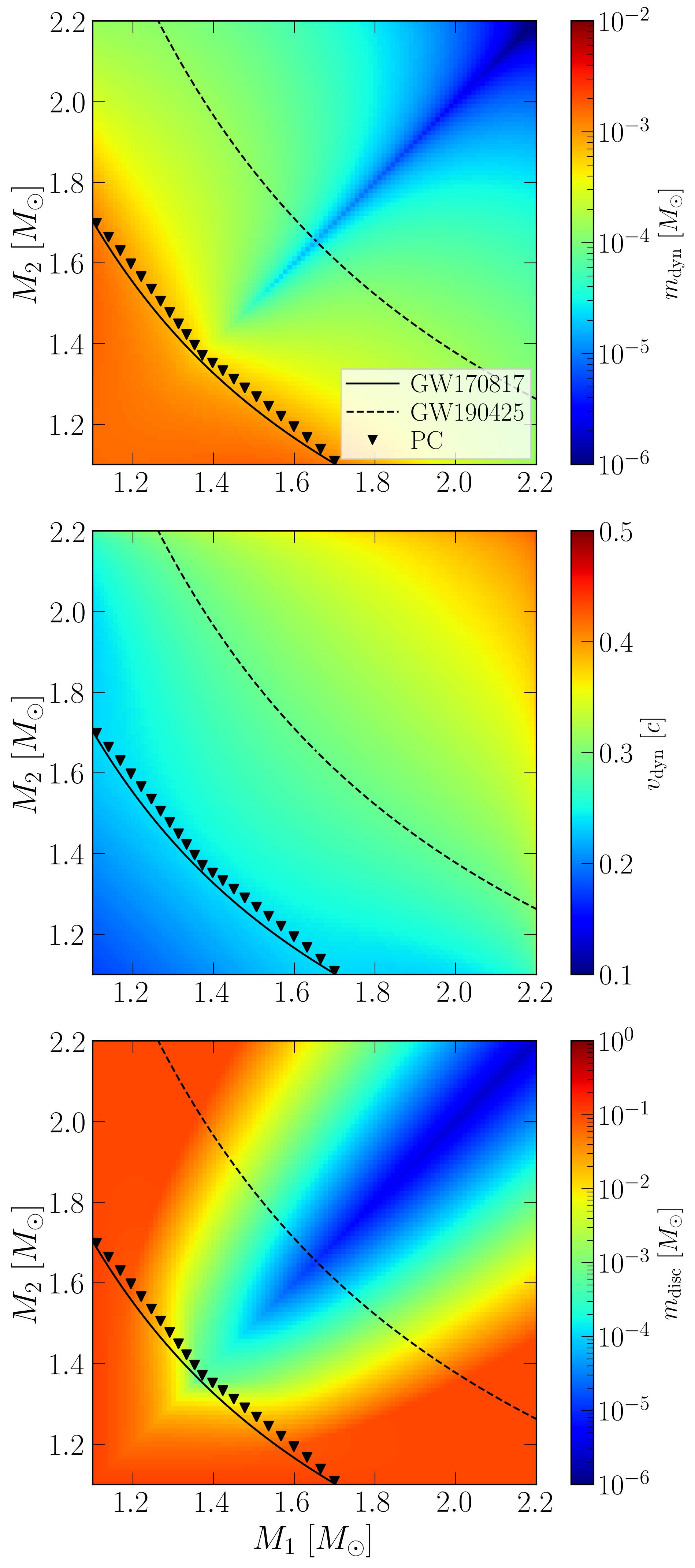}
    \caption{Mass of the dynamical ejecta (top), velocity of the dynamical ejecta (centre), and mass of the disc (bottom), as defined in \refeq{eq:mdyn}-(\ref{eq:mdisc}) for the \ac{EOS} APR4. $x$ and $y$ axes represent the BNS masses $M_1$, $M_2$. The solid and the dashed black lines mark the chirp mass of GW170817 and GW190425, respectively. The triangular markers show the prompt-collapse mass threshold $M_{\rm PC}$.}
    \label{fig:APR4_all_fits}
\end{figure}

\begin{figure}[]
    \centering
    \includegraphics[scale = 0.37]{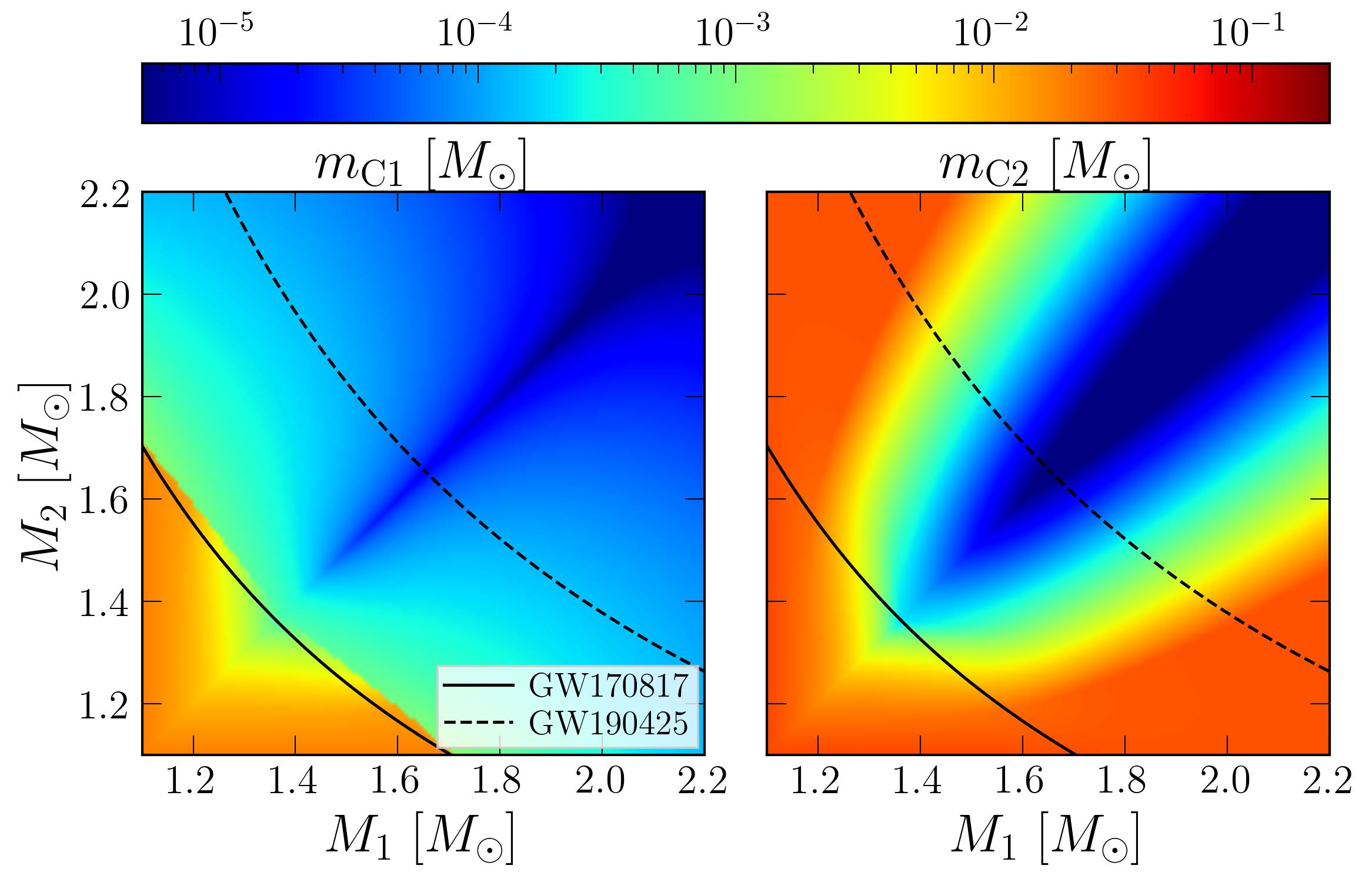}
    \caption{Same as \reffig{fig:BLh_ej_comp} but for the APR4 EOS.}
    \label{fig:APR4_ej_comp}
\end{figure}

In \reffig{fig:APR4_all_fits}, we show $m_{\rm dyn},~v_{\rm dyn}$, and $m_{\rm disc}$, as defined in \refeq{eq:mdyn}-(\ref{eq:mdisc}), in the $M_1 - M_2$ space for the APR4 EOS. See Figs. \ref{fig:BLh_all_fits}-\ref{fig:BLh_ej_comp} for a comparison with BLh. \reffig{fig:APR4_ej_comp} displays the ejecta component masses $m_{\rm C1}$ and $m_{\rm C2}$, as defined in \refeq{eq:mass_C1} and \refeq{eq:mass_C2}, in the $M_1 - M_2$ space for the APR4 \acp{EOS}.

\begin{figure}[]
    \centering
    \includegraphics[scale=0.4]{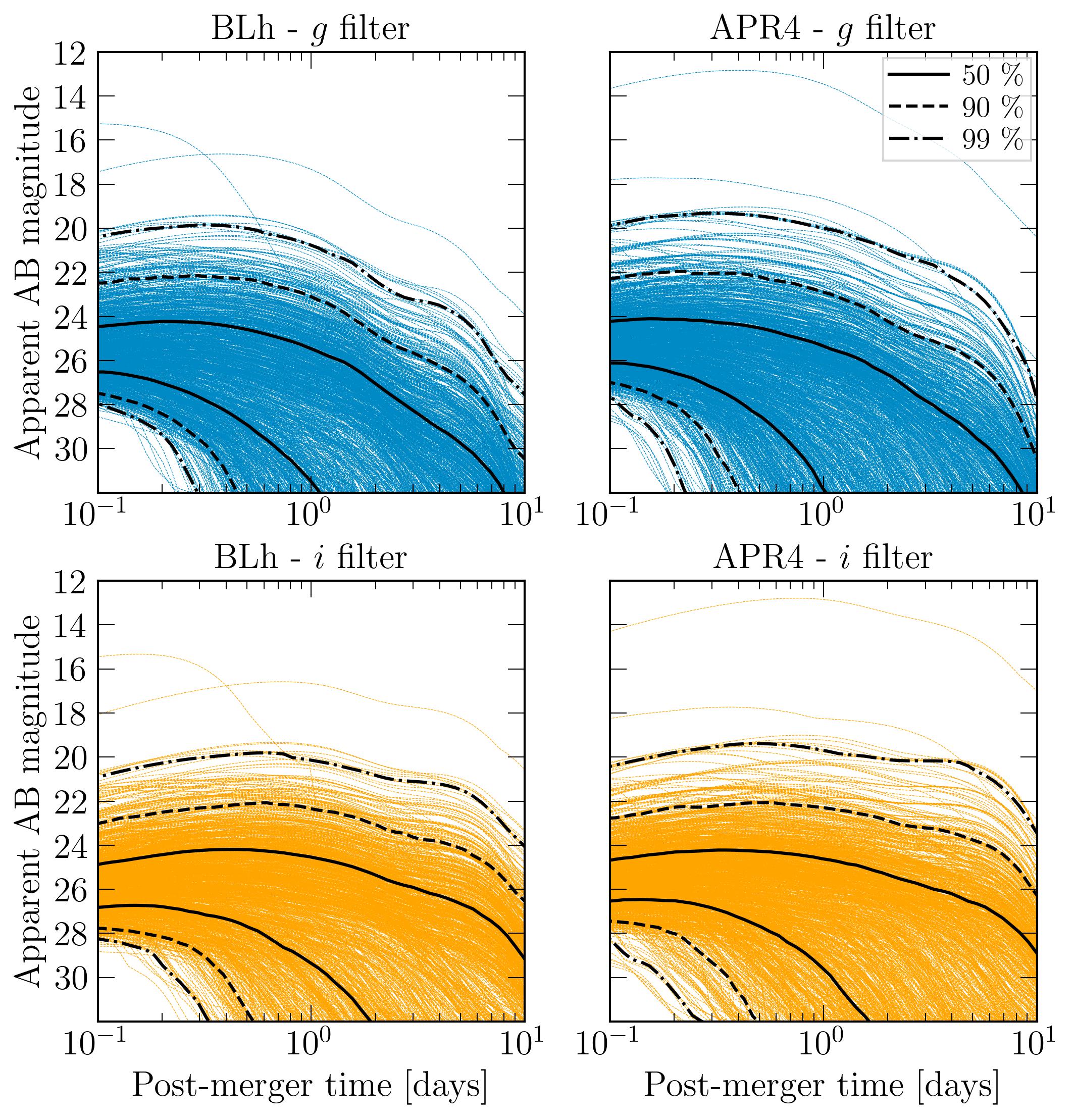}
    \caption{Same as in \reffig{fig:lc_gaussian} in the case of uniform NS mass distribution.}
 \label{fig:lc_uniform}
\end{figure}

\reffig{fig:lc_uniform} presents the KN light curves and their probability distributions in the \textit{g} and \textit{i} filters of Rubin. These light curves are calculated for the BNS mergers detected by ET-triangle within 100 deg$^2$ from populations characterised by a common envelope efficiency $\alpha = 1.0$, uniform NS mass distribution, and BLh (left) and APR4 (right) EOSs. For a comparison with the Gaussian mass distribution, see \reffig{fig:lc_gaussian}.

\begin{figure}[h!]
    \centering
    \includegraphics[scale=0.43]{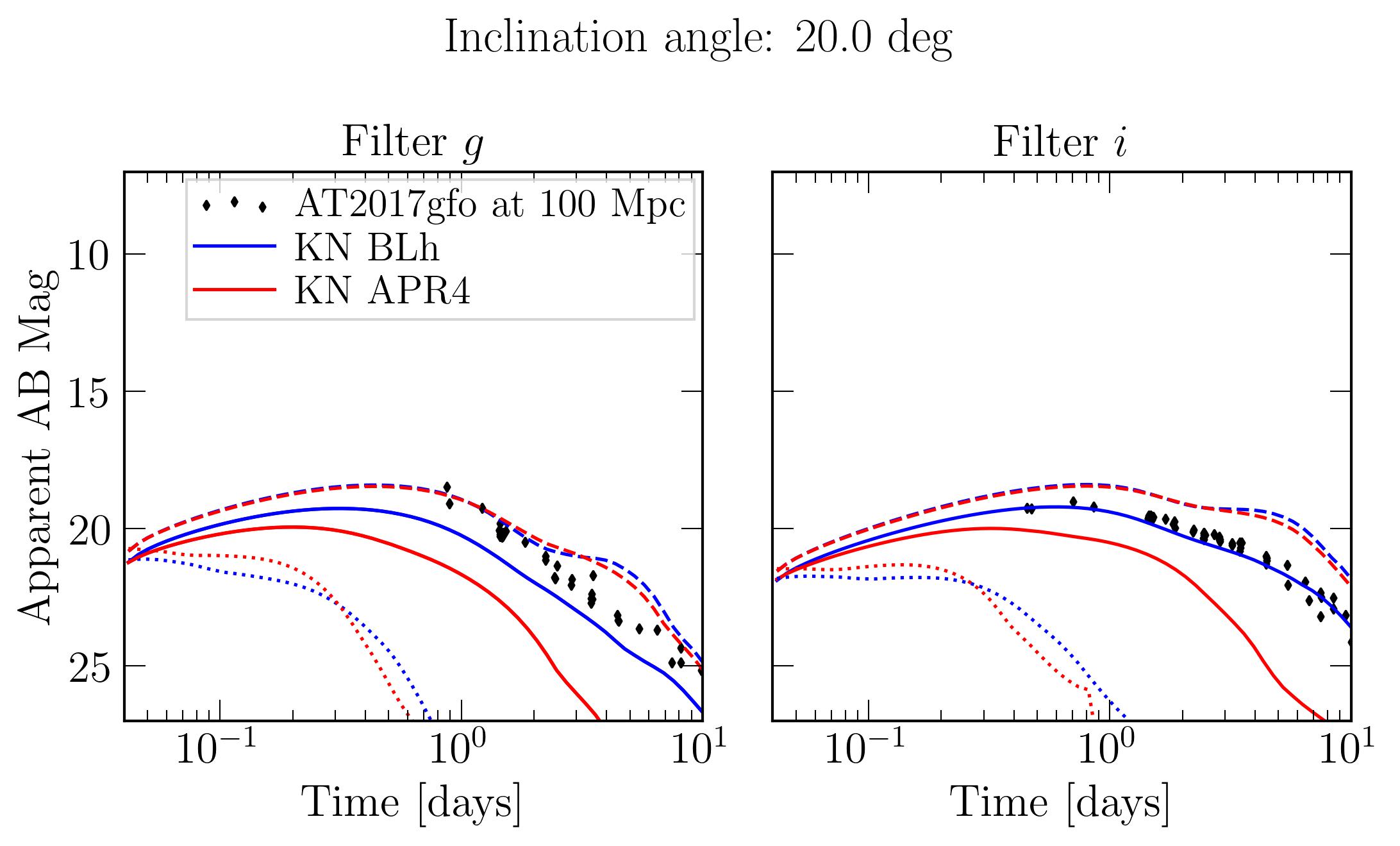}
    \caption{KNe light curves for BLh (blue lines) and APR4 (red lines) EOSs computed in the $g$ and $i$ filter of Rubin at inclination angle 20 deg, assuming luminosity distance $d_{\rm L} = 100$ Mpc. KNe light curves correspond to three different BNS systems. Solid lines correspond to $M_1 = M_2 = 1.33M_\odot$ (the peak of the NS mass distribution); dashed lines correspond to $M_1 = 1.18 M_\odot$, $M_2 = 1.10 M_\odot$ (the BNS producing the most massive disc); dotted lines correspond to $M_1 = M_2 = 1.57 M_\odot$ (the BNS producing the lightest disc). KNe light curves are compared to AT2017gfo (black diamonds) \citep{Villar:2017wcc} placed at $d_{\rm L} = 100$ Mpc.}
 \label{fig:kn_comparison_at2017gfo}
\end{figure}

Finally, in \reffig{fig:kn_comparison_at2017gfo} we compare three relevant cases of our modelled KN light curves for APR4 (red) and BLh (blue) EOSs to AT2017gfo at fixed luminosity distance $d_{\rm L} = 100$ Mpc and inclination angle $\iota = 20$ deg. Specifically, an average KN (solid lines) is modelled using an equal-mass BNS system with masses at the peak of a Gaussian distribution. Bright KN light curves (dashed lines) arise from a BNS system in the Gaussian mass distribution that produces the most massive disc, while faint KN light curves (dotted lines) correspond to a BNS system that results in the least massive disc. We notice that our population of KNe includes light curves that are significantly fainter than AT2017gfo, originating from BNS mergers that undergo prompt collapse into BH.

\subsection{Kilonova modelling reproducing AT2017gfo}\label{app:KNtest}

In \refsec{sec:kn_modeling}, we described how we compute the \ac{KN} light curves given the component masses and the \ac{EOS} of the BNS system. In this section, we evaluate the effectiveness of our model by assessing its ability to replicate the observed light curve of AT2017gfo, the \ac{KN} associated with GW170817. We consider two populations of \ac{BNS} mergers over ten years, one obtained using the APR4 EOS and the other the BLh \ac{EOS}. Both populations adopt a common envelope efficiency $\alpha = 1.0$ and a Gaussian mass distribution (refer to \refsec{sec:population}). From each population, we extract binaries whose chirp mass and mass ratio align with those of GW170817, specifically setting $ M_{\rm chirp} = 1.186 \pm 0.005$ and $q > 0.725$. For every selected binary, we compute the corresponding KN light curve (as detailed in \refsec{sec:kn_modeling}), assuming a luminosity distance of 40 Mpc and an inclination angle of 20 degrees, akin to GW170817. We then identify the brightest and faintest \ac{KN}e light curves from each population and compare them with the observed data of AT2017gfo. 

\begin{figure}[ht!]
    \includegraphics[scale=0.5]{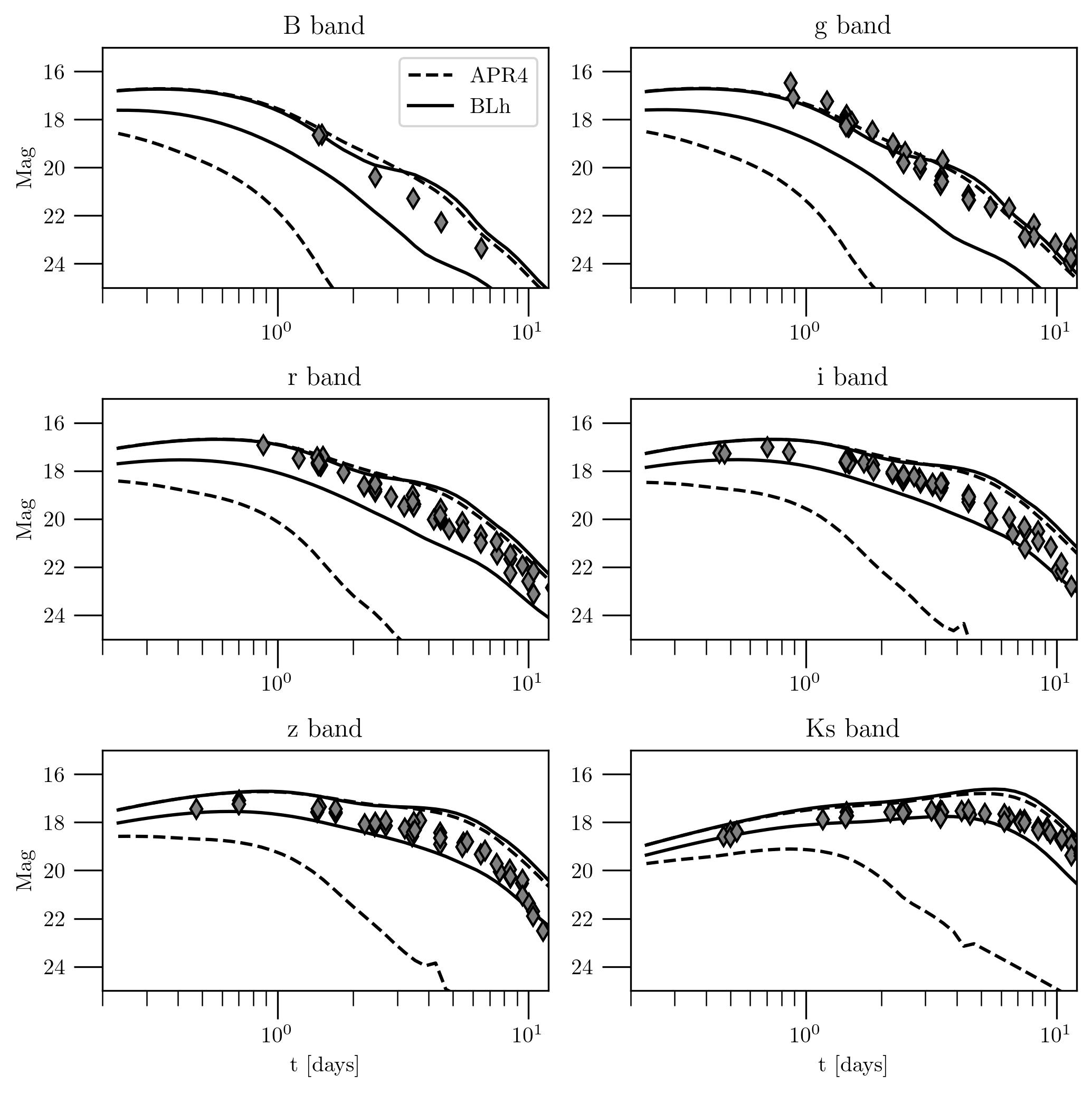}
    \caption{Comparison between AT2017gfo data (apparent AB magnitude) in the \textit{B, g, r, i, z, Ks} bands (diamonds) \citep{Villar:2017wcc} and the \ac{KN} light curves corresponding to the brightest and the faintest \ac{KN} signals obtained by selecting \acp{BNS} with chirp mass $M_{\rm chirp} = 1.186 \pm 0.005$ and binary mass ratio $q > 0.725$ from the population with $\alpha = 1.0$ and Gaussian mass distribution. Dashed and solid lines correspond to the case of APR4 and BLh, respectively.}
    \label{fig:compare_gw170817}
\end{figure}

Our findings are presented in \reffig{fig:compare_gw170817}, where the observed magnitudes of AT2017gfo in the \textit{B}, \textit{g}, \textit{r}, \textit{i}, \textit{z}, and \textit{Ks} bands are depicted with diamonds \citep{Villar:2017wcc}. The lines represent the brightest and the faintest KN light curves for APR4 (dashed lines) and BLh (solid lines). For both the \acp{EOS}, the brightest light curves correspond to BNS component masses $(M_{1}, M_{2}) = (1.58, 1.17) M_{\odot}$, while the faintest to $(M_{1}, M_{2}) = (1.37, 1.37) M_{\odot}$. Notably, the light curve of AT2017gfo falls between our model's brightest and faintest synthetic KN light curves for both the EOSs, except for the \textit{g} band, where the brightest light curves match the observations. This comparison demonstrates that our model is capable of reproducing the observations of AT2017gfo for source parameters consistent with GW170817.

\FloatBarrier

\section{Complements to \ac{GW} results}\label{appD}

\begin{table}[h!]
\centering
\caption{Number of BNS mergers detected in ten years by several networks of current and next generation detectors for the fiducial populations ($\alpha$ =1.0).}
{\small
\begin{tabular}{l | r r | r r}
\hline
\textbf{NETWORKS} &\multicolumn{2}{c}{\textbf{APR4}} &\multicolumn{2}{c}{\textbf{BLh}}\\
& Uniform & Gaussian & Uniform & Gaussian\\
\hline
\hline
ET-$\Delta$ & 191949 & 153726 & 186543 & 153632 \\
\hline
ET-$\Delta$ \textbf{+} LVKI & 198168  & 159203 & 192384 & 159087 \\
\hline
ET-$\Delta$ \textbf{+} 1CE & 316884  & 300098 & 314644 & 300043\\
\hline
ET-$\Delta$ \textbf{+} 2CE & 348424 & 341191 & 347575 & 341164  \\
\hline
\hline
ET-2L & 251997 & 219780 & 247494 & 219676 \\
\hline
ET-2L \textbf{+} LVKI & 255215 & 223361 & 250869 & 223266\\
\hline
ET-2L \textbf{+} 1CE & 332262 & 319490 & 330526 & 319446\\ 
\hline
ET-2L \textbf{+} 2CE & 353506 & 348098 & 352836 & 348076\\
\hline
\hline
LVKI & 1529 & 884 & 1403 & 883 \\
\hline
A\# & 8349  & 4936 & 7645 & 4930 \\
\end{tabular}
}
\tablefoot{The table shows the results for BLh and APR4 EOSs and for the uniform and Gaussian NS mass distributions. The detection threshold is set at network SNR=8.}
\label{tab:detections}
\end{table}

    \begin{figure}[h!]
    \centering
    \includegraphics[scale=0.55]{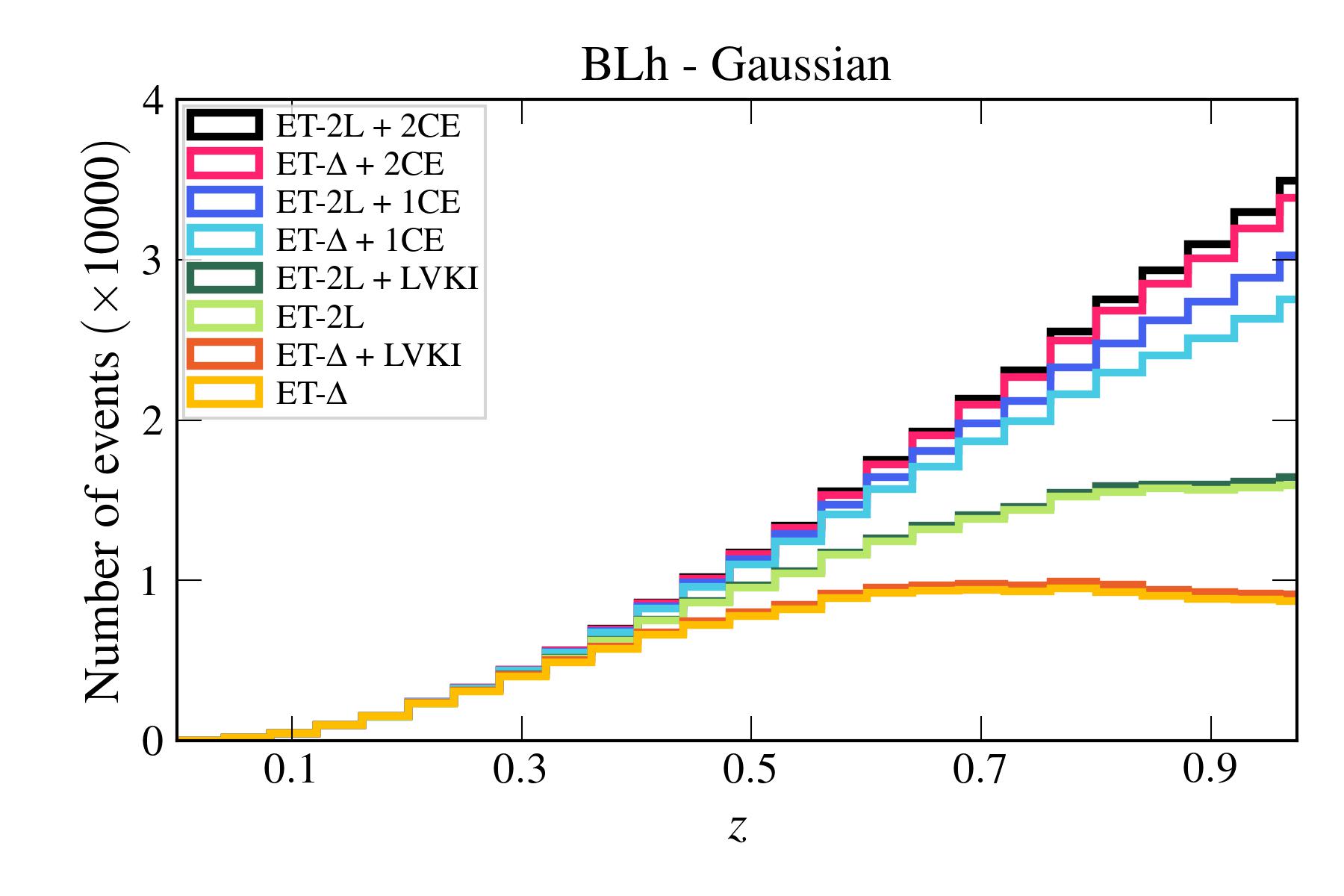}
    \caption{Ten-year detection distribution as a function of redshift for the eight different networks using the fiducial population ($\alpha=1.0$) with BLh EOS and Gaussian mass distribution.
    }
    \label{fig:detections_z_blh_gaussian_alpha1}
    \end{figure}

\begin{table}[ht!]
\centering
\caption{Number of BNS mergers detected by ET triangular shape with 10\,km arms in ten years with a sky localisation better than $100$ deg$^2$, $40$ deg$^2$, $20$ deg$^2$, $10$ deg$^2$, and $1$ deg$^2$.}
\scalebox{0.85}{
\begin{tabular}{l l l r r r r}
\hline
& &
&\multicolumn{2}{c}{\textbf{APR4}}
&\multicolumn{2}{c}{\textbf{BLh}}\\
\hline
& &$\Omega_{90}$ [deg$^2$] &Uniform &Gaussian &Uniform &Gaussian\\
\hline
\hline
\multirow{19}{*}{\rotatebox[origin=c]{90}{\parbox[c]{1.2cm}{\centering $\alpha = 0.5$}}}
&\multirow{4}{*}{\rotatebox[origin=c]{90}{\parbox[c]{1cm}{\centering ET-$\Delta$}}}
&$<100$ &$266$ &$233$ &$272$ &$233$\\
& &$<40$ &$79$ &$62$ &$75$ &$62$\\
& &$<20$ &$29$ &$23$ &$25$ &$23$\\
& &$<10$ &$8$ &$6$ &$8$ &$6$\\
\cline{2-7}
&\multirow{5}{*}{\rotatebox[origin=c]{90}{\parbox[c]{1cm}{\centering ET-$\Delta$ \\\textbf{+}\\ LVKI}}}
&$<100$ &$14946$ &$10031$ &$13873$ &$9797$\\
& &$<40$ &$5105$ &$3159$ &$4659$ &$3074$\\
& &$<20$ &$1808$ &$1149$ &$1673$ &$1131$\\
& &$<10$ &$671$ &$403$ &$592$ &$391$\\
& &$<1$ &$15$ &$9$ &$19$ &$9$\\
\cline{2-7}
&\multirow{5}{*}{\rotatebox[origin=c]{90}{\parbox[c]{1cm}{\centering ET-$\Delta$ \\\textbf{+}\\ 1CE}}}
&$<100$ &$41229$ &$35355$ &$40086$ &$34950$\\
& &$<40$ &$21881$ &$15848$ &$20685$ &$15490$\\
& &$<20$ &$9505$ &$5874$ &$8741$ &$5747$\\
& &$<10$ &$3370$ &$ 2012$ &$3086$ &$1985$\\
& &$<1$ &$99$ &$58$ &$85$ &$58$\\
\cline{2-7}
&\multirow{5}{*}{\rotatebox[origin=c]{90}{\parbox[c]{1cm}{\centering ET-$\Delta$ \\\textbf{+}\\ 2CE}}}
&$<100$ &$62252$ &$59712$ &$61865$ &$59525$\\
& &$<40$ &$54122$ &$50462$ &$53407$ &$50088$\\
& &$<20$ &$44976$ &$40266$ &$43896$ &$39742$\\
& &$<10$ &$33336$ &$ 27980$ &$32070$ &$27341$\\
& &$<1$ &$2620$ &$1587$ &$2413$ &$1548$\\
\hline
\hline
\multirow{19}{*}{\rotatebox[origin=c]{90}{\parbox[c]{1.2cm}{\centering $\alpha = 1.0$}}}
&\multirow{5}{*}{\rotatebox[origin=c]{90}{\parbox[c]{1cm}{\centering ET-$\Delta$}}}
&$<100$ &$1295$ &$1085$ &$1261$ &$1084$\\
& &$<40$ &$400$ &$315$ &$387$ &$315$\\
& &$<20$ &$147$ &$135$ &$154$ &$135$\\
& &$<10$ &$59$ &$49$ &$58$ &$49$\\
\cline{2-7}
&\multirow{5}{*}{\rotatebox[origin=c]{90}{\parbox[c]{1cm}{\centering ET-$\Delta$ \\\textbf{+}\\ LVKI}}}
&$<100$ &$74095$ &$49000$ &$69098$ &$47662$\\
& &$<40$ &$25246$ &$15499$ &$23009$ &$15075$\\
& &$<20$ &$8904$ &$5615$ &$8205$ &$5506$\\
& &$<10$ &$3221$ &$2027$ &$3007$ &$1989$\\
& &$<1$ &$105$ &$71$ &$103$ &$71$\\
\cline{2-7}
&\multirow{5}{*}{\rotatebox[origin=c]{90}{\parbox[c]{1cm}{\centering ET-$\Delta$ \\\textbf{+}\\ 1CE}}}
&$<100$ &$207029$ &$176325$ &$201449$ &$174112$\\
& &$<40$ &$109116$ &$77348$ &$103009$ &$75578$\\
& &$<20$ &$46537$ &$28731$ &$42915$ &$28127$\\
& &$<10$ &$16231$ &$9759$ &$14893$ &$9639$\\
& &$<1$ &$514$ &$309$ &$466$ &$307$\\
\cline{2-7}
&\multirow{5}{*}{\rotatebox[origin=c]{90}{\parbox[c]{1cm}{\centering ET-$\Delta$ \\\textbf{+}\\ 2CE}}}
&$<100$ &$316613$ &$303242$ &$314298$ &$302312$\\
& &$<40$ &$275354$ &$256334$ &$271701$ &$254435$\\
& &$<20$ &$228395$ &$204945$ &$223353$ &$202064$\\
& &$<10$ &$169922$ &$142069$ &$163892$ &$138734$\\
& &$<1$ &$12829$ &$7849$ &$11713$ &$7659$\\
\hline
\end{tabular}
}
\tablefoot{We consider ET operating as a single observatory or included in several network combinations. We distinguish among the different EOSs (APR4 or BLh) and NS mass distributions (uniform or Gaussian) and pessimistic/fiducial populations ($\alpha$ values).}
\label{tab:loc_events_triangle}
\end{table}

\begin{table}[ht!]
\centering
\caption{Number of BNS mergers detected by the 2L ET with 15\,km arms in ten years with a sky localisation better than $100$ deg$^2$, $40$ deg$^2$, $20$ deg$^2$, $10$ deg$^2$, and $1$ deg$^2$ respectively.}
\scalebox{0.85}{
\begin{tabular}{l l l r r r r}
\hline
& &
&\multicolumn{2}{c}{\textbf{APR4}}
&\multicolumn{2}{c}{\textbf{BLh}}\\
\hline
& &$\Omega_{90}$ [deg$^2$] &Uniform &Gaussian &Uniform &Gaussian\\
\hline
\hline
\multirow{19}{*}{\rotatebox[origin=c]{90}{\parbox[c]{1.2cm}{\centering $\alpha = 0.5$}}}
&\multirow{4}{*}{\rotatebox[origin=c]{90}{\parbox[c]{1cm}{\centering ET-2L}}}
&$<100$ &$636$ &$543$ &$623$ &$539$\\
& &$<40$ &$173$ &$144$ &$164$ &$144$\\
& &$<20$ &$64$ &$50$ &$63$ &$50$\\
& &$<10$ &$26$ &$22$ &$26$ &$22$\\
\cline{2-7}
&\multirow{5}{*}{\rotatebox[origin=c]{90}{\parbox[c]{1cm}{\centering ET-2L\\\textbf{+}\\ LVKI}}}
&$<100$ &$21049$ &$15144$ &$19709$ &$14604$\\
& &$<40$ &$6581$ &$4112$ &$6071$ &$3993$\\
& &$<20$ &$2300$ &$1469$ &$2151$ &$1429$\\
& &$<10$ &$819$ &$524$ &$774$ &$516$\\
& &$<1$ &$24$ &$15$ &$26$ &$15$\\
\cline{2-7}
&\multirow{5}{*}{\rotatebox[origin=c]{90}{\parbox[c]{1cm}{\centering ET-2L \\\textbf{+}\\ 1CE}}}
&$<100$ &$36935$ &$33298$ &$36258$ &$32974$\\
& &$<40$ &$22655$ &$17904$ &$21781$ &$17574$\\
& &$<20$ &$11699$ &$7795$ &$10952$ &$7593$\\
& &$<10$ &$4437$ &$2731$ &$4062$ &$2678$\\
& &$<1$ &$128$ &$74$ &$112$ &$73$\\
\cline{2-7}
&\multirow{5}{*}{\rotatebox[origin=c]{90}{\parbox[c]{1cm}{\centering ET-2L \\\textbf{+}\\ 2CE}}}
&$<100$ &$61800$ &$60175$ &$61527$ &$60058$\\
& &$<40$ &$55474$ &$52339$ &$54783$ &$51981$\\
& &$<20$ &$46874$ &$42275$ &$45786$ &$41708$\\
& &$<10$ &$35058$ &$29499$ &$33707$ &$28931$\\
& &$<1$ &$2911$ &$1810$ &$2681$ &$1753$\\
\hline
\hline
\multirow{19}{*}{\rotatebox[origin=c]{90}{\parbox[c]{1.2cm}{\centering $\alpha = 1.0$}}}
&\multirow{4}{*}{\rotatebox[origin=c]{90}{\parbox[c]{1cm}{\centering ET-2L}}}
&$<100$ &$3065$ &$2595$ &$2977$ &$2584$\\
& &$<40$ &$895$ &$760$ &$863$ &$759$\\
& &$<20$ &$336$ &$285$ &$331$ &$284$\\
& &$<10$ &$140$ &$123$ &$138$ &$123$\\
\cline{2-7}
&\multirow{5}{*}{\rotatebox[origin=c]{90}{\parbox[c]{1cm}{\centering ET-2L \\\textbf{+}\\ LVKI}}}
&$<100$ &$105210$ &$75008$ &$98629$ &$72230$\\
& &$<40$ &$32600$ &$20315$ &$29938$ &$19712$\\
& &$<20$ &$11400$ &$7266$ &$10577$ &$7092$\\
& &$<10$ &$4103$ &$2628$ &$3802$ &$2591$\\
& &$<1$ &$134$ &$87$ &$120$ &$85$\\
\cline{2-7}
&\multirow{6}{*}{\rotatebox[origin=c]{90}{\parbox[c]{1cm}{\centering ET-2L \\\textbf{+}\\ 1CE}}}
&$<100$ &$187619$ &$168206$ &$184000$ &$166575$\\
& &$<40$ &$114328$ &$89433$ &$109393$ &$87665$\\
& &$<20$ &$57718$ &$37996$ &$53790$ &$37095$\\
& &$<10$ &$21782$ &$13429$ &$20114$ &$13166$\\
& &$<1$ &$682$ &$435$ &$636$ &$435$\\
\cline{2-7}
&\multirow{5}{*}{\rotatebox[origin=c]{90}{\parbox[c]{1cm}{\centering ET-2L \\\textbf{+}\\ 2CE}}}
&$<100$ &$312610$ &$304213$ &$311070$ &$303455$\\
& &$<40$ &$280242$ &$264039$ &$277022$ &$262179$\\
& &$<20$ &$235691$ &$211974$ &$230760$ &$209108$\\
& &$<10$ &$175625$ &$147446$ &$169663$ &$144339$\\
& &$<1$ &$13958$ &$8619$ &$12886$ &$8400$\\
\hline
\end{tabular}
}
\tablefoot{We consider ET operating as a single observatory or included in several network combinations. We distinguish among the different EOSs (APR4 or BLh) and NS mass distributions (uniform or Gaussian) and pessimistic/fiducial populations ($\alpha$ values).}
\label{tab:loc_events_2l}
\end{table}

\reftab{tab:detections} gives the number of BNS detected in ten years by several configurations of network detectors for the fiducial population ($\alpha=1$). The second and third columns show the numbers for the APR4 EOS and the uniform and Gaussian NS mass distributions, respectively. The fourth and fifth columns show the numbers for the BLh EOS. 

\reffig{fig:detections_z_blh_gaussian_alpha1} shows the number of detections in ten years up to redshift $z=1$ for the fiducial population ($\alpha=1.0$) with the BLh EOS, as in \reffig{fig:detections_z_blh_uniform_alpha1}, but assuming the Gaussian NS mass distribution. The total number of detections by the 2L ET is a factor 40\% larger than the detections by the ET-triangle. When ET-triangle or 2L shape operates with LVKI, an increase in the number of detections of a few percent is present. A much more significant increase takes place when ET operates with the next-generation detectors. For ET-triangle the number of detections doubles, and for ET-2L increases by about 45\% with one CE in the USA, and 60\% with one CE in the USA and one CE in Australia.

\begin{figure}[h!]
    \centering
    \includegraphics[scale=0.35]{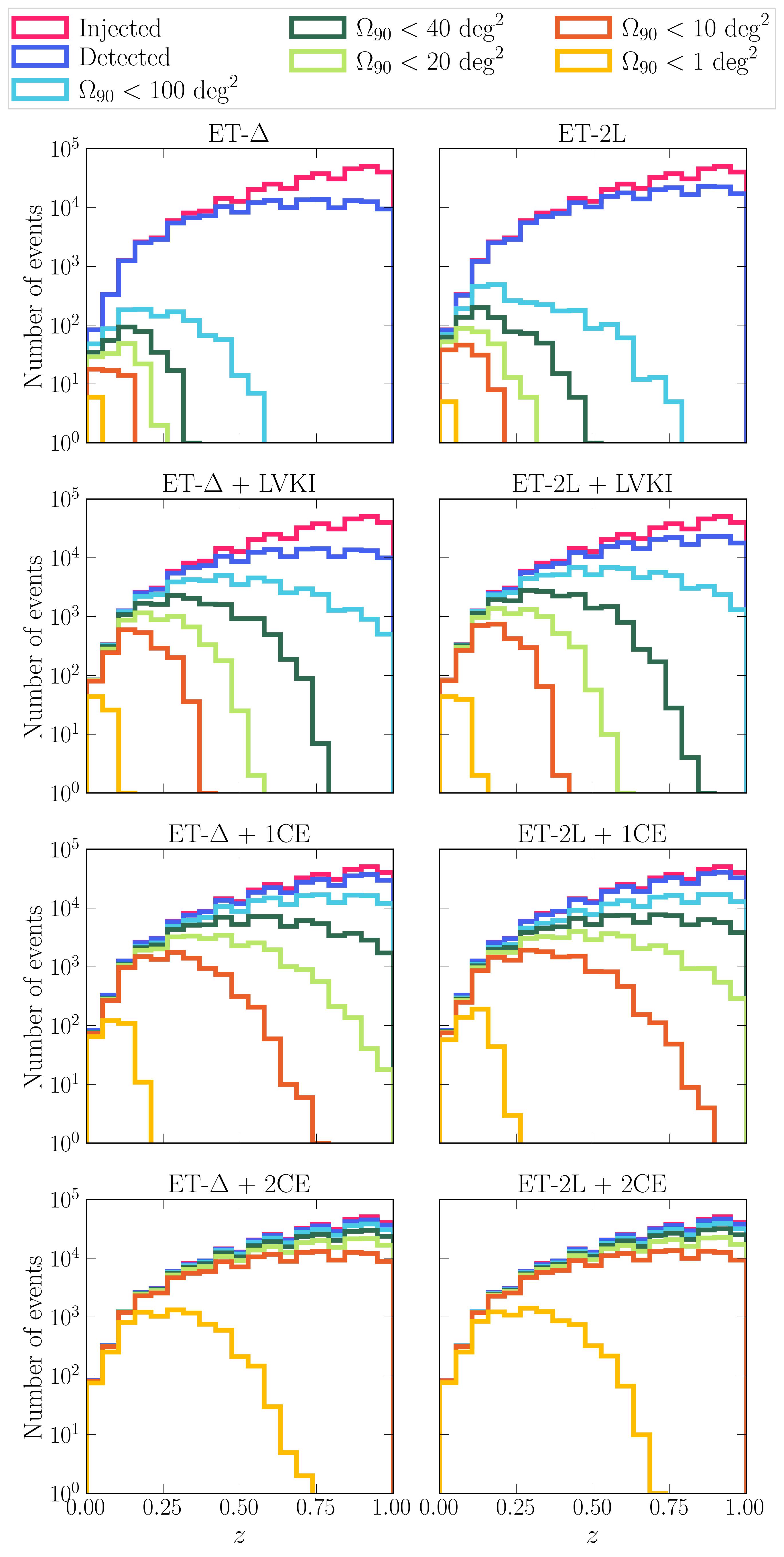}
    \caption{Same as \reffig{fig:skyloc_blh_uni} in the case of the BLh EOS and Gaussian NS mass distribution.}
    \label{fig:skyloc_blh_gauss}
    \end{figure}

\begin{figure}[h!]
    \centering
    \includegraphics[scale=0.35]{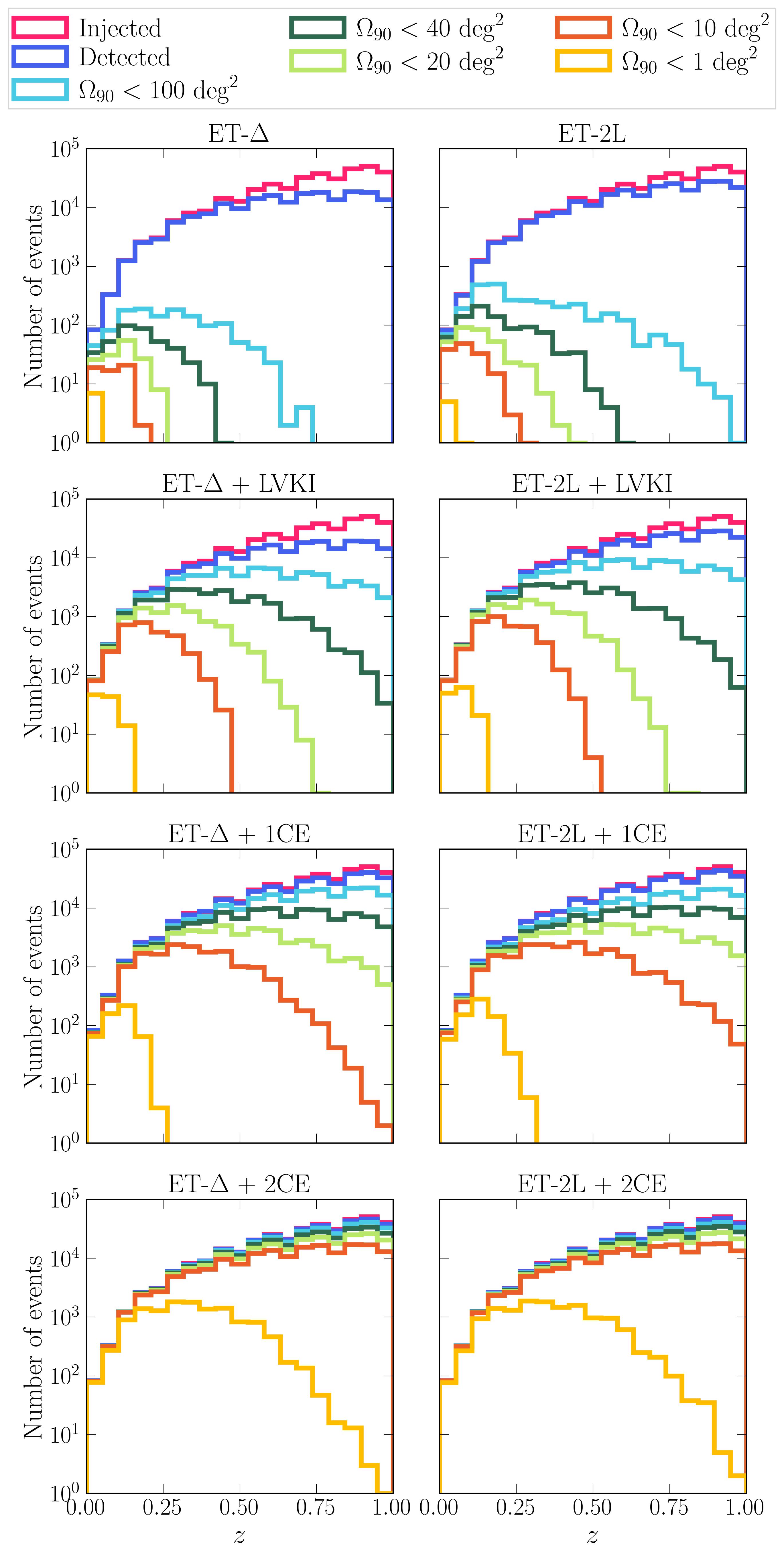}
    \caption{Same as \reffig{fig:skyloc_blh_uni} in the case of the APR4 EOS and uniform NS mass distribution.}
    \label{fig:skyloc_apr4_uni}
    \end{figure}

\begin{figure}[h!]
    \centering
    \includegraphics[scale=0.35]{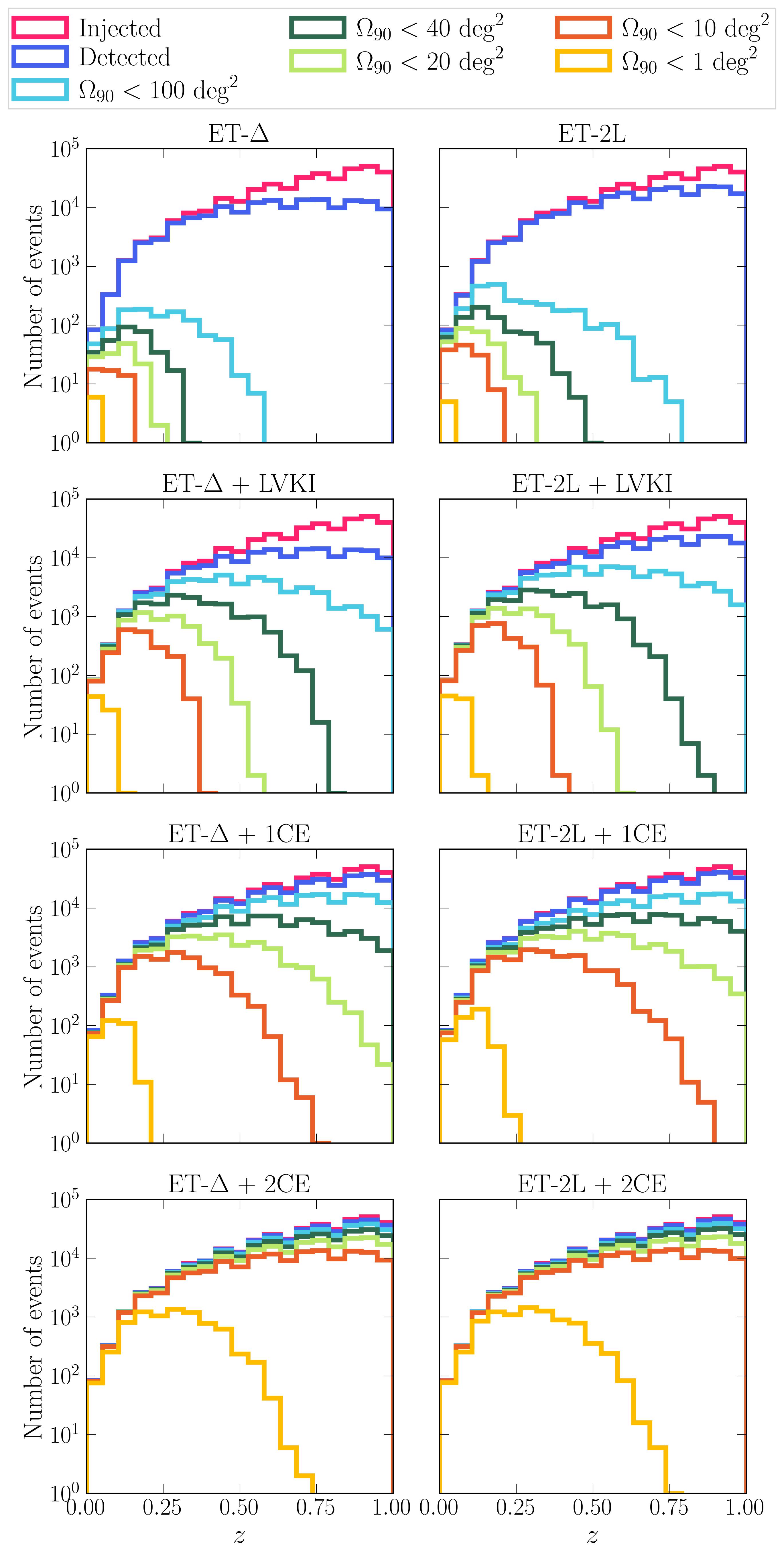}
    \caption{Same as \reffig{fig:skyloc_blh_uni} in the case of the APR4 EOS and Gaussian NS mass distribution.}
    \label{fig:skyloc_apr4_gauss}
    \end{figure}

Figs. \ref{fig:skyloc_blh_gauss}, \ref{fig:skyloc_apr4_uni}, and \ref{fig:skyloc_apr4_gauss} show the number of events localised better than 100, 40, 20, 10, and 1 deg$^2$ by the different detector networks for the fiducial population ($\alpha = 1.0$) assuming the BLh EOSs and Gaussian NS mass distribution, the APR4 EOSs and uniform NS mass distribution, the APR4 EOSs and Gaussian NS mass distribution, respectively.
The left plots refer to ET-triangle and the right ones to ET-2L. 
It is evident from all the plots the importance of the presence of current detectors which significantly increase of the number of detections and of the redshifts reached by well-localised events. These improvements, as expected, are then largely enhanced with networks of next-generation detectors. In \reftab{tab:loc_events_triangle} and \reftab{tab:loc_events_2l}, we list the number of BNS mergers detected within a certain sky-localisation threshold by the triangular and the 2L ET, respectively, operating as a single observatory or in a network of GW detectors.

\begin{figure}[h!]
    \centering
    \includegraphics[scale=0.45]{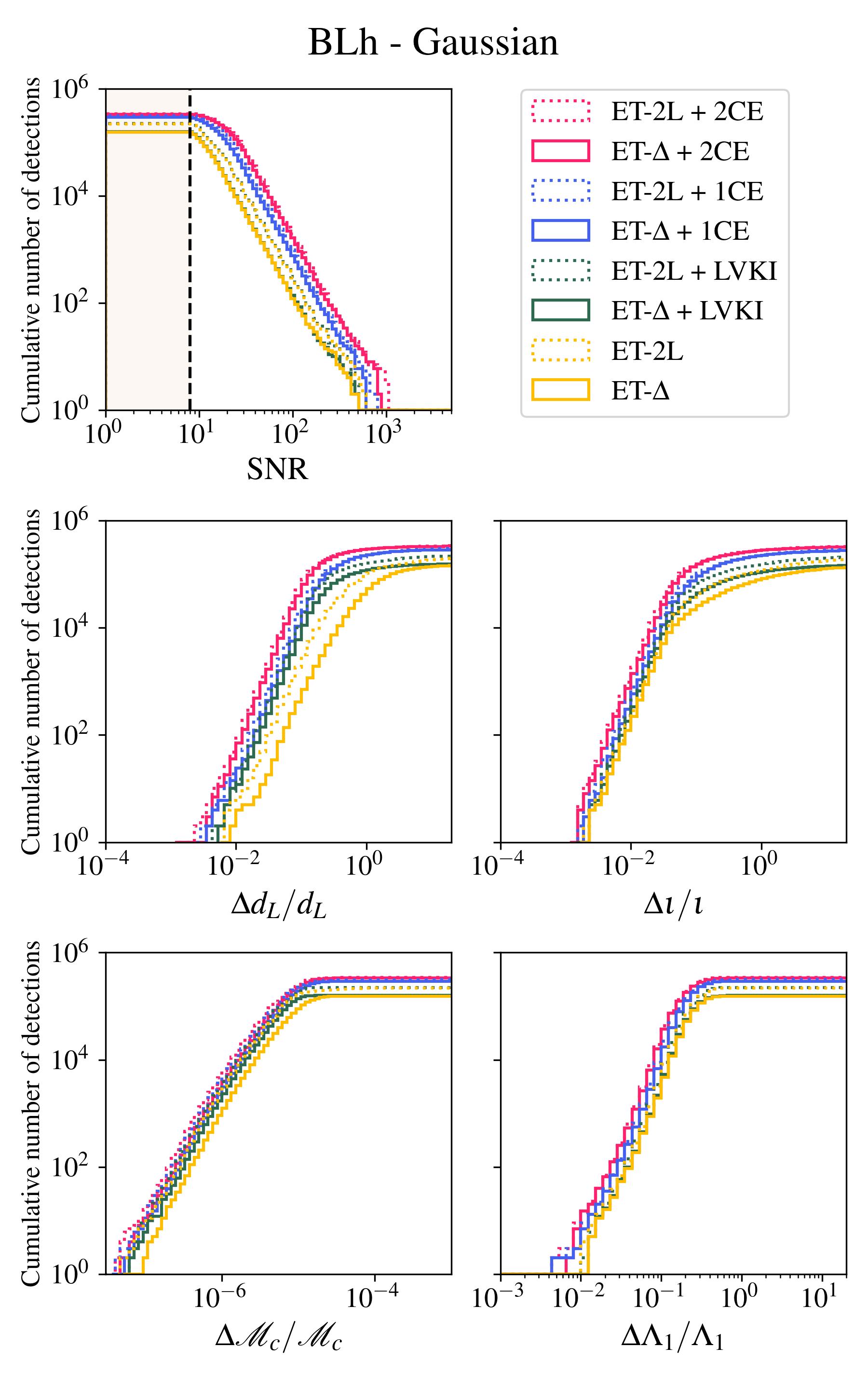}
    \caption{Same as \reffig{fig:rel_errors_blh_uniform_alpha1} for Gaussian mass distribution and BLh EOS.}
    \label{fig:rel_errors_blh_gaussian_alpha1}
    \end{figure}

\begin{figure}[h!]
    \centering
    \includegraphics[scale=0.45]{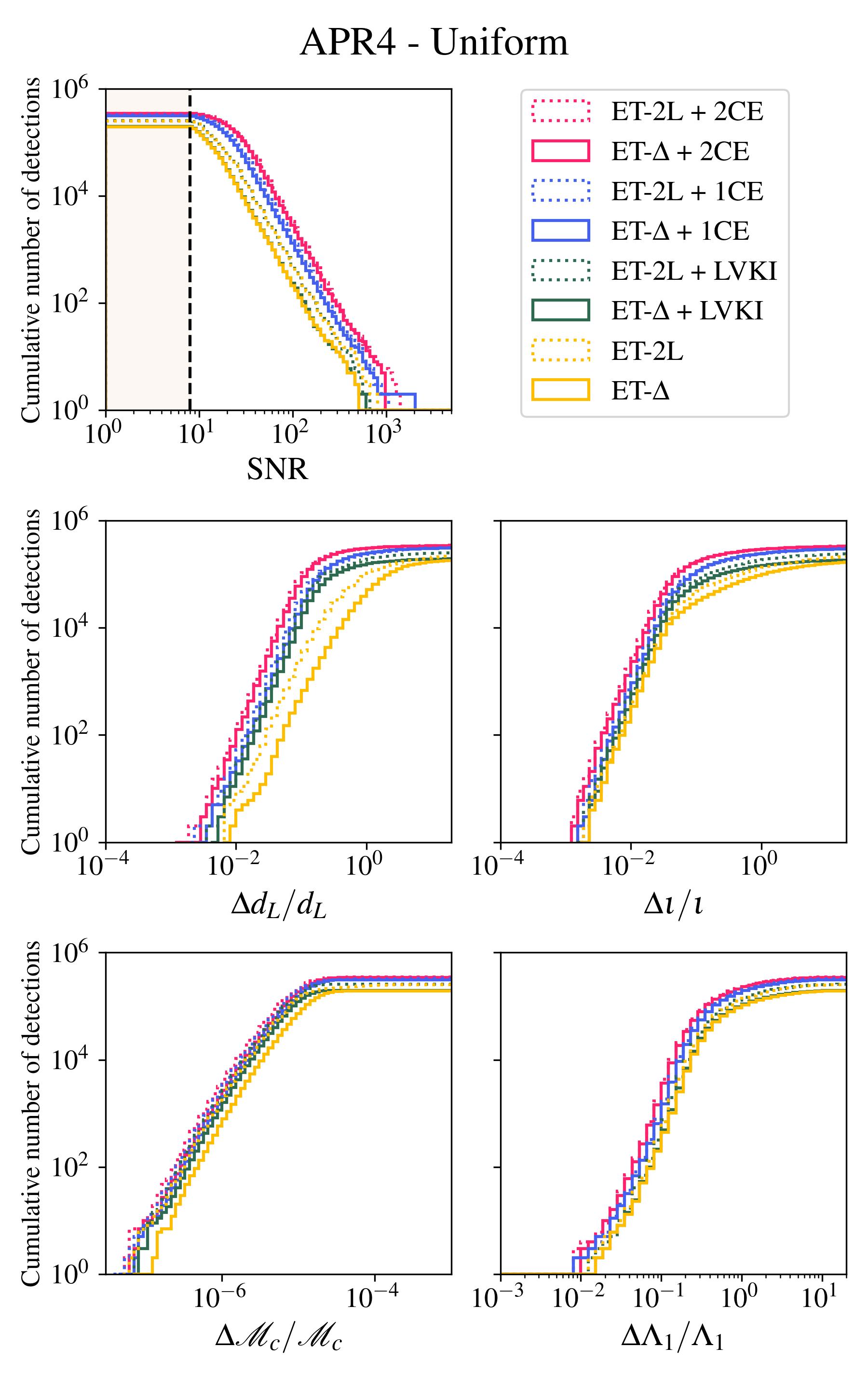}
    \caption{Same as \reffig{fig:rel_errors_blh_uniform_alpha1} for uniform mass distribution and APR4 EOS.
    }
    \label{fig:rel_errors_apr4_uniform_alpha1}
    \end{figure}

    Figs. \ref{fig:rel_errors_blh_gaussian_alpha1} and \ref{fig:rel_errors_apr4_uniform_alpha1} show the distribution of SNR and uncertainties on the source parameters obtained with \texttt{GWFish} for the fiducial population assuming the Gaussian NS mass distribution with BLh EOS and the uniform NS mass distribution with APR4 EOS, respectively.

\FloatBarrier
\section{Complements to joint GW and optical detection results}\label{app:joint_GW_EM}

\begin{figure}[h!]
    \includegraphics[scale=0.245]{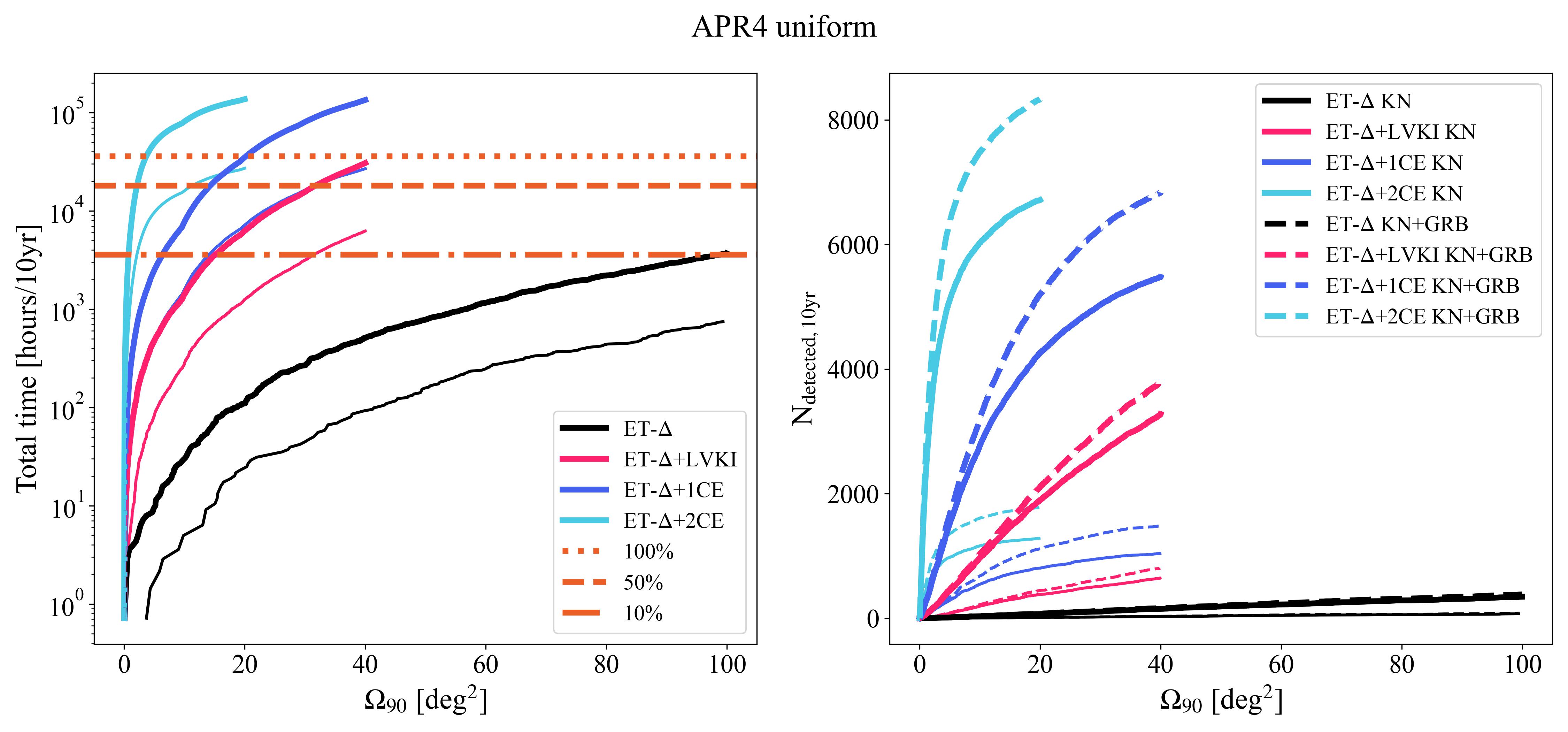}
    \includegraphics[scale=0.245]{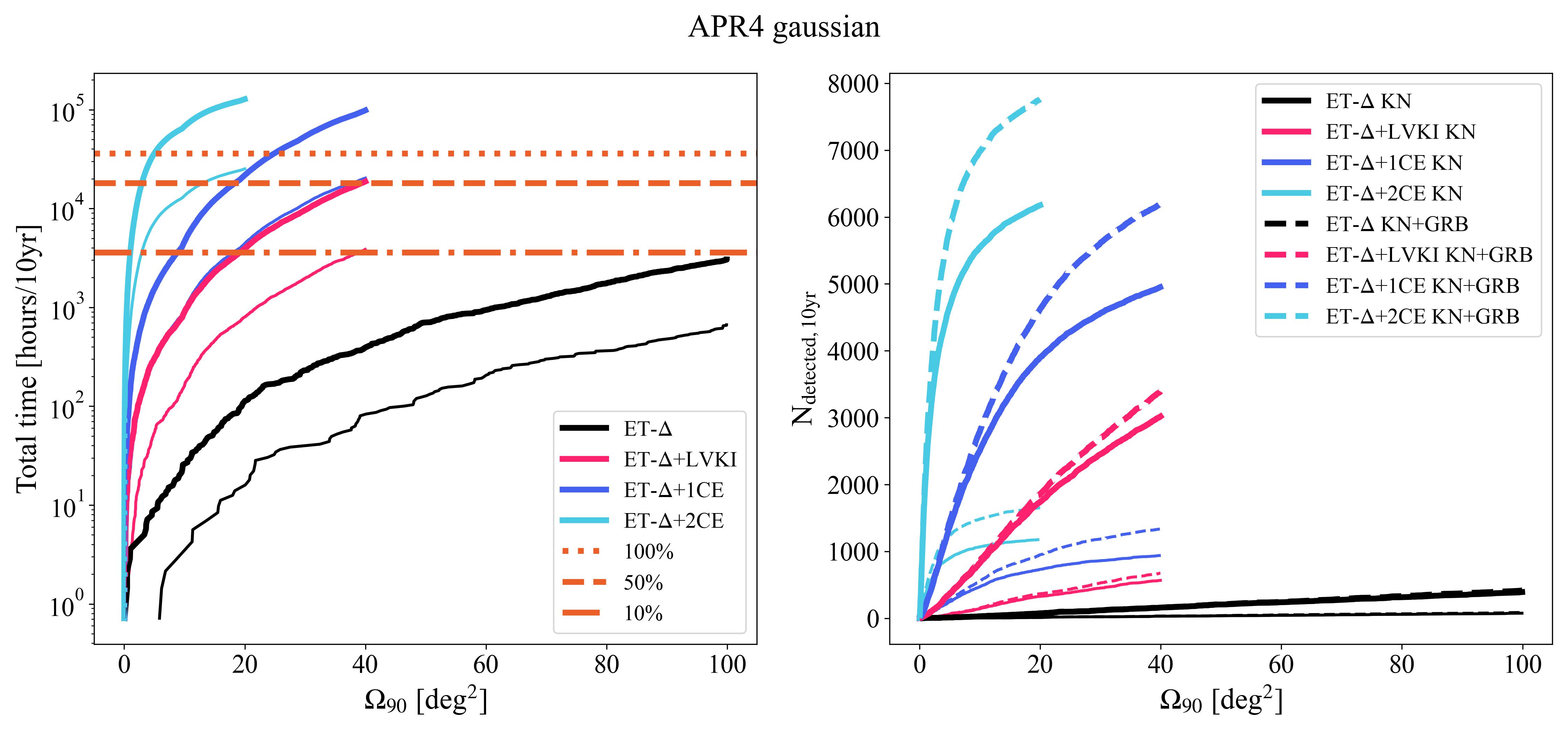}
    \caption{Same as \reffig{fig:time_number_BLh} for the APR4 EOS.}
    \label{fig:time_number_APR4}
    \end{figure}

\begin{figure}[h!]
    \includegraphics[scale=0.245]{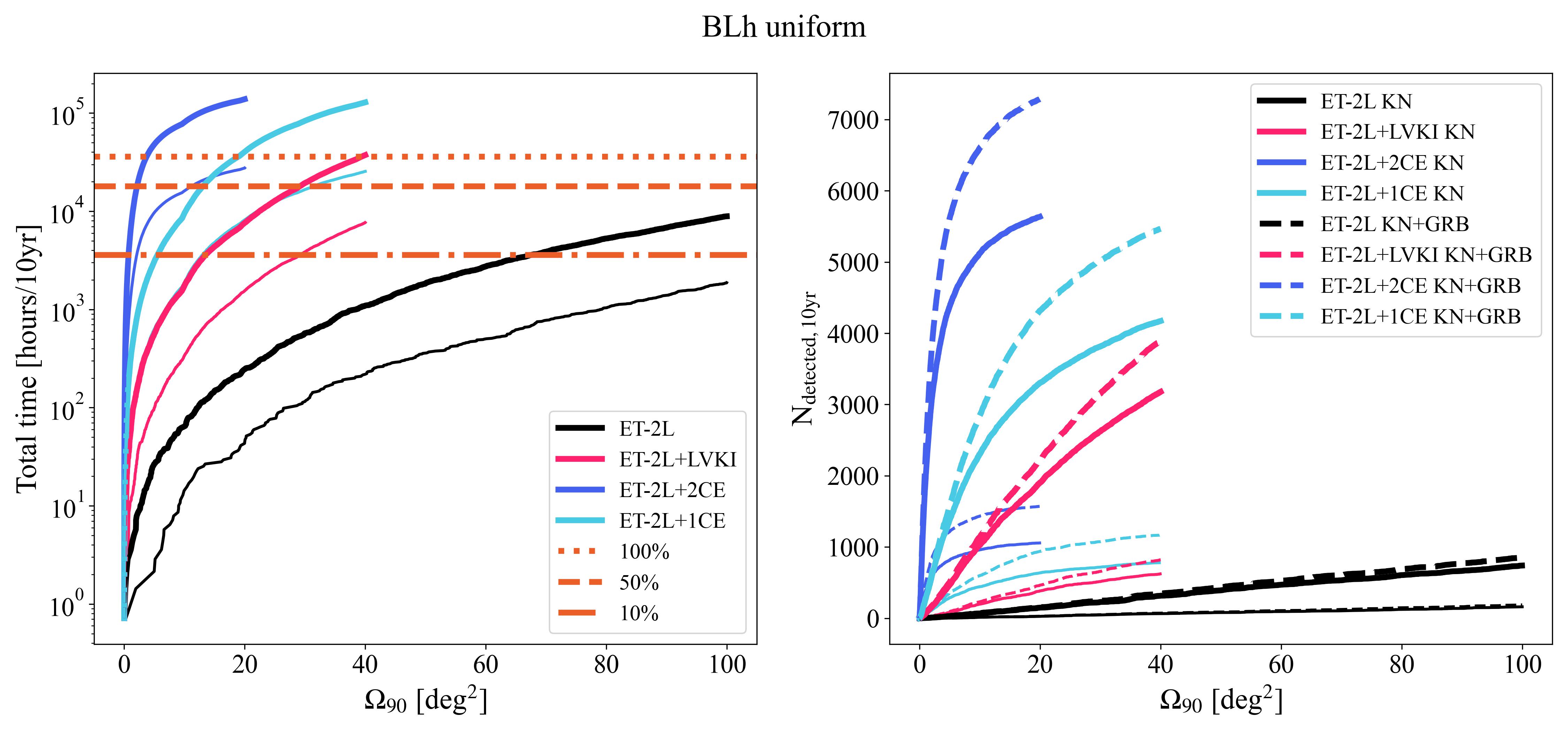}
    \includegraphics[scale=0.245]{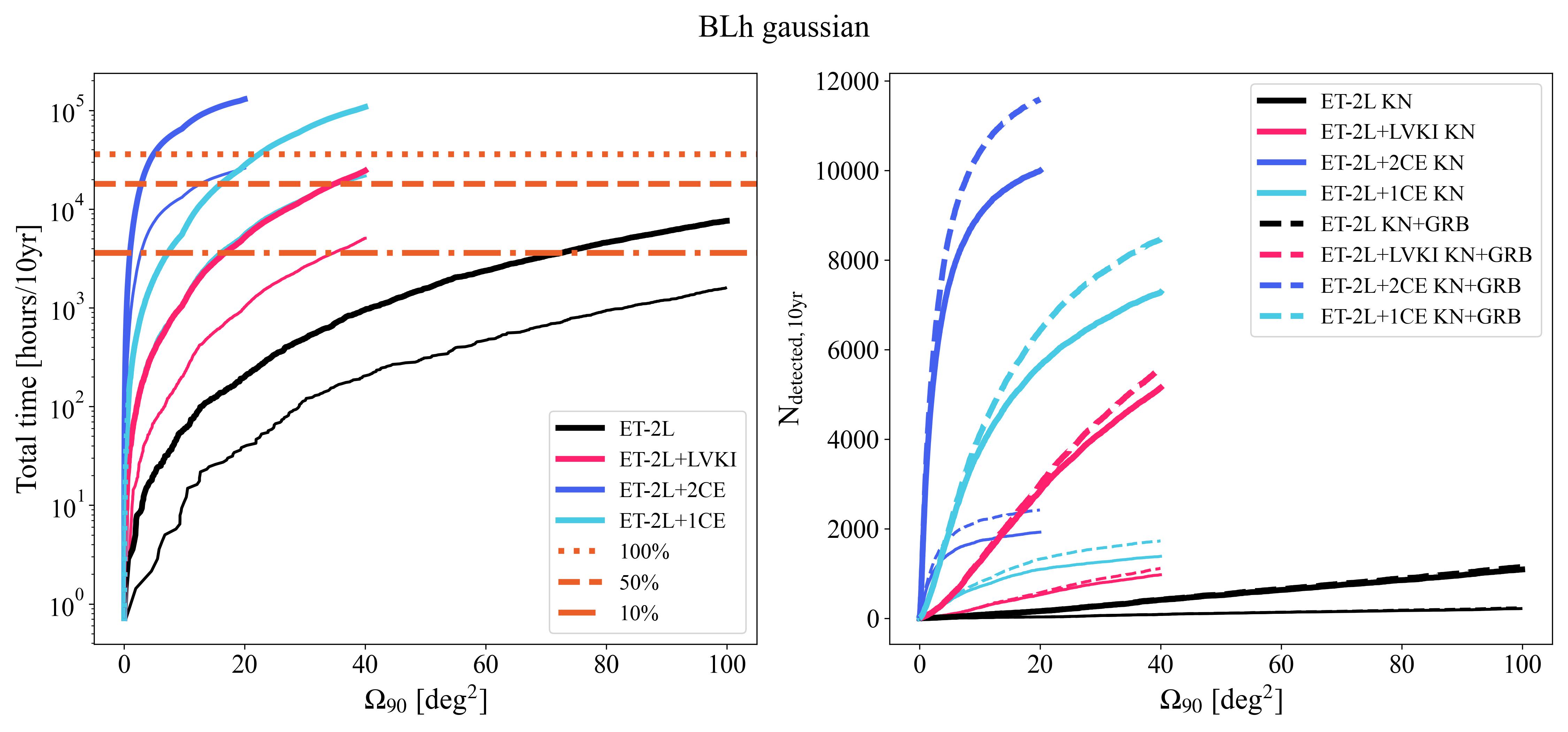} 
    \caption{Same as \reffig{fig:time_number_BLh} for the 2L-shape ET.}
    \label{fig:2L_time_number_BLh}
    \end{figure}

\begin{figure}
    \includegraphics[scale=0.245]{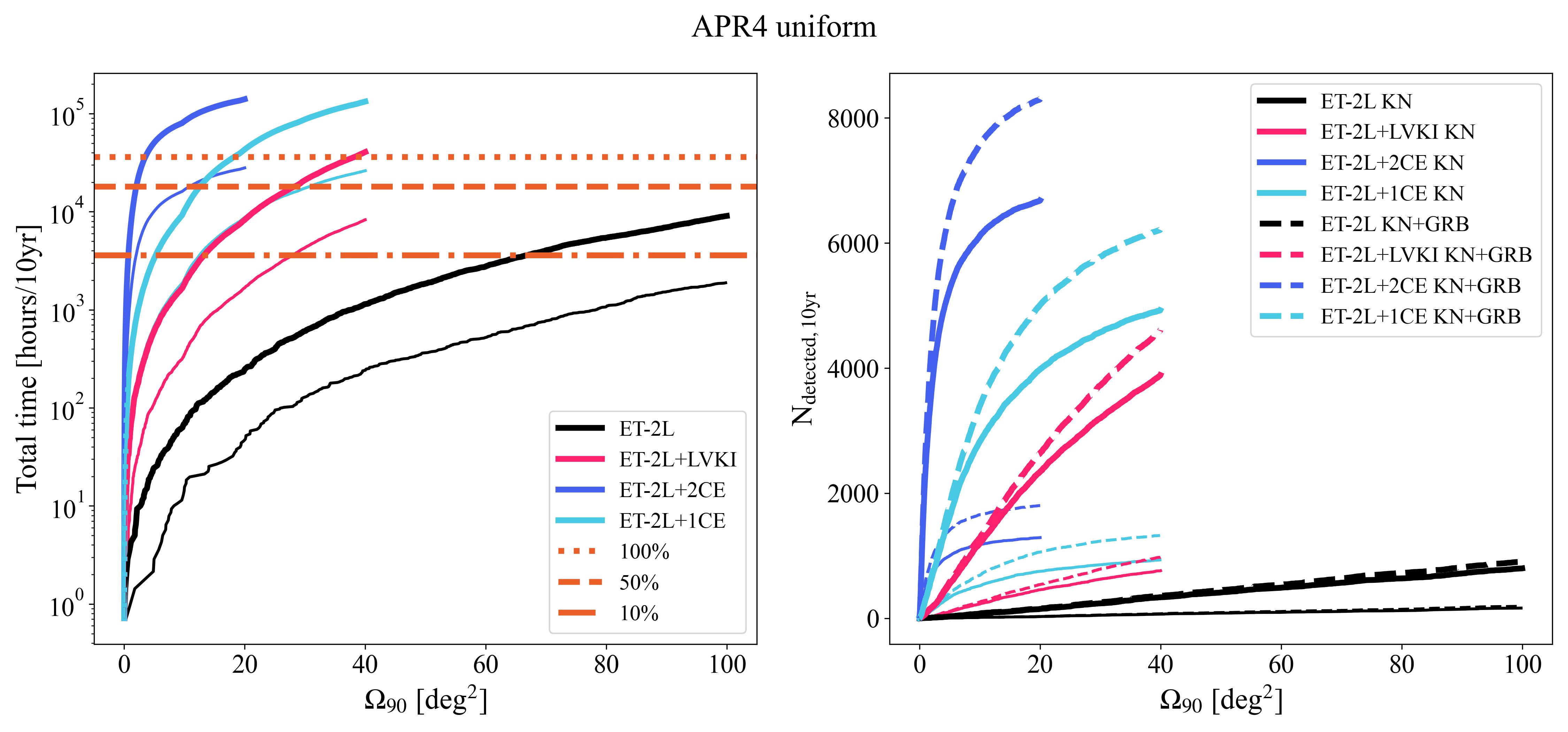}
    \includegraphics[scale=0.245]{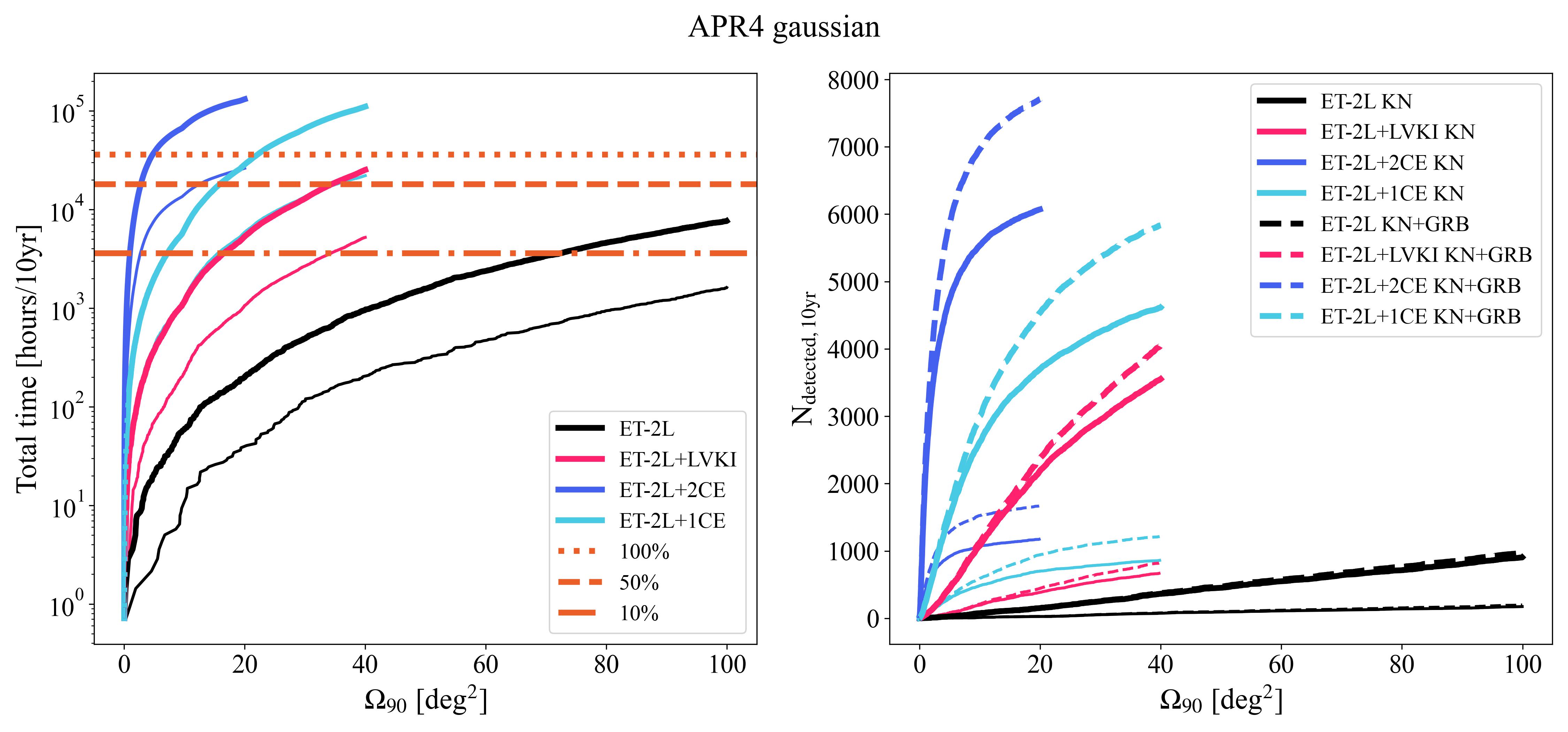}
    \caption{Same as \reffig{fig:time_number_BLh} for the APR4 EOS and 2L-shape ET.}
    \label{fig:2L_time_number_APR4}
    \end{figure}

\begin{figure}[htb]
    \centering
    \begin{minipage}{0.49\columnwidth}
        \centering
        \includegraphics[width=\linewidth]{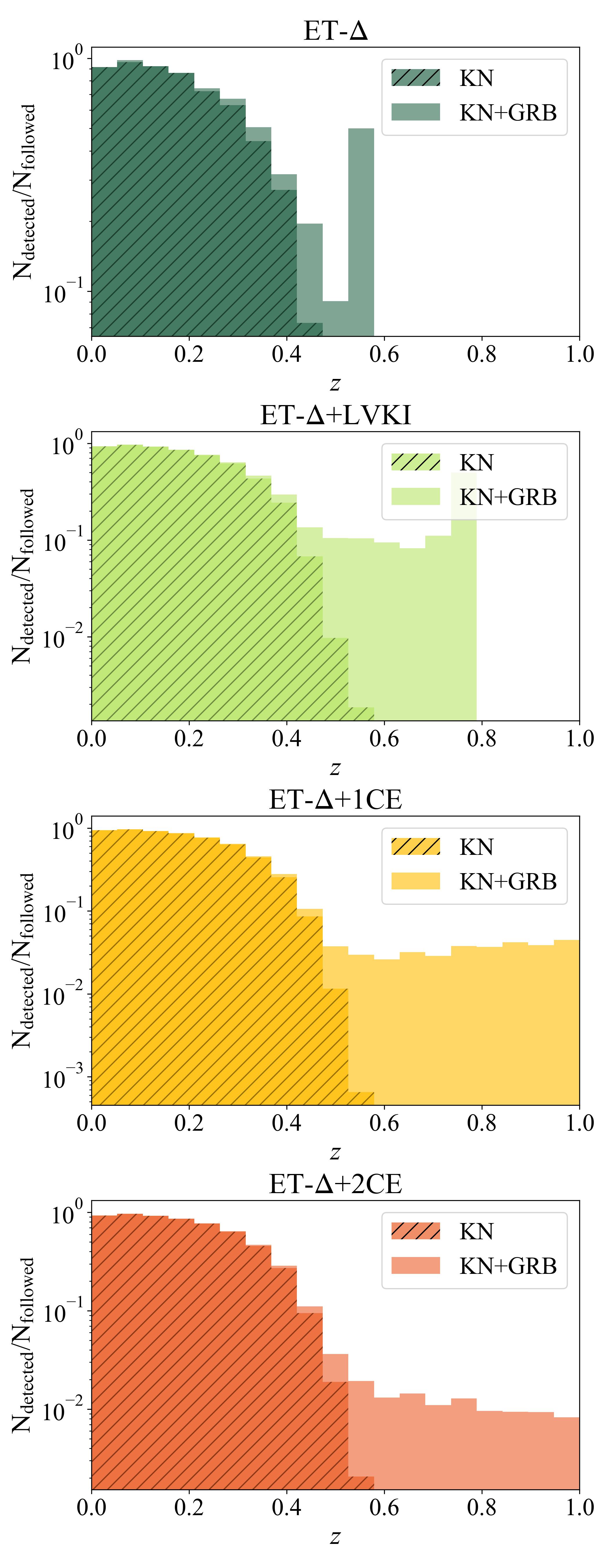}
    \end{minipage}
    \begin{minipage}{0.49\columnwidth}
        \centering
        \includegraphics[width=\linewidth]{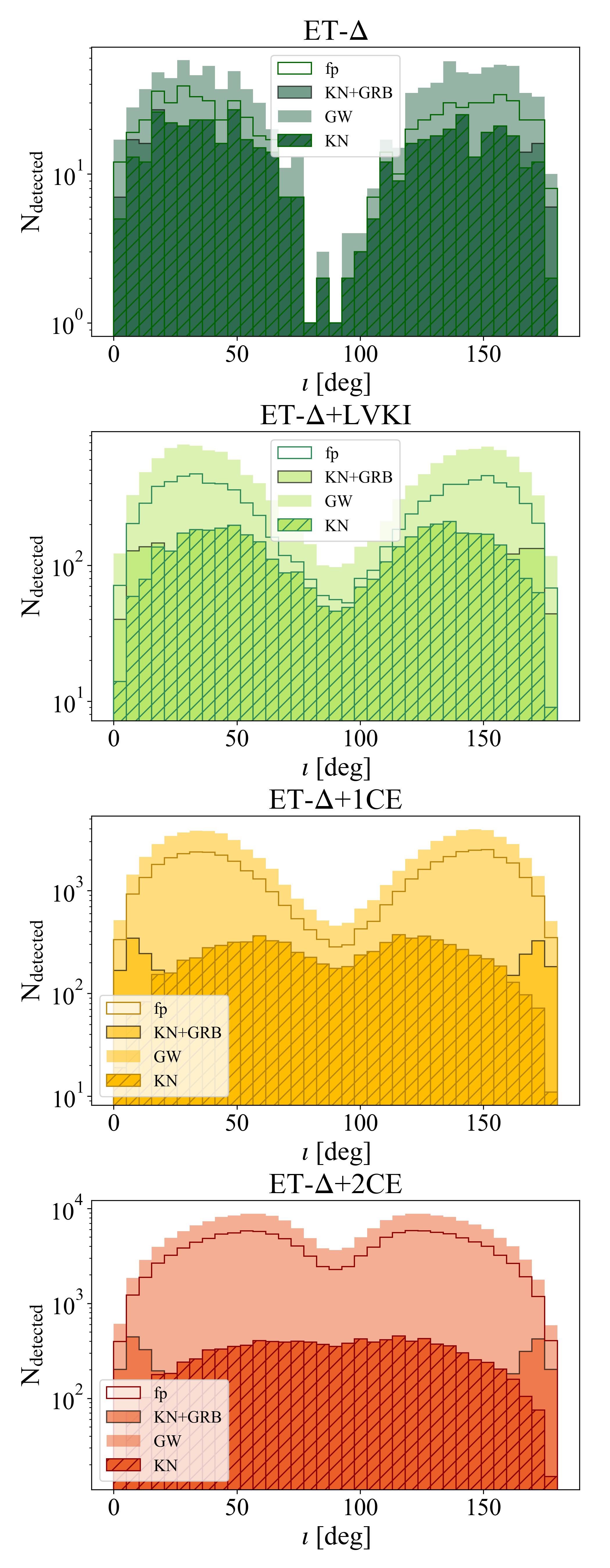}
    \end{minipage}
    \caption{Same as \reffig{fig:eff_theta_blh_uniform} but for the Gaussian NS mass distribution.}
    \label{fig:eff_theta_blh_gaussian}
\end{figure}

\begin{figure}
    \centering
    \begin{minipage}{0.49\columnwidth}
    \centering
    \includegraphics[width=\linewidth]{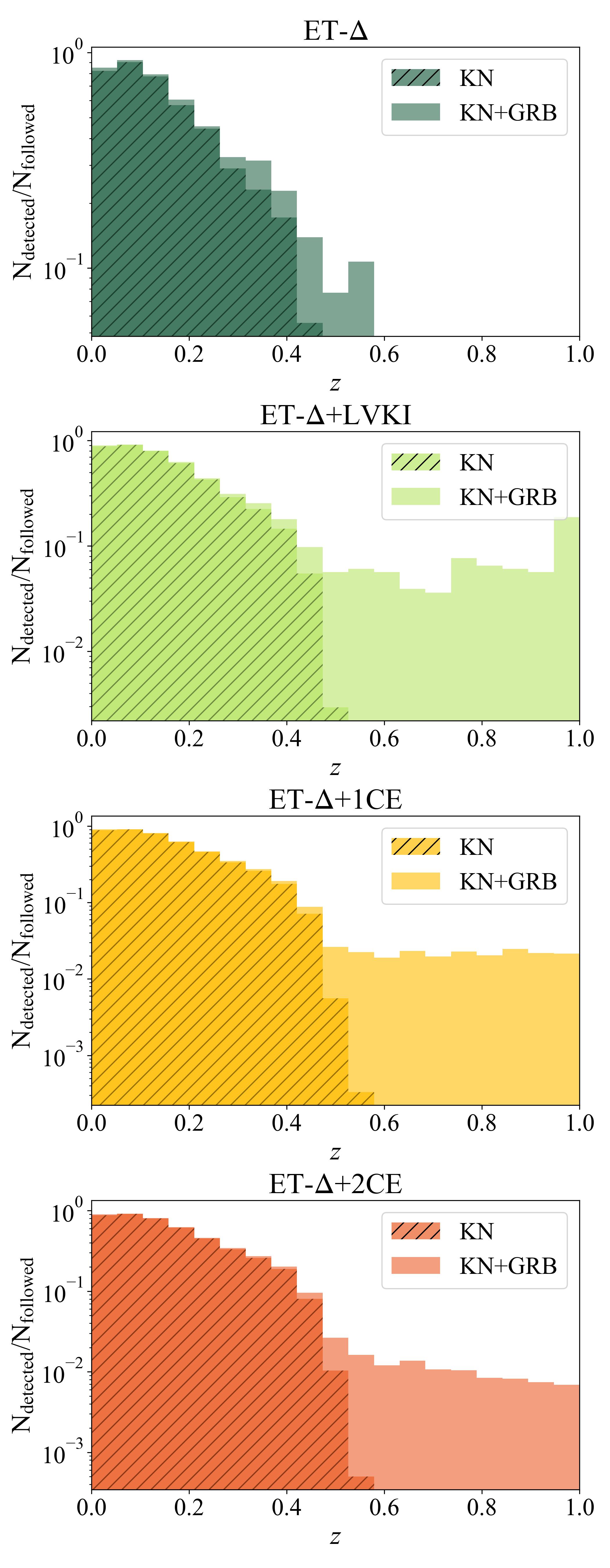}
    \end{minipage}
    \begin{minipage}{0.49\columnwidth}
    \centering
    \includegraphics[width=\linewidth]{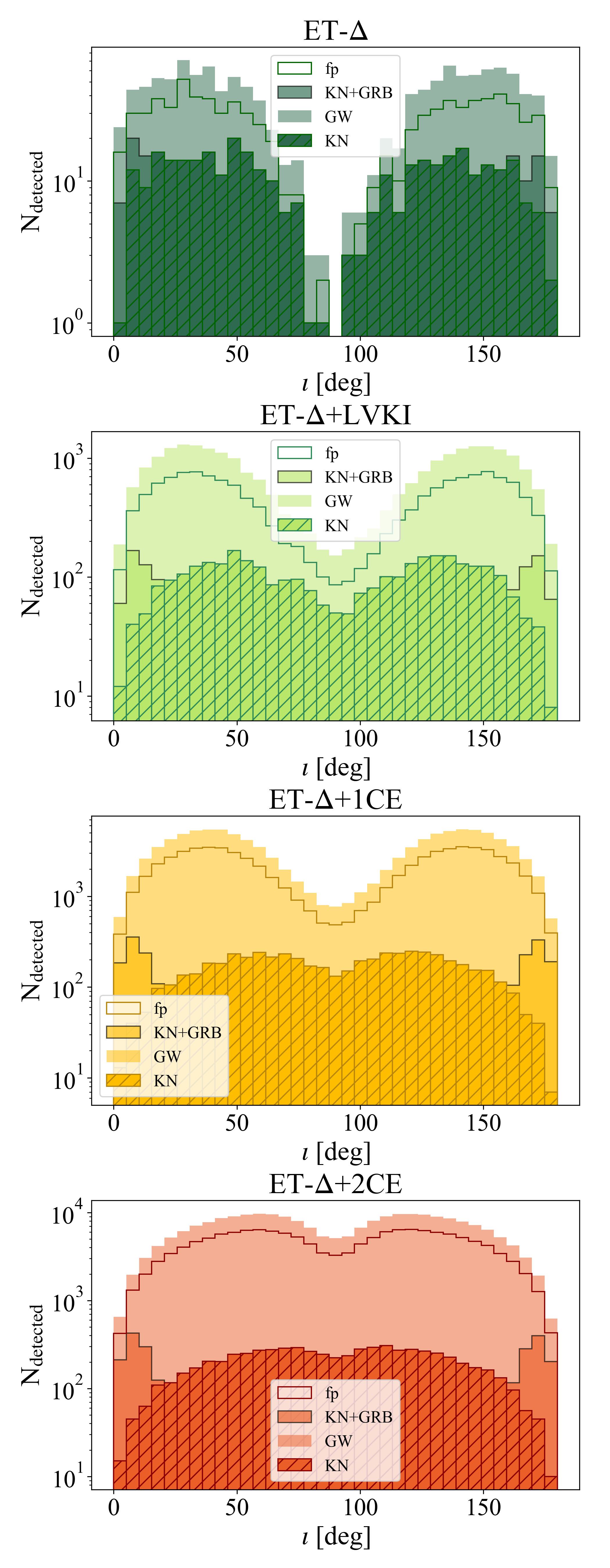}
    \end{minipage}
    \caption{Same as \reffig{fig:eff_theta_blh_uniform} but for the APR4 EOS.}
    \label{fig:eff_theta_apr4_uniform}
\end{figure}

\begin{figure}
    \centering
    \begin{minipage}{0.49\columnwidth}
    \centering
    \includegraphics[width=\linewidth]{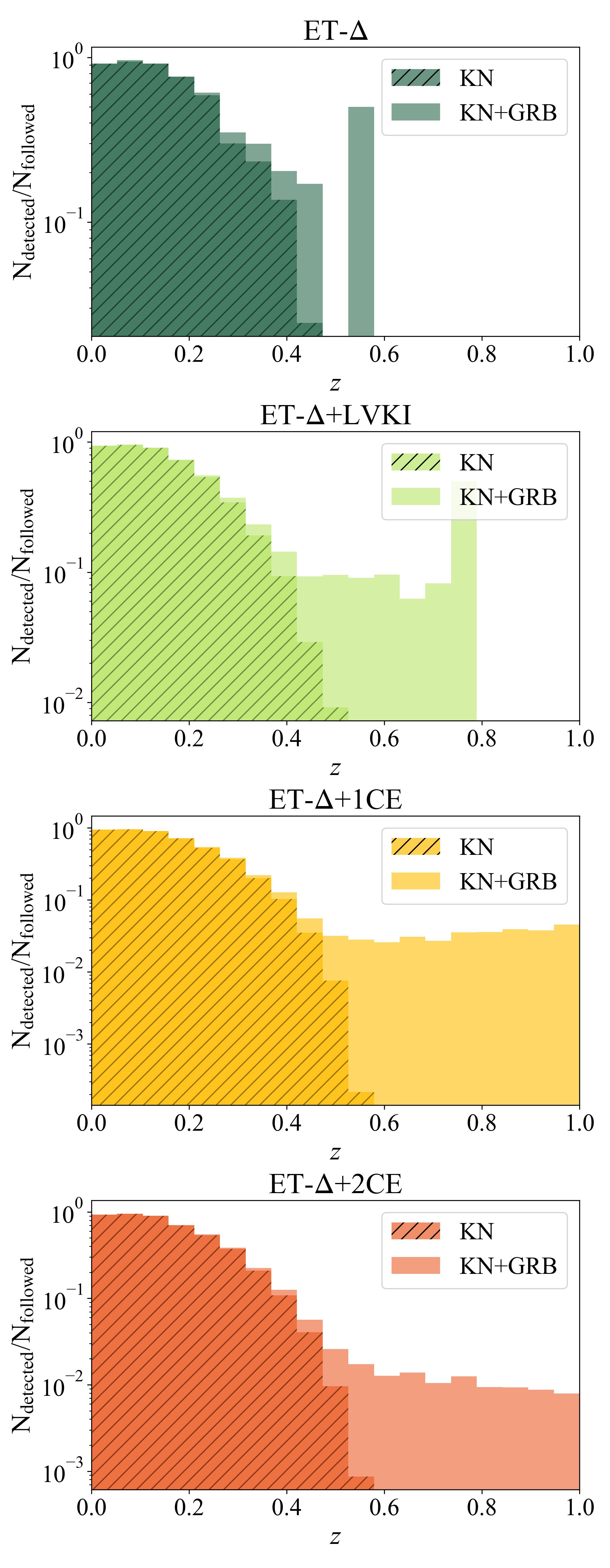}
    \end{minipage}
    \begin{minipage}{0.49\columnwidth}
    \centering
    \includegraphics[width=\linewidth]{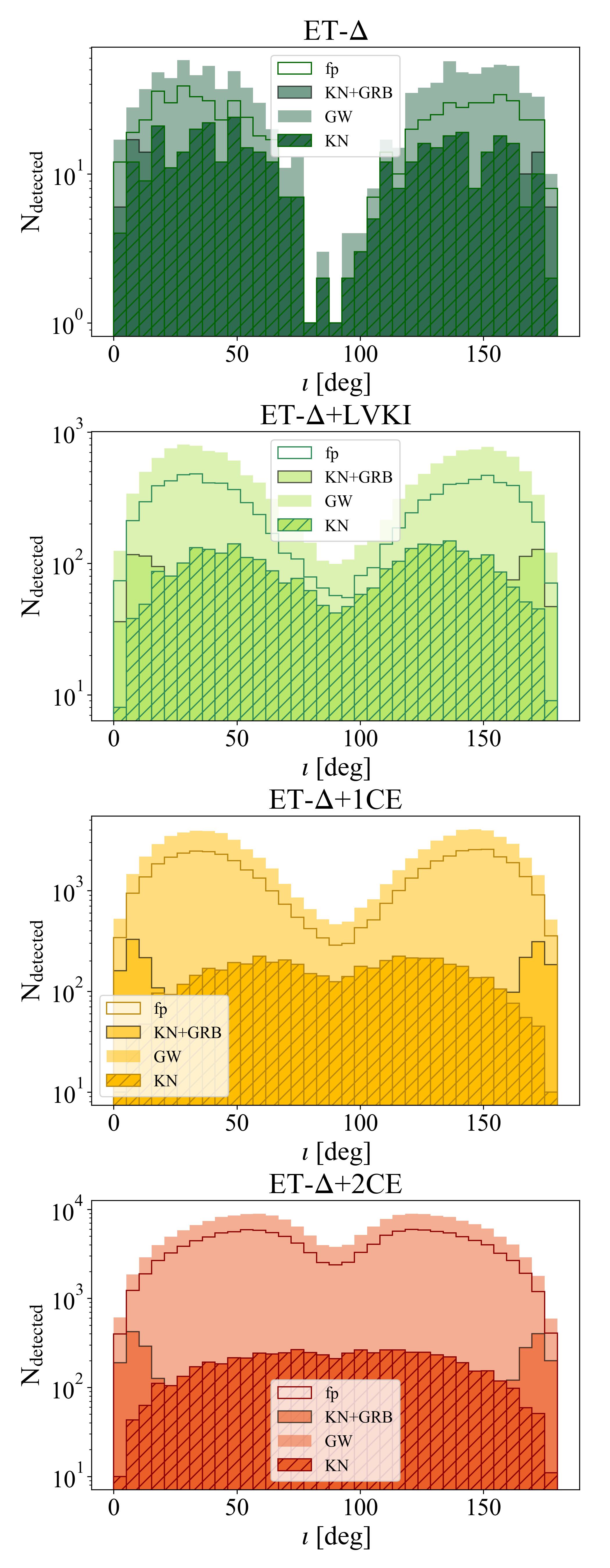}
    \end{minipage}
    \caption{Same as \reffig{fig:eff_theta_blh_uniform} but for the APR4 EOS and Gaussian NS mass distribution.}
    \label{fig:eff_theta_apr4_gaussian}
\end{figure}

\begin{figure}[]
\centering
\includegraphics[scale=0.35]{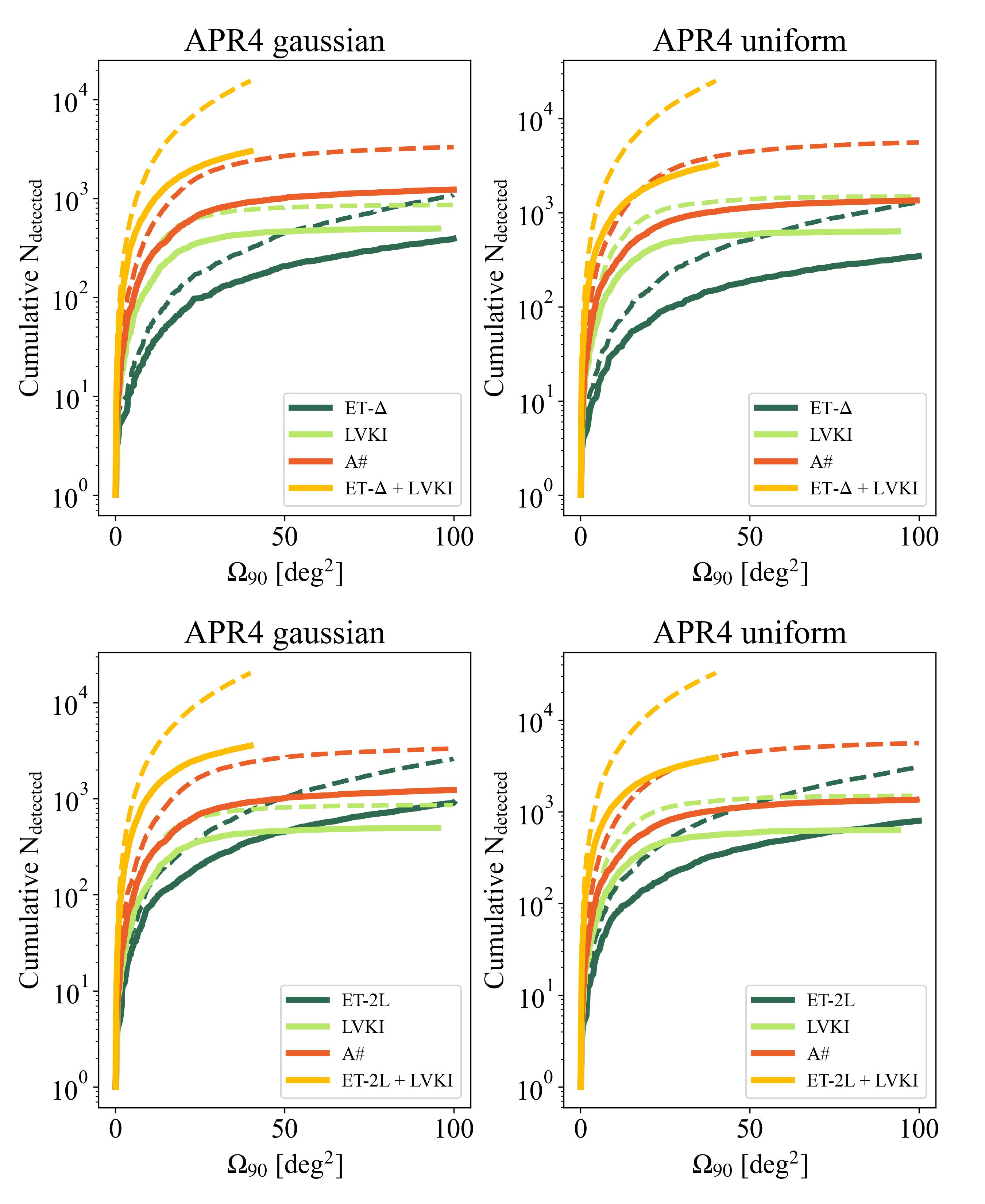} 

\caption{Same as \reffig{fig:skyloc_lvki_blh} but for the APR4 EOS.}
\label{fig:skyloc_lvki_apr4}
\end{figure}

\begin{figure}[h!]
\centering
\includegraphics[scale=0.35]{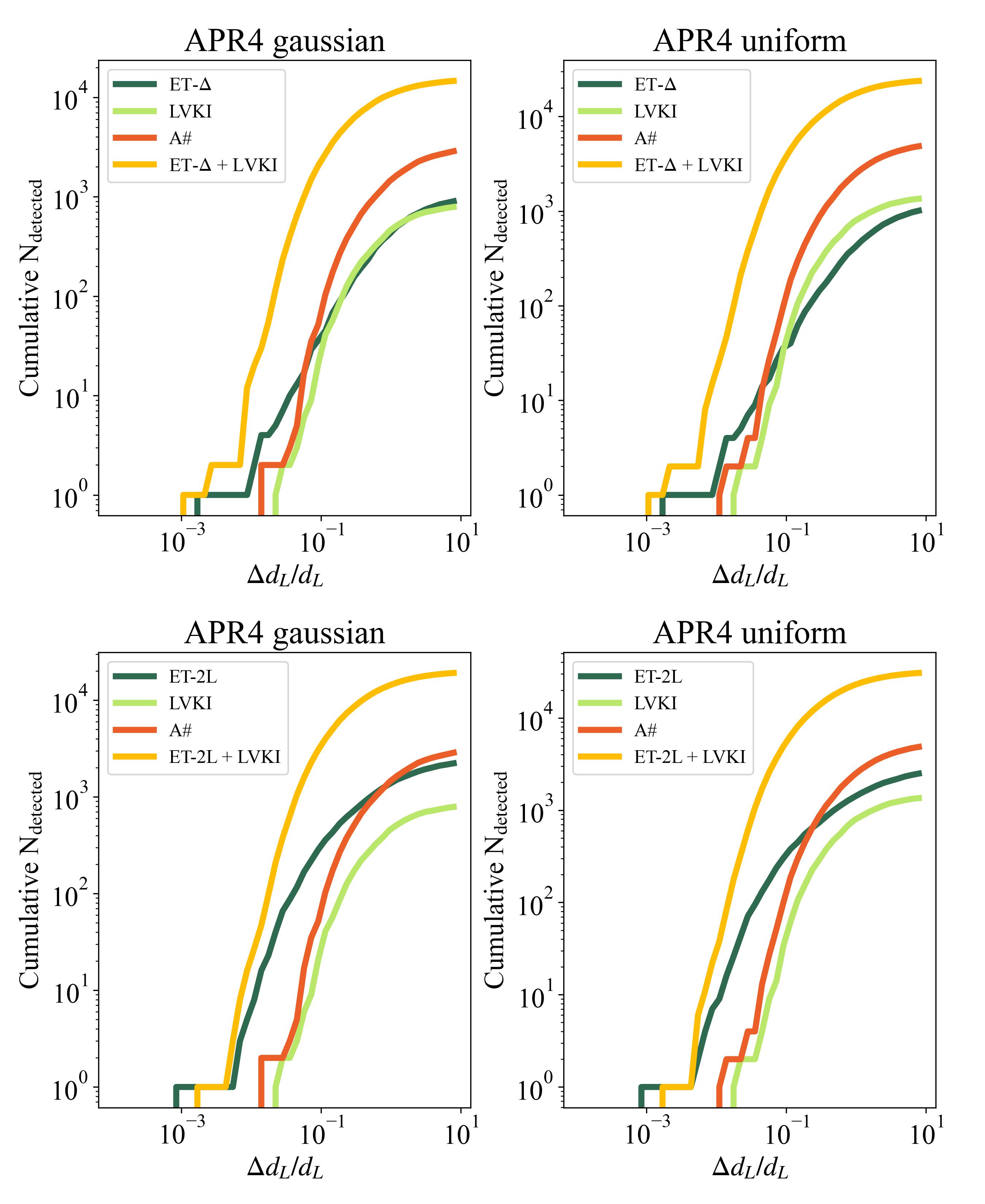} 

\caption{Same as \reffig{fig:ddlvdl_blh} but for the APR4 EOS.}
\label{fig:ddlvdl_apr4}
\end{figure}

\reffig{fig:time_number_APR4}, \reffig{fig:2L_time_number_BLh} and \reffig{fig:2L_time_number_APR4} show 
the time necessary to follow up all the events in the Rubin footprint with a sky localisation smaller than $\Omega_{90}$ (left panels) and the corresponding number of optical detections by Rubin (right panels). \reffig{fig:time_number_APR4} provides  the results obtained for ET-triangle and APR4 EOS, while \reffig{fig:2L_time_number_BLh} and \reffig{fig:2L_time_number_APR4} the results obtained for 2L ET and the BLh and APR4 EOSs, respectively. \reffig{fig:eff_theta_blh_gaussian} shows the joint GW/optical detection efficiency as a function
of redshift for the BLh EOS as in \reffig{fig:eff_theta_blh_uniform} but for the Gaussian NS mass distribution. \reffig{fig:eff_theta_apr4_uniform} and \reffig{fig:eff_theta_apr4_gaussian} show the detection efficiency for the APR4 EOS considering the uniform NS and Gaussian mass distributions, respectively. \reffig{fig:skyloc_lvki_apr4} shows the cumulative number of GW detection and corresponding KN counterparts as a function of the sky localisation for ET, and possible upgrade scenarios of current detectors, LVKI and 3A\#, considering the APR4 EOS. \reffig{fig:ddlvdl_apr4} shows the comparison of the parameter estimation capabilities of ET alone, LVKI, 3A\# and ET+LVKI in terms of the relative error on the luminosity distance, considering the APR4 EOS.

\reftab{tab:rubin_det_2L} lists the number of KN and KN plus GRB optical afterglow detected by Rubin working in synergy with the 2L ET, operating as a single observatory or in a network of GW detectors. 
In \reftab{tab:rubin_2CE}, we report the number of \ac{KN} (\ac{KN} plus GRB optical afterglow) detections obtained by ET operating in a network with two CEs and Rubin, following events with a sky localisation better than $1~{\rm deg^2}$. A comparison with \reftab{tab:rubin_det_triangle} and \reftab{tab:rubin_det_2L}, where the cut on sky localisation is set to $5~{\rm deg^2}$, shows how selecting the better-localised events ($\Omega_{90} < 1~{\rm deg^2}$) considerably reduces the requested telescope time, that stays always below $20\%$ of the observatory's time availability, while maintaining the number of detection around hundreds per year. 

\begin{figure}[]
\centering
\includegraphics[scale=0.35]{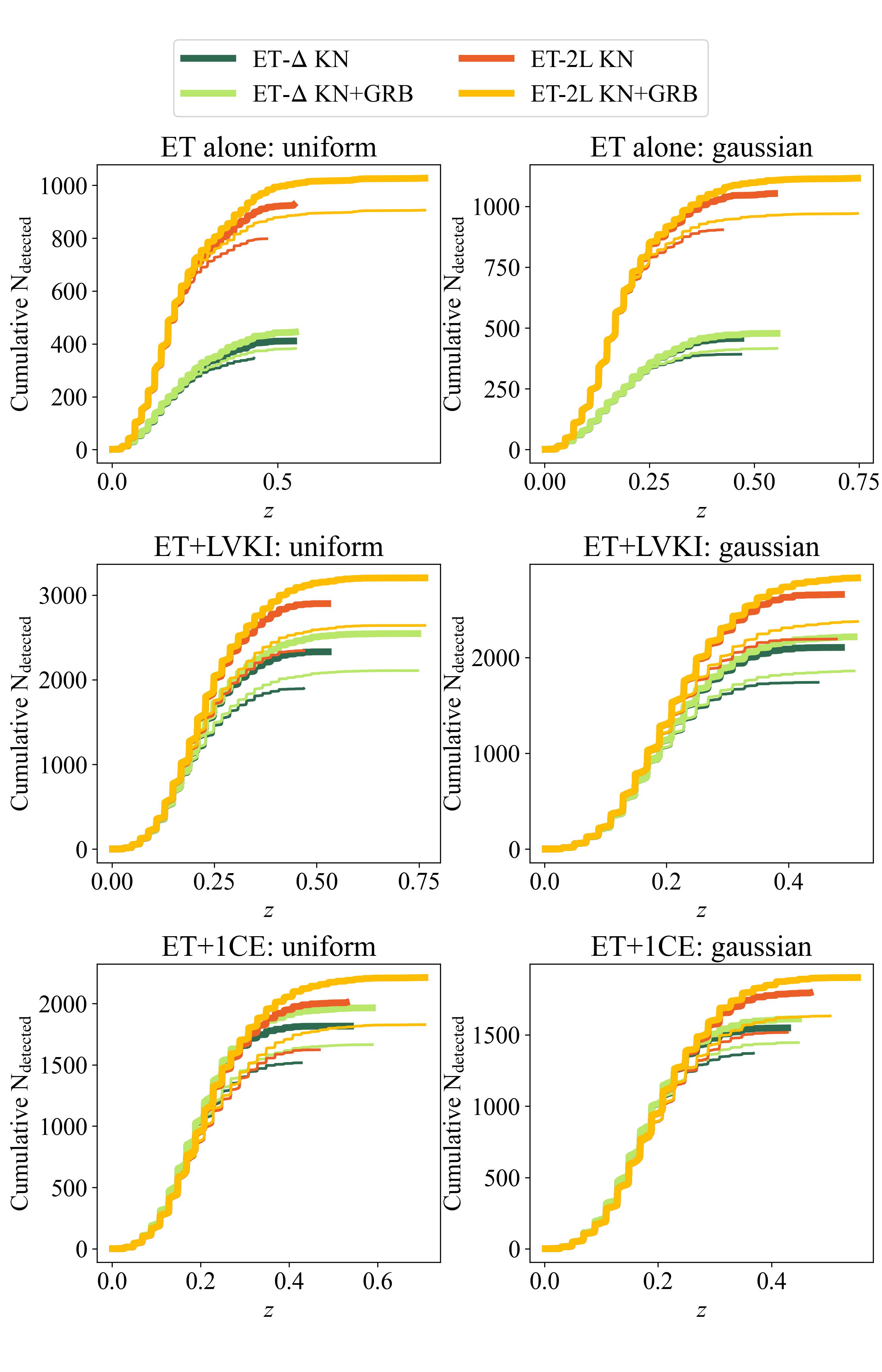} 

\caption{Same as \reffig{fig:deeper_blh} but for the APR4 EOS. 
}
\label{fig:deeper_apr4}
\end{figure}

Finally, in \reffig{fig:deeper_apr4} we compare our nominal strategy of 600 s exposure (thin lines) with deeper exposure of 1200 s (thick lines) in the case of the APR4 EOS. We show the cumulative number of KN and KN+GRB detections as a function of the redshift for ET alone (top panels), ET+LVKI (middle panels), and ET+1CE (bottom panels). Both ET configurations (triangle and 2L) are considered. The left panels show the results obtained for the uniform NS mass distribution, and the right panels the results obtained for the NS Gaussian distribution.

\begin{table*}[]
\centering
\caption{Same as \reftab{tab:rubin_det_triangle} but for the 2L-shaped ET configuration.}

\scalebox{0.8}{

\begin{tabular}{l l l l c r c r c r c r}
\hline
& & &
&\multicolumn{4}{c}{\textbf{APR4}}
&\multicolumn{4}{c}{\textbf{BLh}}\\
\hline
& &$\Omega_{90}$&Transient 
&\multicolumn{2}{c}{Uniform} 
&\multicolumn{2}{c}{Gaussian}
&\multicolumn{2}{c}{Uniform}
&\multicolumn{2}{c}{Gaussian}\\
& &[deg$^2$] & &Followed&Detected &Followed&Detected &Followed&Detected &Followed&Detected\\
& & & &&1ep/2ep &&1ep/2ep &&1ep/2ep &&1ep/2ep\\
\hline
\hline
\multirow{8}{*}{\rotatebox[origin=c]{90}{\parbox[c]{1.2cm}{\centering $\alpha = 0.5$}}}
&\multirow{2}{*}{\rotatebox[origin=c]{90}{\parbox[c]{1.2cm}{\centering ET-2L}}}
&\multirow{2}{*}{\centering 100}
&KN 
&\multirow{2}{*}{\centering 418}&$163/108$ &\multirow{2}{*}{\centering 356}&$174/122$ &\multirow{2}{*}{\centering 409}&$161/112$ &\multirow{2}{*}{\centering 352}&$216/170$\\
& & & KN+GRB &(5.2\%)&$191/131$ &(4.5\%)&$198/144$ &(5.2\%)&$188/134$ &(4.4\%)&$234/187$\\
\cline{2-12}
&\multirow{2}{*}{\rotatebox[origin=c]{90}{\parbox[c]{1.2cm}{\centering ET-2L \\\textbf{+}\\ LVKI}}}
&\multirow{2}{*}{\centering 20}
&KN 
&\multirow{2}{*}{\centering 1384}&$455/269$ &\multirow{2}{*}{\centering 872}&$386/246$ &\multirow{2}{*}{\centering 1293}&$385/237$ &\multirow{2}{*}{\centering 843}&$525/335$\\
& & &KN+GRB &(4.6\%)&$538/336$ &(2.9\%)&$446/301$ &(4.3\%)&$465/295$ &(2.8\%)&$571/390$\\
\cline{2-12}
&\multirow{2}{*}{\rotatebox[origin=c]{90}{\parbox[c]{1.2cm}{\centering ET-2L \\\textbf{+}\\ 1CE}}}
&\multirow{2}{*}{\centering 10}
&KN 
&\multirow{2}{*}{\centering 2736 }&$518/263$ &\multirow{2}{*}{\centering 1680 }&$485/253$ &\multirow{2}{*}{\centering 2511 }&$443/233$ &\multirow{2}{*}{\centering 1641 }&$707/379$\\
& & &KN+GRB &(5.7\%)&$687/366$&(3.5\%) &$593/345$ &(5.2\%)&$599/324$ &(3.4\%)&$807/473$\\
\cline{2-12}
&\multirow{2}{*}{\rotatebox[origin=c]{90}{\parbox[c]{1.2cm}{\centering ET-2L \\\textbf{+}\\ 2CE}}}
&\multirow{2}{*}{\centering 5}
&KN 
&\multirow{2}{*}{\centering 12194 }&$998/442$ &\multirow{2}{*}{\centering 10794 }&$919/395$ &\multirow{2}{*}{\centering 13549 }&$817/361$ &\multirow{2}{*}{\centering 10422 }&$1444/618$\\
& & &KN+GRB &(28.2\%)&$1410/645$ &(21.4\%)&$1285/577$ &(26.9\%)&$1218/541$ &(20.7\%)&$1800/811$\\

\hline
\hline
\multirow{8}{*}{\rotatebox[origin=c]{90}{\parbox[c]{1.2cm}{\centering $\alpha = 1.0$}}}
&\multirow{2}{*}{\rotatebox[origin=c]{90}{\parbox[c]{1.2cm}{\centering ET-2L}}}
&\multirow{2}{*}{\centering 100}
&KN 
&\multirow{2}{*}{\centering 2049}&$798/555$ &\multirow{2}{*}{\centering 1729}&$904/683$ &\multirow{2}{*}{\centering 1999}&$740/545$ &\multirow{2}{*}{\centering 1723}&$1089/883$\\
& & & KN+GRB &(25.1\%)&$906/639$ &(21.1\%)&$971/744$ &(24.6\%)&$855/624$ &(21.1\%)&$1145/936$\\
\cline{2-12}
&\multirow{2}{*}{\rotatebox[origin=c]{90}{\parbox[c]{1.2cm}{\centering ET-2L \\\textbf{+}\\ LVKI}}}
&\multirow{2}{*}{\centering 20}
&KN 
&\multirow{2}{*}{\centering 6912}&$2343/1395$ &\multirow{2}{*}{\centering 4426}&$2192/1351$ &\multirow{2}{*}{\centering 6374}&$1904/1150$ &\multirow{2}{*}{\centering 4327}&$2849/1853$\\
& & &KN+GRB &(22.9\%)&$2641/1605$ &(14.7\%)&$2375/1509$ &(21.1\%)&$2222/1347$ &(14.4\%)&$2992/2023$\\
\cline{2-12}
&\multirow{2}{*}{\rotatebox[origin=c]{90}{\parbox[c]{1.2cm}{\centering ET-2L \\\textbf{+}\\ 1CE}}}
&\multirow{2}{*}{\centering 10}
&KN 
&\multirow{2}{*}{\centering 13682 }&$2826/1465$ &\multirow{2}{*}{\centering 8371 }&$2620/1397$ &\multirow{2}{*}{\centering 12607 }&$2304/1209$ &\multirow{2}{*}{\centering 8215 }&$3770/2043$\\
& & &KN+GRB &(28.7\%)&$3397/1785$ &(17.5\%)&$2999/1647$ &(26.5\%)&$2848/1503$ &(17.2\%)&$4100/2310$\\
\cline{2-12}
&\multirow{2}{*}{\rotatebox[origin=c]{90}{\parbox[c]{1.2cm}{\centering ET-2L \\\textbf{+}\\ 2CE}}}
&\multirow{2}{*}{\centering 5}
&KN 
&\multirow{2}{*}{\centering 70978 }&$5264/2333$ &\multirow{2}{*}{\centering 53349 }&$4742/2144$ &\multirow{2}{*}{\centering 67115 }&$4378/1971$ &\multirow{2}{*}{\centering 51591 }&$7507/3288$\\
& & &KN+GRB &(140.8\%)&$6478/2866$ &(105.8\%)&$5873/2680$ &(133.1\%)&$5616/2518$ &(102.3\%)&$8597/3866$\\
\hline
\end{tabular}
}
\label{tab:rubin_det_2L}
\end{table*}

\begin{table*}[]
\renewcommand{\arraystretch}{1.5}
\caption{Number of KNe and KNe + GRB detected by Rubin, operating in synergy with \ac{ET} + 2 CE for the two observational strategies \textit{1ep} and \textit{2ep} and following up events with sky localisation $\Omega_{90} < 1~{\rm deg^2}$.}\label{tab:rubin_2CE}
\centering
\begin{adjustbox}{center, max width=1.1\textwidth, scale=0.85}
\begin{tabular}{l l l l c r c r c r c r}
\hline
& & &
&\multicolumn{4}{c}{\textbf{APR4}}
&\multicolumn{4}{c}{\textbf{BLh}}\\
\hline
& &$\Omega_{90}$&Transient 
&\multicolumn{2}{c}{Uniform} 
&\multicolumn{2}{c}{Gaussian}
&\multicolumn{2}{c}{Uniform}
&\multicolumn{2}{c}{Gaussian}\\
& &[deg$^2$] & &Followed&Detected &Followed&Detected &Followed&Detected &Followed&Detected\\
& & & &&\textit{1ep}/\textit{2ep} &&\textit{1ep}/\textit{2ep} &&\textit{1ep}/\textit{2ep} &&\textit{1ep}/\textit{2ep}\\
\hline
\hline
\multirow{4}{*}{\rotatebox[origin=c]{90}{\parbox[c]{1.5cm}{\centering $\alpha = 0.5$}}}
&\multirow{2}{*}{\rotatebox[origin=c]{90}{\parbox[c]{1.2cm}{\centering ET-$\Delta$ \\\textbf{+}\\ 2CE}}}
&\multirow{2}{*}{\centering 1}
&KN 
&\multirow{2}{*}{\centering 1609 }&$430/227$ 
&\multirow{2}{*}{\centering 976 }&$362/206$ 
&\multirow{2}{*}{\centering 1503}&$350/196$ 
&\multirow{2}{*}{\centering 953}&$529/293$\\
& & &KN+GRB &(3.2\%)&$528/299$ &(1.9\%)&$424/260$ &(3.0\%)&$448/261$ &(1.9\%)&$582/349$\\
\cline{2-12}
&\multirow{2}{*}{\rotatebox[origin=c]{90}{\parbox[c]{1.2cm}{\centering ET-2L \\\textbf{+}\\ 2CE}}}
&\multirow{2}{*}{\centering 1}
&KN 
&\multirow{2}{*}{\centering 1830 }&$481/251$ &\multirow{2}{*}{\centering 1149 }&$426/235$ &\multirow{2}{*}{\centering 1677 }&$393/214$ &\multirow{2}{*}{\centering 1104 }&$586/337$\\
& & &KN+GRB &(3.6\%)&$605/341$ &(2.3\%)&$501/300$ &(3.3\%)&$502/283$ &(2.2\%)&$652/403$\\
\hline
\hline
\multirow{4}{*}{\rotatebox[origin=c]{90}{\parbox[c]{1.5cm}{\centering $\alpha = 1.0$}}}
&\multirow{2}{*}{\rotatebox[origin=c]{90}{\parbox[c]{1.2cm}{\centering ET-$\Delta$ \\\textbf{+}\\ 2CE}}}
&\multirow{2}{*}{\centering 1}
&KN 
&\multirow{2}{*}{\centering 8162 }&$2266/1266$ &\multirow{2}{*}{\centering 4956 }&$2086/1195$ &\multirow{2}{*}{\centering 7398 }&$1868/1024$ &\multirow{2}{*}{\centering 4824 }&$2848/1673$\\
& & &KN+GRB &(16.2\%)&$2595/1476$ &(9.8\%)&$2301/1359$ &(14.7\%)&$2192/1231$ &(9.6\%)&$3017/1845$\\
\cline{2-12}
&\multirow{2}{*}{\rotatebox[origin=c]{90}{\parbox[c]{1.2cm}{\centering ET-2L \\\textbf{+}\\ 2CE}}}
&\multirow{2}{*}{\centering 1}
&KN 
&\multirow{2}{*}{\centering 8818 }&$2401/1322$ &\multirow{2}{*}{\centering 5502 }&$2176/1250$ &\multirow{2}{*}{\centering 8147 }&$1954/1087$ &\multirow{2}{*}{\centering 5367 }&$3006/1724$\\
& & &KN+GRB &(17.5\%)&$2735/1533$ &(10.9\%)&$2402/1416$ &(16.2\%)&$2301/1306$ &(10.6\%)&$3191/1910$\\

\hline
\end{tabular}
\end{adjustbox}
\tablefoot{Both the triangular and 2L-shaped ET configurations are considered. The analysis is performed using different EOSs (APR4 or BLh), NS mass distributions (uniform or Gaussian), and $\alpha$ values (0.5 and 1.0).
For each scenario, we report the total number of followed events (within Rubin's footprint), the corresponding telescope time required for follow-up (expressed as a percentage below the number of followed events), and the number of detected events. All listed data refer to 10 years of observations.}
\end{table*}

\end{appendix}

\end{document}